\documentclass{aa}  

\usepackage{graphicx}
\usepackage{txfonts}
\usepackage{booktabs}
\usepackage{pdflscape}
\usepackage{longtable}  
\usepackage{placeins}
\usepackage{float}
\usepackage[colorlinks=false, linkcolor=blue]{hyperref}
\usepackage{afterpage} 
\usepackage{orcidlink}

\begin{document}

\title{Reconstructing the orbits of Milky Way dwarf galaxies: A Large Magellanic Cloud perspective}
\titlerunning{Orbit integration of MW dwarfs}
\authorrunning{A. M. Martínez-García et al.}

\author{Alberto Manuel Martínez-García\inst{\ref{IAA}}\orcidlink{0000-0001-9755-3872}
        \and Andrés del Pino\inst{\ref{IAA}}\orcidlink{0000-0003-4922-5131}
		\and Roeland P. van der Marel\inst{\ref{STScI},\ref{JHU}}\orcidlink{0000-0001-7827-7825}
		\and Giuseppina Battaglia\inst{\ref{IAC}, \ref{ULL}}\orcidlink{0000-0002-6551-4294}
		\and Ewa L. Łokas\inst{\ref{CAMK}}\orcidlink{0000-0001-7138-8899}
		\and Eduardo Vitral\inst{\ref{ROE}}\thanks{Royal Society Newton International Fellow}\orcidlink{0000-0002-2732-9717}
		\and Kevin A. McKinnon\inst{\ref{UT}}\orcidlink{0000-0001-7494-5910}
        \and Laura L. Watkins\inst{\ref{AURA}}\orcidlink{0000-0002-1343-134X}
        \and Nitya Kallivayalil\inst{\ref{UV}}\orcidlink{0000-0002-3204-1742}
		\and Sangmo Tony Sohn\inst{\ref{STScI},\ref{KHU}}\orcidlink{0000-0001-8368-0221}	
		\and Guillaume F. Thomas\inst{\ref{ULL},\ref{IAC}}\orcidlink{0000-0002-2468-5521}
		\and Salvador Cardona-Barrero\inst{\ref{ULL},\ref{IAC}}\orcidlink{0000-0002-9990-4055}
		\and Borja Anguiano\inst{\ref{CEFCA}}\orcidlink{0000-0001-5261-4336}
		\and Jairo A. Alzate-Trujillo\inst{\ref{IAA}}\orcidlink{0000-0002-8351-8854}
		\and Francisco Nogueras-Lara\inst{\ref{IAA}}\orcidlink{0000-0002-6379-7593}
		\and Paul Bennet\inst{\ref{STScI}}\orcidlink{0000-0001-8354-7279}
		\and Adrián Hidalgo-Pinilla\inst{\ref{CEFCA}}\orcidlink{0009-0003-1933-4529}
        }

\institute{Instituto de Astrofísica de Andalucía, CSIC, Glorieta de la Astronomía, 18008 Granada, Spain\label{IAA} \email{ammartinez@iaa.csic.es}
          \and Space Telescope Science Institute, 3700 San Martin Drive, Baltimore, MD 21218, USA\label{STScI}
          \and Center for Astrophysical Sciences, The William H. Miller III Department of Physics \& Astronomy, Johns Hopkins University, Baltimore, MD 21218, USA\label{JHU}
          \and Instituto de Astrof\'isica de Canarias, calle Vía L\'actea s/n, E-38205 La Laguna, Tenerife, Spain\label{IAC}
          \and Departamento de Astrof\'isica, Universidad de La Laguna, Avenida Astrof\'isico Francisco S\'anchez s/n, E-38206 La Laguna, Spain\label{ULL}
          \and Nicolaus Copernicus Astronomical Center, Polish Academy of Sciences, Bartycka 18, 00-716 Warsaw, Poland\label{CAMK}
          \and Institute for Astronomy, University of Edinburgh, Royal Observatory, Blackford Hill, Edinburgh EH9 3HJ, UK\label{ROE}
          \and David A. Dunlap Department of Astronomy \& Astrophysics, University of Toronto, 50 St George Street, Toronto, ON M5S 3H4, Canada\label{UT}
          \and AURA for the European Space Agency (ESA), Space Telescope Science Institute, 3700 San Martin Drive, Baltimore, MD 21218, USA\label{AURA}
          \and Department of Astronomy, University of Virginia, 530 McCormick Road, Charlottesville, VA 22904, USA\label{UV}
          \and Dept. of Astronomy \& Space Science, Kyung Hee University, Gyeonggi-do 17104, Republic of Korea\label{KHU}
          \and Centro de Estudios de Física del Cosmos de Aragón (CEFCA), Plaza San Juan 1, 44001 Teruel, Spain\label{CEFCA}
          }

\date{Received XXXXXX; accepted XXXXX}

\abstract
{
The orbital histories of the dwarf satellites of the Milky Way (MW) are key to understanding their evolution and placing their present-day properties in a dynamical context. We present the results of the orbit integration of 72 dwarfs in the vicinity of the MW, based on accurate 6D phase-space coordinates from the literature  and a suite of six realistic, time-evolving gravitational potentials that account for the mutual interaction of the MW and the Large Magellanic Cloud (LMC). We provide the largest catalogue of orbital parameters for MW dwarfs to date, in terms of both galaxy sample size and the range of potentials explored. We also assess the binding status of the dwarfs and estimate their infall times, finding that the majority of them have spent the last 5 Gyr within the MW virial radius. From the reconstructed orbits, we identify ten likely LMC satellites, several of which have experienced very close passages within the LMC stellar disc. For the Small Magellanic Cloud (SMC), we find that its most recent pericentre about the LMC ($\sim$8 kpc, $\sim$170 Myr ago) is consistent with predictions from the direct collision scenario proposed to explain the LMC's offset and tilted bar. We also note a broad temporal coincidence between previous SMC pericentres and star formation rate peaks reported in both Magellanic Clouds, suggesting a causal connection. Finally, we identify Grus II and Tucana IV as possible MW satellites recently captured by the LMC, based on their pronounced orbital deflections and velocities relative to the LMC.
}


\maketitle

\section{Introduction}
Dwarf galaxies account for the vast majority of galaxies in the Universe. In the vicinity of our own Galaxy, the Milky Way (MW), about 70 dwarfs have been identified as part of its satellite population (e.g. \citealt{Doliva-Dolinsky2025, Tan2025b, Pace2024}). This census is expected to continue growing in the near future thanks to current and upcoming wide-field surveys, such as the Ultraviolet Near-Infrared Optical Northern Survey (UNIONS; \citealt{Gwyn2025}) and the Legacy Survey of Space and Time (LSST; \citealt{Ivezic2019}), as well as space-based observatories like \textit{Euclid} (\citealt{EuclidCollaboration2022}) and the \textit{Nancy Grace Roman Space Telescope} (\textit{Roman}, \citealt{Spergel2015}). Forecasts suggest that these programmes will lead to the discovery of dozens of additional MW satellites (e.g. \citealt{Newton2023, Nadler2024, Tsiane2025, Ahvazi2025}), reinforcing the status of the MW as the best laboratory for studying  dwarf galaxies, given their proximity and abundance. Understanding the present-day properties of these dwarfs, however, requires knowledge of their past dynamical evolution, for which reconstructing their orbital histories is essential.

Orbit integration provides information on the shape of satellite trajectories, their pericentre distances, and the frequency of pericentric passages, thereby constraining  the strength of interactions with the MW and with other galaxies. Such interactions are critical for interpreting observed features such as truncations or peaks of star formation (e.g. \citealt{Rusakov2021, Massana2022, Bennet2024, Bennet2025}) or present-day internal kinematics (\citealt{Battaglia2022b,MartinezGarcia2023a, MartinezGarcia2023b}). Beyond individual galaxies, orbital reconstructions also enable tests of whether apparent satellite groupings share a common origin (\citealt{Julio2024}) and whether anisotropies in the MW satellite distribution, such as the Vast Polar Structure (VPOS; \citealt{Pawlowski2021}),  represent transient or long-lived configurations (\citealt{Sawala2023b,Taibi2024, Kumar2025, MartinezGarcia2025}). Orbit integration thus serves as a powerful bridge between the observed present-day properties of dwarf galaxies and their environmental and dynamical  histories within the MW halo. However, despite its importance, orbit reconstruction remains challenging, as it is highly sensitive to both the accuracy of the initial phase-space coordinates and the assumed Galactic potential.

The initial phase-space coordinates correspond to the present-day values of the three-dimensional positions and velocities of the satellites.  
Deriving these quantities involves different observational challenges. Positions on the sky (right ascension and declination) can be measured with high precision and have been available for decades (\citealt{Wilson1955}), although refinements are still being reported (e.g. \citealt{Munoz2018, delPino2021}). Distances to MW satellites are considerably more uncertain, typically inferred from standard candles such as RR Lyrae stars or the tip of the red giant branch (e.g. \citealt{MartinezVazquez2015, MartinezVazquez2021}), yet they can also be inferred from their 3D kinematics (\citealt{Vitral2024, Vitral2025}), with associated uncertainties often comparable to, or even larger than, the physical size of the satellites themselves (see \citealt{McConnachie2012, Pace2024}). 

For the velocity components, line-of-sight velocities (LOSVs) can be obtained through spectroscopy, but proper motions (PMs) are far more difficult to measure, requiring high-precision astrometry over long time baselines. Early pioneering efforts relied on photographic plates to derive the PMs of the MW satellites (\citealt{Jones1994, ScholzIrwin1994, Schweitzer1995}), while high-accuracy PMs began to emerge with the advent of the Hubble Space Telescope (\textit{HST}; \citealt{Piatek2002, Piatek2003, Piatek2005, Kallivayalil2006b, Kallivayalil2006, Sohn2013, Sohn2017, Julio2024, Vitral2024, Vitral2025}). Another major player that has revolutionized the field over the last few years is the \textit{Gaia} mission (\citealt{GaiaCollaboration2016}), which has provided accurate PMs for individual stars in the majority of known MW satellites, and thus enabled their systemic PMs to be derived (\citealt{Fritz2018, McConnachie2020a, McConnachie2020b, Li2021, Vitral2021, Pace2022, Battaglia2022}). While \textit{Gaia} has opened a new era in the study of satellite kinematics, \textit{HST} continues to play a crucial role by measuring motions of the faintest and most distant satellites inaccessible to \textit{Gaia} (\citealt{Bennet2025}). However, both observatories have proved to be complementary tools, as several studies have combined their data to improve PMs accuracy by extending the time baselines of the observations (\citealt{Massari2018, Massari2020, delPino2022, Warfield2023, Warfield2026, McKinnon2024}). \textit{Gaia} observations have also been paired with \textit{JWST} (\citealt{Libralato2023, Libralato2024b}) and \textit{Euclid} (\citealt{Libralato2024}), while prospects for further astrometric improvements through its combination with the 
upcoming \textit{Roman Space Telescope} have also been discussed \citep{McKinnon2026}.

After 6D phase-space coordinate uncertainties, the second dominant source of error in orbital reconstructions arises from the assumed Galactic potential (see \citealt{Pace2022}). Although progressively refined, estimates of the MW’s total mass remain loosely constrained, with values around ($0.5$-$2$)$\times10^{12}$ M$_{\odot}$ depending on tracer populations and modelling techniques (see \citealt{Wang2020, BobylevBaykova2023, Bayer2025} and references therein). The shape and radial profile of the dark matter (DM) halo are likewise uncertain, while the baryonic components (such as the bar and disc) add further complexity. 
The problem is compounded by the MW’s massive neighbour, the Large Magellanic Cloud (LMC). With a mass of  $\sim$($1$–$2$)$\times10^{11}$ M$_{\odot}$ (see e.g. \citealt{Kallivayalil2013, Penarrubia2016, Laporte2018, Erkal2021, Shipp2021, Vasiliev2021,Sawala2023, Vasiliev2024, Watkins2024} and references therein), the LMC  significantly perturbs the Galactic potential, affecting halo stars, stellar streams, and satellites alike (e.g., \citealt{Vasiliev2023}). Properly accounting for this influence is therefore essential for realistic orbit integration. Different methods have been developed to include the effect of the LMC in the  orbit reconstruction, ranging from using static MW + LMC potentials (e.g. \citealt{Patel2020}) to fully dynamical models tracing the mutual interaction via N-body simulations (\citealt{Vasiliev2021, Vasiliev2023, Vasiliev2024}).

In recent years, several works on orbit integration have included the effect of the LMC (e.g. \citealt{Patel2020, Battaglia2022, CorreaMagnus2022, Pace2022, Vasiliev2024}), consistently finding a strong impact on the reconstructed trajectories. This clearly supports the idea  that its inclusion is essential for accurate orbit reconstruction. Yet, the variety of MW+LMC models explored to date remains limited, leaving ample room for further exploration of alternative realistic configurations and their effects on satellite dynamics.

In this paper, we present the results of the orbit integration of 72 dwarf galaxies in the vicinity of the MW, based on accurate phase-space coordinates from the literature and a suite of six time-evolving gravitational potentials that account for the mutual interaction between the MW and the LMC. This allows us to provide the largest catalogue of dwarf galaxy orbital parameters with respect to the MW to date, covering the widest range of realistic MW+LMC configurations for which such parameters have been reported so far.
We then exploit the explicit LMC component of our potentials to go beyond a MW-centric view of satellite dynamics, assessing possible associations between individual dwarfs and the LMC. For those identified as likely satellites, we derive their orbital parameters relative to LMC. We further investigate the mutual interaction of the LMC with the Small Magellanic Cloud (SMC), and the possibility that the LMC has recently captured any of the MW satellites.

\section{Methods}
\label{sec:methods}
\subsection{Sample of dwarf galaxies and initial conditions}

 \begin{figure*}[h]
    \centering
    \includegraphics[width=\textwidth]{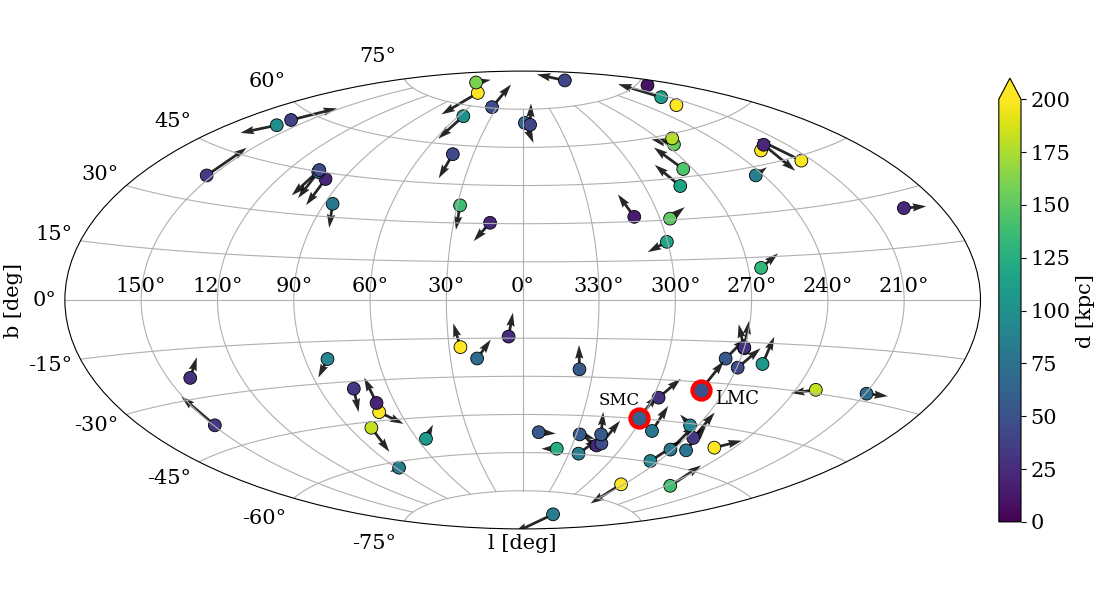}
    \caption{Distribution of the dwarf galaxy sample in Galactic coordinates. Each point represents the position of a  galaxy, colour-coded by its heliocentric distance. Arrows indicate the direction and relative magnitude of the PM vector in Galactic longitude and latitude, corrected for the reflex motion induced by the Sun's motion with respect to the Galactic centre. The SMC, part of the sample, and the LMC, included for reference, are each outlined in red for identification. LMC phase-space coordinates were adopted from \citet{vanderMarel2002, GaiaCollaboration2021LMC, Pietrzynski2019}.}
    \label{fig:sample}
 \end{figure*}
 
In this work, we focus on a sample of dwarf galaxies in the vicinity of the MW for which  3D positions (i.e., right ascension, declination and distance), and 3D systemic velocities (PMs and LOSVs) are available in the literature. We limit the sample to galaxies within 500 kpc of the MW. Galaxies located beyond this threshold are highly unlikely to be associated with the MW-system, and thus the integration of their orbits would require modelling the whole Local Group (see \citealt{McConnachie2021, Bennet2024, Bennet2025}).
When gathering kinematic information for MW dwarf galaxies, we prioritise works that have derived accurate systemic PMs for large numbers of galaxies, as this ensures we have a dataset as homogeneous and consistent as possible. Where a galaxy is not covered by such works, we supplement the sample with measurements from individual studies.

Our sample primarily relies on \citet{Battaglia2022}, the study that has derived systemic PMs for the largest number of Local Group galaxies to date (a total of 73) using the \textit{Gaia} Early Data Release 3 (EDR3; \citealt{GaiaEDR3}), and that also provides us with an exhaustive compilation of coordinates, distances, and LOSVs from the literature. We select a total of 50 dwarfs with full phase-space coordinates available and located within 500 kpc of the MW. We note that for Pisces II and Tucana V, we adopt the PMs derived using a spectroscopic prior (Table B.3), rather than the astrometry-only derivations we adopted for the rest of the sample (Table B.2), following the recommendation of \citet{Battaglia2022}, who flagged the latter as not reliable enough to derive accurate orbits for these two galaxies.

We complement this sample by including  other recently discovered dwarfs around the MW for which their positions and velocities have been reported. This is the case for Aquarius III (\citealt{Cerny2025}), Bootes V (\citealt{Smith2023, Cerny2023}), Eridanus IV (\citealt{Cerny2021}), Leo VI (\citealt{Tan2025a}), and Pegasus IV (\citealt{Cerny2023b}).
We also include Centaurus I, Pictor II, and Pegasus III, three galaxies that are part of the \citet{Battaglia2022} dataset but could not be included from that source: the first two lacked LOSVs measurements at the time, which have since been published (\citealt{Heiger2024, Pace2025}), while the PM of Pegasus III was flagged as unreliable for orbit integration, and we instead adopt the measurement reported in \citet{Pace2022}. Finally, we add to our sample two other well known MW satellites, the Sagittarius dSph galaxy and the SMC. For all these additional galaxies we obtained their phase-space coordinates from the \textit{Local Volume Database}\footnote{The phase-space coordinates of the galaxies we selected from \citet{Pace2024} were originally reported in the following studies: Aquarius III \citep{Cerny2025}, Bootes V \citep{Smith2023, Cerny2023}, Centaurus I \citep{MartinezVazquez2021, Heiger2024, Casey2025}, Eridanus IV \citep{Cerny2021, Heiger2024, Casey2025}, Leo VI \citep{Tan2025a}, Pegasus III (\citealt{Kim2016, Pace2022, Richstein2022, Geha2026}), Pegasus IV \citep{Cerny2023b, Geha2026}, Pictor II \citep{Pace2025}, Sagittarius \citep{McConnachie2012, An2024}, and the SMC \citep{Cioni2000, Harris2006, Munoz2018, Zivick2018}.} (\citealt{Pace2024}).

Very recently, \citet{Cerny2026} published an extensive catalogue of phase-space coordinates for nearly twenty ultra-faint compact satellites, systems whose nature is still unclear, and it is debated whether they are globular clusters (GC) or ultra-faint dwarfs (UFD). From this sample, we discard the systems with conclusive evidence of being GCs (labelled as `Definite Star Cluster' and `Very Likely Star Cluster' in Table 5 of \citealt{Cerny2026}), and keep the rest  for orbital integration. For simplicity, we will refer to all the considered systems as dwarf galaxies throughout the text, but we do recommend the reader to be aware that the nature of some of them, in particular the very faint and low mass ones, is still unclear (e.g. see \citealt{Arroyo-Polonio2026}). 

Our final sample contains a total of 72 dwarf galaxies. In Figure~\ref{fig:sample} we display their distribution in Galactic coordinates.
For all the dwarfs except the SMC, the PMs were derived using \textit{Gaia} EDR3/DR3 (\citealt{Battaglia2022, Pace2024, Cerny2026}).
The astrometry of \textit{Gaia} EDR3/DR3 is known to suffer from systematic effects on both large and small angular scales (\citealt{Lindegren2020}). These systematics can bias the measured PMs, thereby affecting the accuracy of orbit integrations. Accounting for such effects is therefore essential to ensure the reliability of our results. We follow a procedure analogous to that described in Section 5 of \citet{Battaglia2022}, which we outline below, to correct for large-scale systematics and to incorporate the impact of small-scale systematics into the PM uncertainties.
The PM of the SMC, however, was derived using \textit{HST} \citep{Zivick2018}, as compiled in \citet{Pace2024}. Since \citet{Pace2024} only reports the random uncertainties for this measurement, we combine the random and systematic components provided in the original study in quadrature, to obtain a more comprehensive estimate of the total uncertainty.

To assess the effect of the large-scale systematics in \textit{Gaia} EDR3 astrometry, we calculate the average PMs of the quasi-stellar objects (QSOs) in the vicinity of each dwarf. This astrometric zero-point is then subtracted from the reported systemic PMs. \citet{Battaglia2022} already calculated these zero-points using the QSO dedicated table of \textit{Gaia} EDR3, but they are not available for data from \citet{Pace2024} or \citet{Cerny2026}. Regardless of the source of data, we calculate the zero-point using the newly available QSO candidate table released with $Gaia$ DR3 (\texttt{qso\_candidate}, \citealt{GaiaDR3QSO}). We search for QSO candidates within 7 degrees around the galaxies (a region large enough to encompass a sizeable sample of QSOs, $\sim10^3$) and then we apply the recommended selection that provides a 95\% purity sample as described in \citealt{GaiaDR3QSO}) (see their ADQL query in Table 11). In short, we require sources to meet at least one of the following criteria:  being a source used in the derivation of the Gaia Celestial Reference Frame (\texttt{gaia\_crf\_source=`true'}), having no close neighbours detected by the Extended Object analysis (\texttt{host\_galaxy\_flag}~$< 6$), being classified as a quasar by the Discrete Source Classifier (\texttt{classlabel\_dsc\_joint=`quasar'}), or being identified as an AGN by the variability analysis module (\texttt{vari\_best\_class\_name=`AGN'}).
We impose some additional criteria to the selected QSO candidates to further refine the samples. First, we only consider QSO candidates with 5-parameter solutions, as they are reported to be more precise than the 6-parameter (\citealt{Fabricius2020, GaiaEDR3, Lindegren2020b}). We further restrict the sample to QSO candidates with  $G < 19$ in order to reduce the statistical errors, and also impose \texttt{ruwe} $< 1.4$, \texttt{ipd\_gof\_harmonic\_amplitude} $< 0.2$, \texttt{ipd\_frac\_multi\_peak} $\leq 2$ to screen out possible non-single objects (\citealt{Fabricius2020}). The resulting samples are further refined by recursively filtering potential outliers at $5\sigma$ in the PM space.
Finally, the astrometric PM zero-point and its associated uncertainty are computed as the uncertainty-weighted average of the QSO sample PMs for each galaxy, and subsequently subtracted from the corresponding galaxy's systemic PM.

As for the small-scale systematics, we account for their impact on the uncertainties of the systemic PMs using the Equation 2 from \citet{VasilievBaumgardt2021}, taking as angular separation the half-light radius of the galaxies. As the zero-points obtained through the QSO samples are also affected by these small-scale systematics, we proceed likewise over the 7 deg scale used.

\subsection{Potentials}
\label{sec:pot}

\begin{table*}[t]
    \centering
    \caption{Initial parametrizations of the MW and LMC models}
    \label{tab:models}

    \begin{tabular}{lccc lccc}
    \toprule

      & \multicolumn{3}{c}{MW halo} 
      & \multicolumn{3}{r}{LMC halo} \\

    \cmidrule(lr){2-4}
    \cmidrule(lr){6-8}

    Model 
    & $r_{\rm vir}$ 
    & $M_{\rm vir}$ 
    & $M_{\rm tot}$ 
    
    & Model 
    & $r_{\rm vir}$ 
    & $M_{\rm vir}$ 
    & $M_{\rm tot}$  \\

    & [kpc]
    & [$10^{11}\,M_\odot$]
    & [$10^{11}\,M_\odot$]
    &
    & [kpc]
    & [$10^{11}\,M_\odot$]
    & [$10^{11}\,M_\odot$]\\

    \cmidrule(lr){1-4}
    \cmidrule(lr){5-8}
    V23 & 243 &  8.1 & 8.5  & V23 & 137 & 1.4 & 1.5 \\
    M10 & 260 & 10.0 & 11.8 & L2 & 150 & 1.9 & 2.0\\
    M11 & 268 & 11.0 & 12.9 & L3 & 169 & 2.8 & 3.0  \\

    \bottomrule
    \end{tabular}

    \tablefoot{Virial radius ($r_{\rm vir}$ ), virial mass ($M_{\rm vir}$), and total mass ($M_{\rm tot}$) of the DM haloes of the different MW and LMC models, prior to their interaction. All MW models share the same baryonic component, consisting of an exponential disc of $5 \times 10^{10}\, M_\odot$ and a spherical bulge of $1.2 \times 10^{10}\, M_\odot$, that are held fixed throughout the interaction. The LMC has no baryonic component.}
\end{table*}

We use a suite of six time-evolving gravitational potentials that consider both the MW and the LMC, as presented in \citet{Vasiliev2023, Vasiliev2024}, allowing us to explore the orbits of MW dwarf galaxies under a wide range of realistic MW+LMC configurations. These potentials were derived from N-body simulations in which the DM haloes of the MW and the LMC evolve  under their mutual gravitational influence, and are represented by multipole expansions.

The first potential, referred to hereafter as V23, was presented in \citet{Vasiliev2023}. It consists of a MW model formed by an exponential disc of total mass $5 \times 10^{10}$ M$_{\odot}$, a spherical bulge of $1.2 \times 10^{10}$ M$_{\odot}$, and a a triaxial DM halo of $8 \times 10^{11}$ M$_{\odot}$. The MW baryonic component remains unchanged over time. The LMC is modelled by a truncated Navarro-Frenk-White (NFW, \citealt{NFW1996}) profile with initial mass $1.5 \times 10^{11}$ M$_{\odot}$, scale radius of 10.84 kpc, and truncation radius of 108.4 kpc, with no associated baryonic component. The stellar component of the LMC is not considered since its contribution to the total mass is subdominant ($\sim 3 \times 10^9$ M$_{\odot}$, \citealt{vanderMarel2002}) 
In this model, the LMC performs a single pericentre  passage about the MW \footnote{The trajectory of the LMC is derived from the simulation of its interaction with the MW, in which the LMC initial conditions are iteratively adjusted until its present-day position and velocity match the observational constraints to within 0.2 kpc and 1 km s$^{-1}$, respectively. The adopted constraints are $(\alpha, \delta) = (81^{\circ}, -69.75^{\circ})$, $\mu_{\alpha} = 1.8$ mas yr$^{-1}$, $\mu_{\delta} = 0.35$ mas yr$^{-1}$ \citep{Vasiliev2018}, $d = 50$ kpc \citep{Freedman2001}, and $v_\mathrm{los} = 260$ km s$^{-1}$ \citep{vanderMarelKallivayalil2014}. We address the reader to \citealt{VasilievBelokurov2020, Vasiliev2021} for a more detailed description and discussion on this procedure.}. The masses of the MW and LMC haloes evolve over time, as a result of the interaction between them, especially during the pericentre. 
This model is the same as the evolving triaxial MW + $1.5 \times 10^{11}$ M$_{\odot}$ LMC potential described in \citet{Vasiliev2021}, that successfully fits the Sagittarius stream and that was also used in \citet{Battaglia2022} for orbit integration, with the exception that it allows us to explore longer integration times thanks to the longer simulation time of the MW-LMC interaction. 

\begin{figure}
    \centering
    \includegraphics[width=\linewidth]{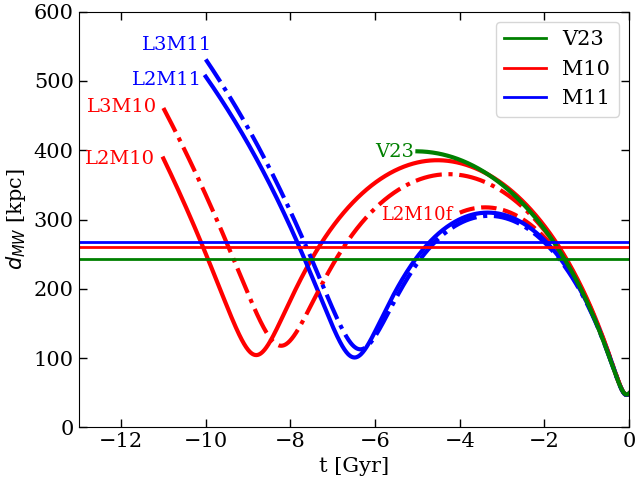}
    \caption{LMC trajectories in the different potentials. Each curve shows the Galactocentric distance of the LMC as a function of time for a given potential, colour coded by the MW model combination. Horizontal lines represent the MW virial radius for the different MW models.}
    \label{fig:LMCtraj}
\end{figure}

The rest of the potentials are combinations of the MW and LMC models presented in \citet{Vasiliev2024}, in which the LMC trajectory has two pericentres\footnote{Similarly to V23, the initial conditions of the LMC are iteratively adjusted for each MW+LMC combination until the present-day position and velocity match the observational constraints to within $\lesssim 1$ kpc and $1$ km s$^{-1}$. The adopted constraints are $(\alpha, \delta) = (81.28^{\circ}, -69.78^{\circ})$, $\mu_{\alpha} = 1.858$ mas yr$^{-1}$, $\mu_{\delta} = 0.385$ mas yr$^{-1}$ \citep{GaiaCollaboration2021LMC}, $d = 49.6$ kpc \citep{Pietrzynski2019}, and $v_\mathrm{los} = 262.2$ km s$^{-1}$ \citep{vanderMarel2002} (see Appendix A of \citealt{Vasiliev2024} for full details on the implementation, which is  an improved version from that of V23).}. Although the first PMs measurements for the LMC favoured a first-infall scenario (\citealt{Kallivayalil2006, Besla2007}), which became the standard paradigm in the literature, subsequent measurements (\citealt{Kallivayalil2013, GaiaDR2kin, GaiaCollaboration2021LMC}) revealed lower tangential velocities. Combined with current uncertainties in both the MW and LMC potentials, a two-passage scenario with a previous pericentre at a distance of $\sim 100$~kpc around $5\text{--}10$~Gyr ago cannot be excluded and remains consistent with observational constraints (\citealt{Vasiliev2023, Vasiliev2024}).
The DM haloes of both the MW and the LMC consist of  truncated NFW profiles. The LMC initial virial\footnote{The virial radius ($r_{\mathrm{vir}}$) and virial mass ($M_{\mathrm{vir}}$) are defined by the condition that the mean enclosed density, $\rho = 3 M_{\mathrm{vir}} / (4 \pi r_{\mathrm{vir}}^3)$, equals $\Delta_c \approx 100$ times  the critical density of the Universe  \citep{BryanNorman1998}, $\rho_{\mathrm{crit}} = 3 H_0^2 / (8\pi G)$, assuming $H_0 = 70$ km s$^{-1}$ Mpc$^{-1}$.} masses are $1.92 \times 10^{11}$ M$_{\odot}$ (hereafter, model L2) and $2.76 \times 10^{11}$ M$_{\odot}$ (L3) respectively, and decrease with time due to the interaction with the MW, mostly during their first pericentre.
The MW models have initial virial masses of $10.0 \times 10^{11}$ M$_{\odot}$ (M10) and $11.0 \times 10^{11}$ M$_{\odot}$ (M11), and their mass profiles match the reported observational constrains at the present time (see \citealt{Wang2020} and Figure 1 of \citealt{Vasiliev2024}). The baryonic component of both MW models is the same as the one described for V23. The four resulting combinations of MW and LMC models, namely L2M10, L2M11, L3M10, and L3M11, where each label is formed by the LMC model name followed by the MW model name, allow us to explore a wide range of realistic MW+LMC configurations. Lastly we include a variant of the model L2M10 with a single passage (L2M10first). In Table~\ref{tab:models} we present a summary of the initial parametrizations of the different MW and LMC models. In Figure~\ref{fig:LMCtraj} we show the Galactocentric distance of the LMC across our potential suite, alongside the corresponding MW virial radius for each parametrization. The trajectories are very similar over the last $\sim 3~\mathrm{Gyr}$, with the LMC crossing the MW virial radius for the first or most recent time between $t \approx -1.5~\mathrm{Gyr}$ (model V23) and $t \approx -2.0~\mathrm{Gyr}$ (model L3M11). At lookback times earlier than $4~\mathrm{Gyr}$, the differences between trajectories become more pronounced, grouping primarily by the underlying MW halo mass (M10 vs.\ M11). In the two-passage configurations, the previous pericentre  occurs $\sim 6-9~\mathrm{Gyr}$ ago ($d_{\mathrm{MW}} \approx 100~\mathrm{kpc}$). We note that our integrations in this work are restricted to the last $5~\mathrm{Gyr}$ (see Section~\ref{sec:orbintmeth}).

The time-evolving potentials adopted in this work are derived from $N$-body simulations of the MW--LMC interaction, which capture key dynamical processes such as the dynamical friction-induced wake and the collective response of the MW halo. Interactions of MW and LMC analogues in cosmological simulations have been reported to display greater structural complexity and stronger halo responses due to the background cosmological environment, halo assembly history, and filamentary accretion \citep{DarraghFord2025}. While these additional complexities represent a source of uncertainty not fully captured by our potentials, we do not expect them to significantly impact the derived orbits, given that observational phase-space uncertainties largely remain the main source of error in orbit integration (see \citealt{Pace2022}, and Section~\ref{sec:orbparam}).

\subsection{Orbit integration method}
\label{sec:orbintmeth}
For the orbit integration procedure we made use of the code \textsc{Agama} (\citealt{AGAMA2018, AGAMA2019}). We embedded it within a Monte Carlo (MC) scheme to account  for all known sources of uncertainty and obtain robust orbital histories. For each galaxy, under each potential, the MC performs the following procedure  throughout $10^3$ iterations.

The phase-space coordinates of the galaxy at the present time are sampled from random normal distributions centred on their nominal values and with dispersion equal to their uncertainties (when the uncertainties are asymmetric, we take the mean uncertainty). They are then converted into Galactocentric Cartesian coordinates using Astropy (\citealt{Astropy2013, Astropy2018}). For this conversion, we consider the following solar parameters: $R_{0} = 8.122 \pm 0.021$ kpc (distance from the Sun to the Galactic centre; \citealt{GRAVITY2018}), $z_{\odot} = 20.8 \pm 0.3$ pc (height of the Sun above the Galactic midplane; \citealt{BennettBovy2019}), $V_{R, \odot} = -12.9 \pm 3.0$ km s$^{-1}$ (radial velocity of the Sun with respect to the Galactic centre; positive values are directed outwards from the centre), $V_{\phi, \odot} = 245.6 \pm 1.4$ km s$^{-1}$ (tangential velocity of the Sun in the direction of Galactic rotation), and $V_{Z, \odot} = 7.78 \pm 0.09$ km s$^{-1}$ (vertical velocity of the Sun relative to the Galactic midplane, directed towards the North Galactic Pole; \citealt{DrimmelPoggio2018}). We note that for each iteration, these parameters are also sampled from random normal distributions, allowing us to propagate  their uncertainties as well.

With the resulting sampled coordinates, we proceed to integrate under the corresponding potential using the method \texttt{orbit} of \textsc{Agama}. The integration is performed in 1500 equal intervals between 0 and $-5$ Gyr. In the case of potential L2M10first, it is only possible to integrate over the last 4 Gyr, since the simulated time for this potential does not reach earlier times
(see \citealt{Vasiliev2024}). We do not rewind further in time since cosmological simulations predict that the majority of the mass of MW-like galaxies is generally assembled before look-back time $\sim5$ Gyr, with usually no major mergers afterwards (\citealt{Santistevan2020, Sotillo-Ramos2022}). Therefore, we consider it is sensible to limit our integration to this time bracket. 
The outcome of the integration procedure is a set of $10^3$ orbits for each galaxy and potential.

\subsection{Estimation of the impact of dynamical friction}
\label{sec:dfest}
Dynamical friction (DF) is the drag force experienced by a dwarf galaxy moving through the DM halo of its host. As a satellite travels within the halo, it perturbs the surrounding DM distribution, generating a trailing overdensity, or wake, which exerts a force opposing its motion. This results in a gradual orbital decay, where the apocentre distances are reduced and the orbital period is shortened over time \citep{BinneyTremaine1987}.

To assess the impact of DF on our orbit reconstructions, we follow a procedure analogous to that of \citet{Patel2020}, which allows us to compute the DF acceleration terms generated by both the MW and LMC haloes acting on a given dwarf at a given time (full implementation is detailed in Appendix~\ref{sec:DF}). We use these terms to backward-integrate the nominal orbit of each galaxy, combining them with the gravitational acceleration from the MW+LMC potential. We restrict this analysis to the nominal orbits (i.e. that obtained from the exact phase-space coordinates, excluding observational uncertainties), since the effect of DF is expected to be relevant only for the most massive MW satellites. Consequently, the dominant source of uncertainty in the orbital reconstruction remains that associated with the phase-space coordinates, already explored through the MC sampling described in Section~\ref{sec:orbintmeth}.

Since DF scales with the total mass of the dwarf, for each galaxy we integrate the nominal orbit twice, adopting two different masses to bracket the DF effect. The choice of masses varies depending on the type of dwarf. For the SMC, by far the most massive galaxy in our sample, we adopt the two models of \citet{Patel2020}: SMC1 ($5 \times 10^{9}$ M$_{\odot}$) and SMC2 ($3 \times 10^{10}$ M$_{\odot}$). While SMC1 is consistent with recent SMC mass estimates \citep{DeLeo2024}, SMC2 is significantly more massive and is included to bracket an extreme scenario and establish an upper limit on the DF effect. For the classical dwarf spheroidal galaxies (dSph, namely Carina, Draco, Fornax, Leo I, Leo II, Sagittarius, Sculptor, Sextans, and Ursa Minor), we explore masses of $10^9$ and $10^{10}$ M$_{\odot}$, and for the UFDs $10^8$ and $10^9$ M$_{\odot}$ \citep{Bullock2017, Sales2022}.

\section{Results and discussion}
\label{sec:resuts}

\subsection{Orbits with respect to the MW}
\label{sec:orbparam}
\begin{figure*}[h]
    \centering
    \includegraphics[width=\textwidth]{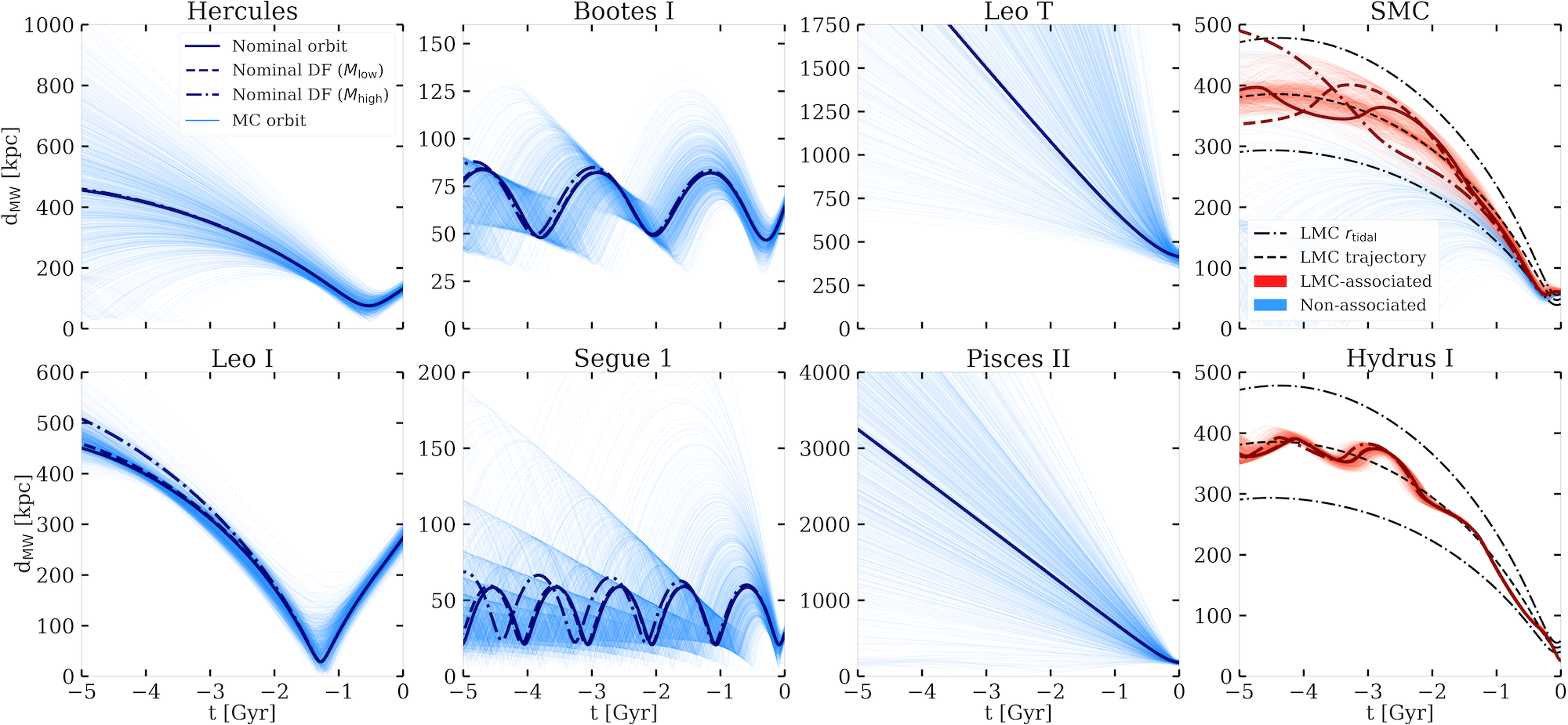}
    \caption{Reconstructed orbits for a selection of galaxies, under the potential L2M10. Each panel displays the evolution of the Galactocentric distance as a function of time for a given galaxy. Thin coloured lines represent the individual trajectories derived with the MC scheme. The thicker solid line shows the nominal trajectory, and is obtained by integrating assuming the phase-space coordinates of the corresponding galaxy are exact (i.e. neglecting observational uncertainties). Analogous dashed and dash-dotted lines represent the nominal orbit integrated accounting for DF, for the lowest and highest satellite masses explored, respectively. Specifically, the masses adopted are $10^8$ and $10^9$ M$_{\odot}$ for UFDs, $10^9$ and $10^{10}$ M$_{\odot}$ for dSphs, and $5 \times 10^{9}$ and $3 \times 10^{10}$ M$_{\odot}$ for the SMC (see Section~\ref{sec:dfest}).  For those galaxies for which we find evidence of an association with the LMC (see Section~\ref{sec:lmcsat}), we superimpose in their panels the LMC trajectory and the extent of the tidal radius using dashed and dash-dotted black lines, respectively. For such galaxies, red lines indicate trajectories that comply with the criterion of being considered associated with the LMC (Section~\ref{sec:lmcsat}), whereas blue ones are for those not meeting the condition.}
    \label{fig:orb_sample}
\end{figure*}

In Figure~\ref{fig:orb_sample}, we show the trajectories derived under potential L2M10 for a selection of galaxies, as representative examples of  the different behaviours that can be observed in the whole sample of dwarfs.
Each panel represents, for a given galaxy, the Galactocentric distance ($d_{\rm MW}$) as a function of time for the ensemble of orbits obtained through the MC realizations. For reference, we  include nominal orbit, which serves as a visual guide to the orbit evolution. We also include the nominal orbits which were derived accounting for the effect of DF. 
The orbits of the galaxies  show a wide variety of behaviours, ranging from some galaxies having performed a single and recent pericentre around the MW (e.g. Hercules or Leo I), some others having a consistent record of passages around it (Bootes I or Segue 1), some being on their first infall (Leo T or Pisces II), and some others displaying trajectories that suggest a potential association with the LMC (the SMC or Hydrus I). The examples show how some galaxies have relatively well-constrained orbits (i.e., the majority of MC realizations yield similar trajectories), while others show a significant spread in the range of reconstructed orbits (e.g., Leo T or Pisces II), which is usually due to the large uncertainties in their initial conditions.

The impact of DF on the reconstructed orbits is very limited, with the exception of the SMC. For first-infall galaxies the effect is negligible, as they have only recently entered, or not yet reached (as in the case of Leo T), the MW virial radius. Galaxies that have spent the full integration time within the MW halo, such as Bootes I, Segue 1, and Hydrus I for the LMC, show a mild effect only for the highest satellite masses tested, which nonetheless remains subdominant with respect to the uncertainties in the phase-space coordinates. Only for the SMC, by far the most massive satellite in our sample, DF has noticeable impact on the reconstructed orbit.

In Figures~\ref{fig:traj1},~\ref{fig:traj2}, and~\ref{fig:traj3} of Appendix~\ref{app:orb} we show the orbits obtained with the potential L2M10, for all the studied galaxies. Analogous plots for the rest of potentials can be found in the publicly accessible repository associated with this work\footnote{\url{https://github.com/ammg-astro/OrbitIntegration}}. Potential L2M10 was chosen as the reference potential throughout this work, as its MW and LMC model masses represent intermediate values within the explored range. Results for other potentials are discussed where relevant.

\subsubsection{Orbital parameters with respect to the MW}
\label{sec:orbparamMW}
We derived orbital parameters with respect to the MW from the reconstructed trajectories. For each galaxy and gravitational potential, we identified the relative extrema of the Galactocentric distance (i.e., pericentres and apocentres) and the times at which they occurred across the corresponding set of MC realizations. 
In Table~\ref{tab:orbparam}, we provide the most recent pericentre ($r_{\rm{peri}}$) and apocentre ($r_{\rm{apo}}$) with respect to the MW, and the time at which they took place ($t_{\rm{peri}}$ and $t_{\rm{apo}}$, respectively) for all the galaxies, based on potential L2M10. The final orbital parameters are reported as the median values among realizations, with associated uncertainties given by the 16th and 84th percentiles. Full tables for the rest of potentials are available in the repository. In Appendix~\ref{app:rec} we compile a series of recommendations for the potential users of our catalogue.

We only provide the parameters for the most recent pericentre and apocentre, since studies of orbital integration in cosmological simulations have shown that these are the ones that can be more robustly estimated (\citealt{Souza2022, Santistevan2024}).  We note that in some cases, for a given galaxy and potential the sets of orbits calculated in the MC scheme can show large differences between them (see e.g. Horologium II in Appendix~\ref{app:orb}, in some of the trajectories it seems to be bound to the MW, in others follows a trajectory similar to that of the LMC, whereas in others it may be in its first infall). Thus, we also include in the Table the fraction of MC realizations for which we detect the pericentre ($f_{\rm{peri}}$) or the apocentre ($f_{\rm{apo}}$). Values of the fraction equal or close to one imply that the majority of the realizations contain those particular extrema (and thus it is well determined), whereas values equal or close to 0 imply that the extrema rarely take place in the MC realizations, and thus it should not be considered robustly determined.

We note that the impact of neglecting DF in the derivation of the orbital parameters is very limited. As shown in Appendix~\ref{app:orb}, the nominal orbits integrated accounting for the effect of DF assuming the lower of the two adopted masses (see Section~\ref{sec:dfest}) display nearly negligible differences with respect to those obtained without DF. When adopting the higher mass, mild deviations can be observed in the orbits of some galaxies, mostly towards very early times. However, even in those cases the orbits reconstructed including the effect of DF remain within the spread of the MC realizations, indicating that the uncertainty associated with the phase-space coordinates dominates over the effect of DF. Furthermore, since we report only the most recent pericentre and apocentre, these usually take place before the effect of DF can have an impact. We therefore conclude that DF does not significantly affect our orbital parameter estimates.

\begin{figure*}
    \centering
    \includegraphics[width=\linewidth]{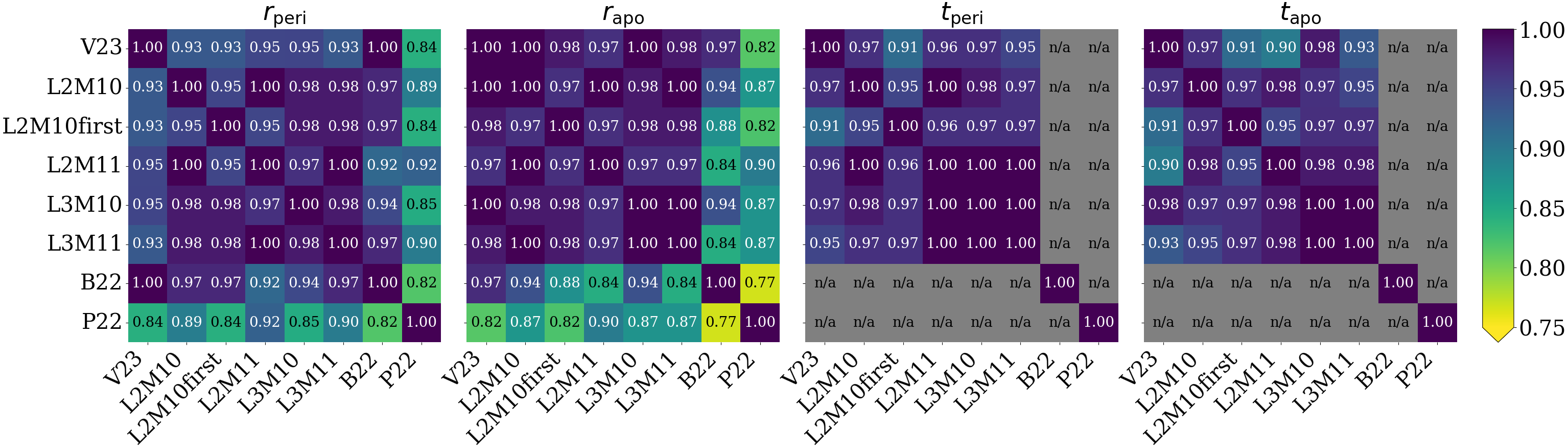}
    \caption{Compatibility of the orbital parameters derived with the different gravitational potentials, between them and with other works. Each panel shows a confusion matrix quantifying the pairwise agreement between the corresponding orbital parameters derived from the six MW+LMC potentials used in this work, and also with those reported in \citet{Battaglia2022} (B22) and \citet{Pace2022} (P22). Each matrix entry, and its colour, represents the fraction of galaxies whose corresponding orbital parameter agrees within $1\sigma$ for a given pair of models. From left to right, the panels show: (1) Galactocentric distance at the most recent pericentre, (2) Galactocentric distance at the most recent apocentre, (3) time of the most recent pericentric passage, and (4) time of the most recent apocentric passage. \citet{Battaglia2022}  and \citet{Pace2022} do not report values for $t_{\rm peri}$ and $t_{\rm apo}$; thus, the corresponding matrix elements are shown as n/a.
    }
    \label{fig:confussion_matrix_self}
\end{figure*}

\begin{figure}
    \centering
    \includegraphics[width=\linewidth]{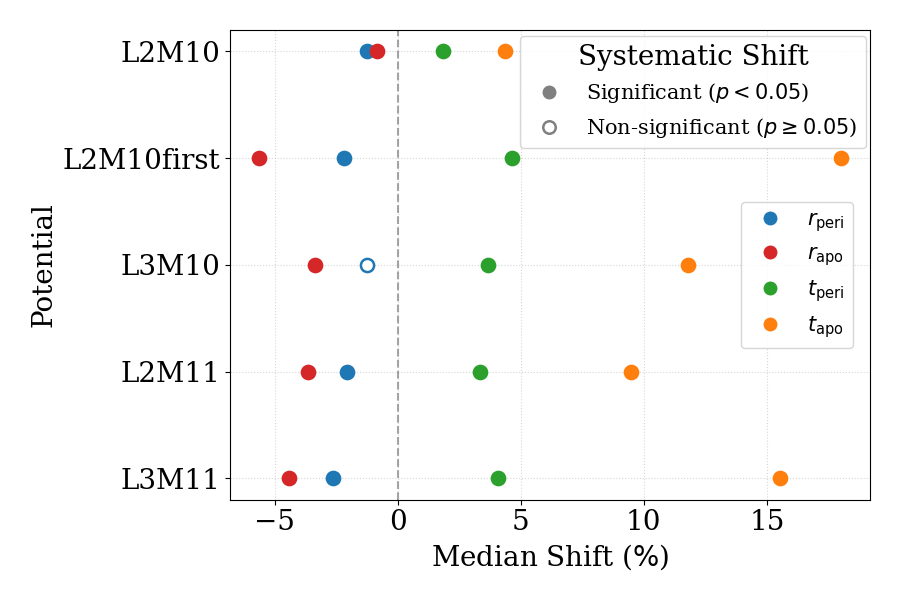}
    \caption{Median relative shift of the orbital parameters. Points represent the sample  median shift of a given orbital parameter between the one derived with the corresponding potential marked on the left, and the one derived with V23. The different orbital parameters are colour coded. Filled points represent parameters for which the systematic shift is statistically significant, based on the Wilcoxon
signed-rank tests, whereas empty points represent those not statistically significant. }
    \label{fig:systematicshifts}
\end{figure}

We checked the compatibility between the parameters derived under the different potentials and also compared them with those reported in other works. To illustrate this, Figure~\ref{fig:confussion_matrix_self} shows confusion matrices that summarize the agreement between the parameters obtained for the different potentials, and also with the ones reported by \citet{Battaglia2022} and \citet{Pace2022}, two works that used MW+LMC potentials to backward-integrate the orbits of a large number of MW satellites. 
For a given orbital parameter, each matrix element gives the fraction of galaxies for which the estimates obtained with two different potentials agree within $1\sigma$. 
A value of 1 means perfect agreement, while lower values indicate larger discrepancies. We note that cases where a parameter is not available for a certain galaxy in one or both potentials are excluded from the comparison. When comparing with other works, only galaxies common to both datasets are considered.

Starting with the comparison between the results obtained in this work, overall, the orbital parameters exhibit very good consistency across the six tested potentials, with at least 90\% of the cases showing agreement within $1\sigma$. Parameters derived with the V23 potential tend to display a slightly lower level of compatibility with the rest of potentials, although the difference is not significant. The discrepancies observed among potentials arise from the different parametrizations. Potential satellites of the LMC (see Section~\ref{sec:lmcsat}), such as Carina II, Carina III, Hydrus I or Pictor II tend to show inconsistencies at the $1\sigma$ level between some potentials; however, we note that  the apocentres and pericentres of these galaxies with respect to the MW are not representative, since their trajectories follow that of the LMC.
In Section~\ref{sec:orbparamlmc}, we provide the orbital parameters of the likely LMC satellites with respect to their host. In any case, the differences in the derived parameters across the various potentials are small, and the results remain consistent in most cases.

Regarding the comparison with previous studies, we find overall good agreement with the orbital parameters reported by \citet{Battaglia2022} and \citet{Pace2022}. We note that these works only provide values for $r_{\rm peri}$ and $r_{\rm apo}$; consequently, the confusion matrices for $t_{\rm peri}$ and $t_{\rm apo}$ show the label 'n/a' when compared with these studies. We also note that both works, like ours, did not account for the impact of the DF in the estimation of orbital parameters.  In general, the reported pericentric and apocentric distances are consistent with those derived in this work.
The agreement is particularly good for with \citet{Battaglia2022} across all tested potentials. In particular, comparisons with the V23 potential yield fully compatible results for $r_{\rm peri}$ and nearly so for $r_{\rm apo}$. This is expected, as the dataset used by \citet{Battaglia2022} closely matches ours and the potential  is similar to V23. For the remaining potentials, the agreement remains good, especially for $r_{\rm peri}$, and the slight differences can be explained by the fact that these potentials consist of heavier MW and LMC models.

The results of \citet{Pace2022}, while still broadly consistent, show a slightly  lower level of compatibility with our measurements and with those of \citet{Battaglia2022}. The discrepancies are most pronounced when compared with the V23 potential, for which the compatibility drops to 82\% for $r_{\rm apo}$, the lowest value among all comparisons. 
These differences are likely driven by the different parametrizations of the MW-LMC potential adopted, since the initial conditions they used are generally compatible with the ones used here and in \citet{Battaglia2022}. The potential adopted in \citet{Pace2022} consists of an LMC of $1.38\times10^{11}$ M$_{\odot}$ and a MW of $1.4\times10^{12}$ M$_{\odot}$ \citep{McMillan2017}. The LMC mass is somewhat below those of our models, yet comparable to that of V23. The MW mass, however, is higher than in any of our models, and particularly  with respect to V23, for which the difference reaches $\sim 75\%$. This mass discrepancy is the most likely explanation for the  larger incompatibilities found when comparing the results of \citet{Pace2022}  with ours, and in particular with those derived with potential V23.

Overall, the orbital parameters derived with our suite of potentials show mutual compatibility within $1\sigma$. However, despite this formal agreement, we find statistically significant systematic shifts in orbital parameters derived with different models, as confirmed by Wilcoxon signed-rank tests applied to all pairwise potential combinations. In Figure~\ref{fig:systematicshifts} we show the median relative shift in each orbital parameter with respect to V23. For a given orbital parameter ($P$), we calculate the relative shift for each galaxy $i$ as:

\begin{equation}
\Delta P_i (\%) = \left( \frac{P_{i, \mathrm{model}} - P_{i, \mathrm{V23}}}{|P_{i, \mathrm{V23}}|} \right).
\end{equation}

The recovered values of $\Delta P_i$ are then median-averaged across the galaxy sample. We adopt V23 as baseline since it represents the lightest MW and LMC configuration in our suite. The median shifts reveal systematic trends tied to the assumed mass configuration: higher MW masses (and to a lesser extent higher LMC masses within the same MW configuration) systematically reduce both pericentric and apocentric distances and shift the timing of the most recent orbital passages to more recent lookback times.

Finally, we evaluated the main source of uncertainty in the  orbital parameters by comparing the impact of phase-space coordinates uncertainties in their derivation ($\sigma_{\mathrm{obs}}$) against the sensitivity to the assumed MW--LMC potential ($\sigma_{\mathrm{pot}}$). For a given orbital parameter and galaxy, we computed the mean of the upper and lower 1$\sigma$ uncertainties and median-averaged this across all six potentials; $\sigma_{\mathrm{obs}}$ is then computed as the median value across all galaxies. Similarly, we computed the standard deviation of the nominal value of the parameter across the six potentials for each galaxy, taking the median across the galaxy sample to obtain $\sigma_{\mathrm{pot}}$. We find that $\sigma_{\mathrm{obs}}$ consistently exceeds $\sigma_{\mathrm{pot}}$ by factors of $\sim 4$--$7$ for the different orbital parameters. This confirms that present-day observational uncertainties in phase-space coordinates, rather than the choice of MW--LMC potential, represent the primary source of error in the reconstructed orbits. Consequently, future improvements in PM measurements will be essential for refining satellite orbital parameters.

\subsubsection{MW-LMC system satellites}
\label{sec:MWLMCsys}
\begin{figure*}
    \centering
    \includegraphics[width=\linewidth]{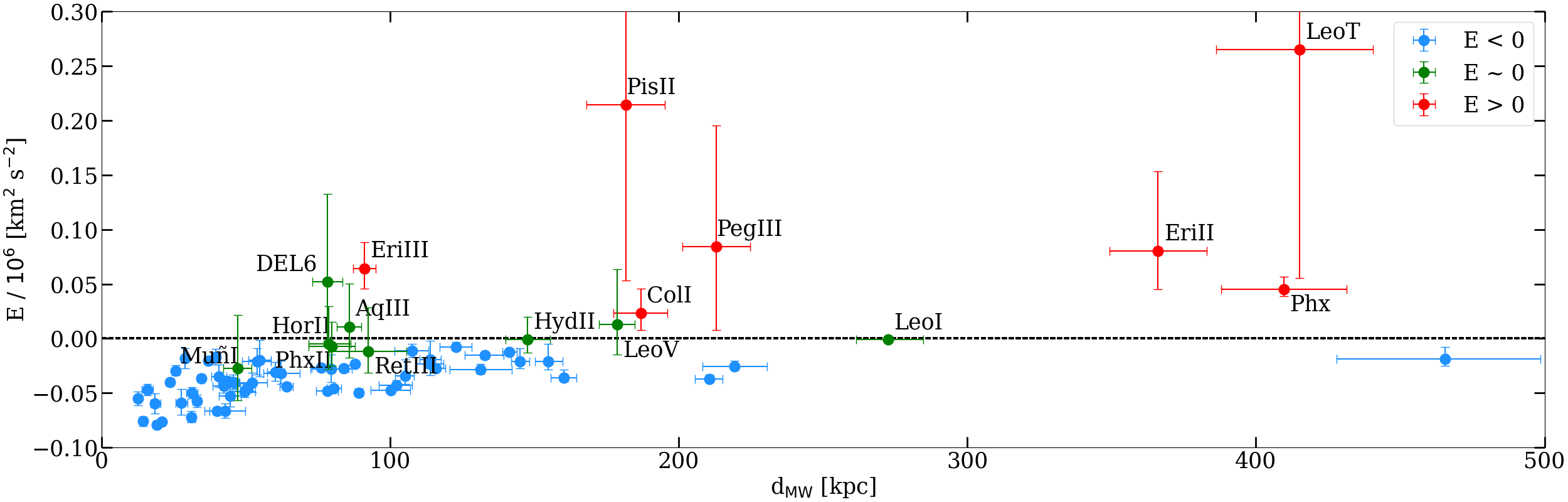}
    \caption{Specific orbital energy derived under potential L2M10, compared to the current Galactocentric distance of the galaxies. Each point with error bar represents the specific energy of the corresponding galaxy and its Galactocentric distance at $t = 0$ and the associated uncertainties. The  horizontal dashed line represents $E = 0$; that is, the boundary between bound and unbound systems.}
    \label{fig:boundtoMWLMC}
\end{figure*}

In addition to the orbital parameters, we  estimated the specific orbital energy of the dwarfs with respect to the MW-LMC system at present time to ascertain which of them are currently bound to it ($E < 0$).  In Figure~\ref{fig:boundtoMWLMC}, we show the specific energy of the different galaxies, calculated with potential L2M10, compared to their Galactocentric distance. 

The values of the specific energy can be found in Table~\ref{tab:orbparam}, alongside the fraction of MC realizations with  $E<0$ at present time ($f_{\mathrm{bound}}$), for each galaxy. For this fraction, values close to 1 imply that the majority of orbits derived in the MC scheme are bound. On the contrary, values close to 0 imply that nearly no orbit is bound, regardless of the sampling of the uncertainties of the phase-space coordinates.

We find that the majority of the dwarf galaxies in our sample are bound to the MW+LMC system, as expected. However, a number of systems show an ambiguous binding status, with specific energies consistent with zero at the 1$\sigma$ level, and $f_{\mathrm{bound}}$ values $\sim0.2-0.7$. These include Aquarius III, DELVE 6, Horologium II, Hydra II, Leo I, Leo V, Muñoz 1, Phoenix II, and Reticulum III. For these galaxies, the uncertainty in the specific energy is typically larger than for clearly bound systems.
These large uncertainties are primarily a result of the generally substantial uncertainties in their present-day phase-space coordinates, particularly in their PMs. As a result, we cannot robustly determine whether these galaxies are bound to the MW+LMC system or are on unbound trajectories; only future improvements in the determination of their velocity components will be able to clarify their dynamical status.

We also identify several systems with specific energies consistent with $E > 0$, namely Columba I, Eridanus II, Eridanus III, Leo T, Pegasus III, Phoenix, and Pisces II, indicating that they are likely not bound to the MW+LMC system. Some of these galaxies are among the most distant in our sample, with Galactocentric distances exceeding $\sim$200 kpc in several cases, and some are located beyond the MW virial radius.
At such large distances, accurate phase-space measurements are challenging, leading to generally large uncertainties, particularly in the PMs (e.g. Leo T, Pisces II or Pegasus III). The very low values of $f_{\mathrm{bound}}$ indicate that, even considering the uncertainties in their phase-space coordinates, almost none of the MC realizations correspond to orbits bound to the MW--LMC system. 

The classification of galaxies as bound, unbound, or ambiguous shows mild variations with  the assumed potential, driven primarily by the MW mass. For the lowest MW mass explored, in model V23, we find a slightly larger number of unbound and ambiguous systems (11 and 9, respectively) compared to the rest of heavier MW models. Conversely, for model L2M11, several galaxies that appear unbound or ambiguous under potential V23, become clearly bound, with 7 unbound galaxies and 8 ambiguous cases remaining. The rest of potentials show very similar values, with 7 unbound and 9 ambiguous. A particularly illustrative case is Leo I, which is unbound under V23, ambiguous under L2M10, and bound under L2M11, highlighting how sensitively its dynamical status depends on the assumed MW mass (see \citealt{BoylanKolchin2013}). The MW mass therefore is the key factor in determining the bound status of the satellite population. Our models span a MW mass range of $\sim0.8$--$1.1 \times 10^{12}$ M$_{\odot}$, which corresponds to the mid-to-low end of current estimates ($\sim0.5$-$2\times10^{12}$ M$_{\odot}$, \citealt{Wang2020, BobylevBaykova2023, Bayer2025}). Therefore, exploring more massive MW configurations, together with improved PM measurements, may shift some of the ambiguous and unbound galaxies identified here towards a bound status.

\subsubsection{Infall times and accretion history}
\label{sec:t_infall}

We also estimated the infall times ($t_{\rm infall}$) of the dwarfs in our sample, a key quantity to determine how long a galaxy has been subject to the environmental effects of the MW and to what extent these may have shaped its present-day properties. For a given potential, we derived $t_{\rm infall}$ from the nominal orbit of each dwarf, as the time at which it first crosses the MW virial radius.

For galaxies whose nominal orbit was already within the virial radius at the earliest integration time ($t = -5$ Gyr, or $t = -4$ Gyr for L2M10first), no infall time can be assigned, as it predates our integration window. To quantify how robustly this applies, we computed $f_{\rm vir}$, the fraction of MC orbits that lie within the virial radius at the earliest integration time. Values close to 1 indicate the galaxy was already inside for the majority of the sampled orbits, while values close to 0 indicate it was beyond it. The values of $t_{\rm infall}$ and $f_{\rm vir}$ obtained for potential L2M10 are provided in Table~\ref{tab:orbparam}.

We find that the majority of the dwarfs were already within the virial radius of the MW at the earliest integration time (48 out of 72 dwarfs, for potential L2M10). For the rest of galaxies, we observe that on average, they usually crossed it $\sim1-2$ Gyr ago, with the exception of NGC 6822 ($t_{\rm infall} \sim -4.5$ Gyr). These late infallers largely consist of LMC satellites, which approached the MW as a compact group rather than as individual systems. Their infall time is therefore driven by the orbital history of the LMC itself, rather than by their individual dynamics with respect to the MW,

Among the remaining late infallers, many dwarfs have large PM uncertainties, including Aquarius III, Leo V, Pisces II, Pegasus III, and Reticulum III. For these galaxies, poorly constrained PM components tend to bias their orbits towards high specific energies (see Section~\ref{sec:MWLMCsys}) and likely translate into  artificially late infall times. Improved PM measurements would therefore likely result in earlier and better constrained infall time estimates for these galaxies.

Beyond the effect of PM uncertainties, the number of late infallers also depends on the assumed MW mass. For potentials with heavier MW configurations, the number of late infallers is further reduced (11 for L2M11 and 12 for L3M11). This is a result of the enhanced gravitational pull of the MW, combined with the fact that the LMC and its satellites are already within the MW virial radius at the earliest integration time for those potentials involving M11 (see Figure~\ref{fig:LMCtraj}). We also note that the infall times are stable across potentials, with a slight tendency to decrease with the MW mass.  As discussed in Section~\ref{sec:MWLMCsys}, our models explore the mid-to-low end of the plausible MW mass range, and still heavier configurations would likely further reduce the number of late infallers and infall times. 

At present, five galaxies in our sample lie close to or beyond the MW virial radius: Eridanus II, Leo I, Leo T, NGC 6822, and Phoenix. Of these, Eridanus II, Leo T, and Phoenix show no evidence of ever having crossed it and NGC 6822 has done so in only $\sim$40--50\% of the MC realizations, always in a very shallow  approach by the MW. The ambiguous behaviour of NGC 6822 has also been noted by \citet{Bennet2024}, who similarly find no conclusive evidence to classify it as a backsplash galaxy. Leo I, on the other hand, crossed the virial radius $\sim2.7$ Gyr ago, reached within $\sim50$ kpc to the MW, and at present it is again in the near vicinity of the virial radius, with $\sim60-85\%$ of the MC realizations finding it already beyond it (depending on the potential), on its way to becoming a backsplash galaxy in the near future (\citealt{Bennet2024}). However, it is still unclear whether Leo I will visit the MW again in the future or will abandon it definitively, given its ambiguous binding status (see Section~\ref{sec:MWLMCsys}).

Together, our results suggest that the majority of MW dwarfs have spent at least the last 5 Gyr within the MW virial radius, with only a minority of systems crossing it on average 1--2 Gyr ago. We note that these conclusions are robust to the inclusion of DF: repeating the analysis using nominal orbits integrated with DF (Section~\ref{sec:dfest}) yields no significant changes. These findings are in tension with those of \citet{Hammer2021, Hammer2023}, who proposed that the majority of MW satellites were accreted within the last $\sim$2--3 Gyr. Their conclusions were based on the assumption that the infall time of MW dwarfs, can be inferred from their present-day orbital energy. Since they found many satellites to have high orbital energies and angular momenta, they argued that a large fraction of the MW dwarf population entered the virial radius only recently.

We identify two likely explanations for this discrepancy. A key difference between our work and that of \citet{Hammer2021, Hammer2023} is that our analysis reconstructs the orbital histories of the satellites directly through backward orbit integration, explicitly accounting for the time-evolving potential of the LMC, rather than inferring infall times from present-day orbital energies alone. The inclusion of the LMC potential modifies the specific energy of the dwarfs, reducing them, mainly for its satellites and neighbouring dwarfs. Additionally, the use of time-evolving potentials renders energy a non-conserved quantity. Consequently, present-day energies do not necessarily provide a reliable tracer of infall time, since present and past interactions with the LMC can both increase and decrease the energy of individual systems in ways that are not captured under the assumption of a static MW-only potential. In addition, the assumed MW mass in the energy calculation plays an important role: higher MW masses reduce the specific energies of the dwarfs, favouring earlier infall times. For reference, \citet{Hammer2021} adopted a MW mass of $8.1\times10^{11}$ M$_{\odot}$ (see their Figure 6), comparable to our lightest MW model (V23), which yields systematically higher orbital energies than our more massive MW+LMC configurations.

\subsection{Orbits of the LMC satellites}
\label{sec:lmcsat}
The use of potentials that include the LMC allows us to explore whether some galaxies currently observed in the vicinity of the MW are instead members of the LMC satellite system. Using the reconstructed trajectories, we identify candidate LMC satellites, characterize their orbital properties with respect to their host, and assess the outcome of interactions between the LMC and some nearby dwarf galaxies.

\subsubsection{Potential LMC satellites}
\label{sec:potlmcsat}
We identify candidate LMC satellites by examining dwarf galaxy trajectories in the LMC rest frame, where actual satellites are expected to display recurrent passages about the LMC. We estimate the probability of a dwarf being an LMC satellite ($p_{\rm{LMC}}$) under a given potential by calculating the fraction of MC realizations in which the galaxy completes at least two pericentres, or one pericentre and one apocentre, within the LMC's tidal radius. This ensures that the candidate recurrently orbits the LMC within the region where its gravitational influence is dominant, effectively discarding galaxies on temporary, unbound flybys.

The LMC tidal radius is approximated by identifying the position along the line between the MW and the LMC where the gravitational forces from both galaxies exerted on a test particle are balanced. This radius is recalculated for each potential and integration time. We adopt this approximation since the LMC's region of influence has a complex irregular geometry when using potentials that account for the mutual interaction between the LMC and MW, which significantly complicates the calculation.

In Table~\ref{tab:LMCprob} we show the probability of galaxies being an LMC satellite under each potential. The table includes only those galaxies with $p_{\rm{LMC}} > 0.1$. We set this threshold to keep only galaxies with high probabilities of being LMC satellites.

\begin{figure*}
    \centering
    \includegraphics[width=\linewidth]{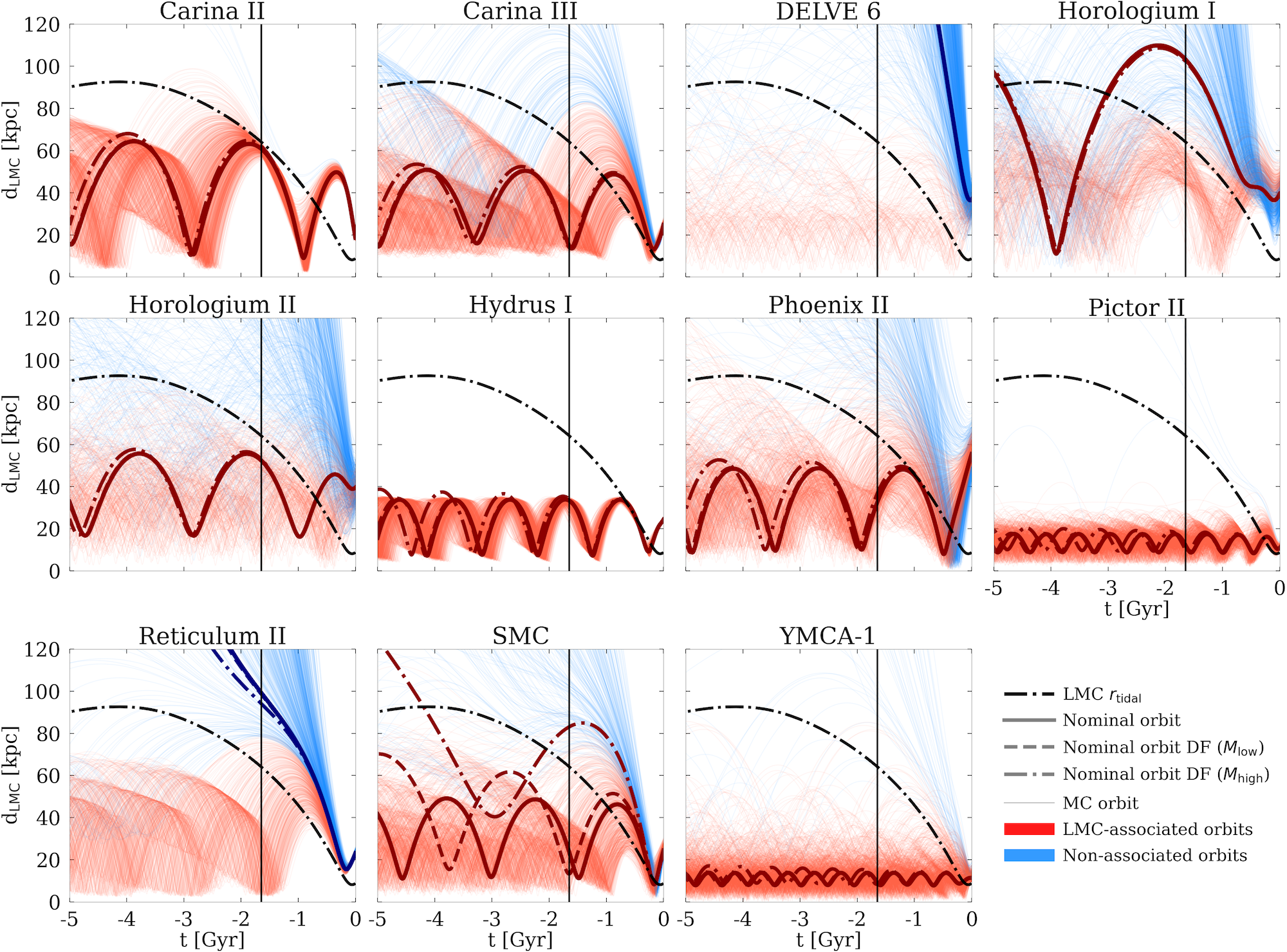}
    \caption{Trajectories relative to the LMC for galaxies with $p_{\rm{LMC}} > 0.1$, computed using the L2M10 potential. Each panel shows, for a given galaxy, the distance to the LMC ($d_{\rm LMC}$) as a function of time for all MC realizations. We also include in each panel, as a visual aid, the nominal orbit of the corresponding dwarf with respect to the LMC, derived from the nominal values of the phase-space coordinates (not considering the associated uncertainties), represented as a thicker darker solid line.  Analogous dashed and dash-dotted lines represent the nominal orbit integrated accounting for DF, for the lowest and highest satellite masses explored, respectively ($10^8$ and $10^9$ M$_{\odot}$ for UFDs, and $5 \times 10^{9}$ and $3 \times 10^{10}$ M$_{\odot}$ for the SMC, see Section~\ref{sec:dfest}). Trajectories that satisfy the criterion for compatibility with LMC satellite membership (see Sect.~\ref{sec:potlmcsat}) are shown in red, while non-compliant trajectories are shown in blue. The dash-dotted black lines indicate the tidal radius of the LMC and the solid  vertical black lines indicate the time the LMC crossed the MW virial radius.}
    \label{fig:orbitLMC}
\end{figure*}

Figure~\ref{fig:orbitLMC} shows the orbits of these galaxies in the LMC rest frame, distinguishing MC realizations that meet our satellite criteria from those inconsistent with LMC association. We include the LMC tidal radius for reference.  We note that the derived probabilities may be slightly underestimated; for some galaxies (e.g., see Carina III), certain orbits highly consistent with LMC satellite status are not classified as such, likely due to our conservative tidal radius approximation described above. We further assess the robustness of our criterion and the derived probabilities in Appendix~\ref{app:LMCprobtest}, where we explore alternative association criteria and find no significant differences in the resulting classifications.

\begin{table}[]
    \centering
    \fontsize{10}{12}\selectfont
    \setlength{\tabcolsep}{1.5pt} 
    \caption{LMC satellites probability}
    \begin{tabular}{lcccccc}
    \toprule
       Satellite &   V23 & L2M10 & L2M10first & L2M11 & L3M10 & L3M11 \\
       \multicolumn{1}{c}{(1)}  & (2) & (3) & (4) & (5) & (6) & (7) \\
    \midrule
    \midrule
     Carina II & 0.998 & 0.995 & 0.999 & 0.977 & 1.0 & 0.995 \\
    Carina III & 0.903 & 0.839 & 0.926 & 0.677 & 0.928 & 0.925 \\
    DELVE 6    & 0.174 & 0.139 & 0.171 & 0.107 & 0.192 & 0.180 \\
  Horo. I & 0.665 & 0.376 & 0.534 & 0.339 & 0.629 & 0.496 \\
 Horo. II & 0.485 & 0.373 & 0.445 & 0.292 & 0.522 & 0.460 \\
      Hydrus I & 1.0   & 1.0   & 1.0   & 1.0   & 1.0   & 1.0   \\
    Phoenix II & 0.783 & 0.671 & 0.808 & 0.712 & 0.834 & 0.758 \\
     Pictor II & 0.997 & 0.996 & 0.999 & 0.995 & 0.998 & 0.999 \\
  Ret. II & 0.743 & 0.436 & 0.673 & 0.368 & 0.802 & 0.626 \\
          SMC  & 0.925 & 0.752 & 0.867 & 0.740 & 0.937 & 0.815 \\
        YMCA-1 & 0.920 & 0.958 & 0.955 & 0.908 & 0.950 & 0.953 \\

    \bottomrule
    \end{tabular}
    \tablefoot{Probability of association with the LMC, under the different potentials, for galaxies with $p_{\rm{LMC}} > 0.1$. Columns represent, from left to right (1) the names of the galaxies  and (2-7) probabilities for the different potentials.}
    \label{tab:LMCprob}
\end{table}

Hydrus I displays the highest probability of being an LMC satellite, consistently achieving $p_{\rm{LMC}} = 1$ across all potentials. It follows a tight, compact orbit around the LMC, remaining mostly well within the tidal radius.

Carina II, Carina III, Pictor II, the SMC, and YMCA-1 also exhibit high probabilities (typically above 0.9) with similarly well-defined trajectories around the LMC, making them very likely satellites. In comparison with Hydrus I, their orbits are slightly less constrained  and in the case of Carina III, they do not get as close to the LMC (in Section~\ref{sec:orbparamlmc}, we present and discuss their orbital parameters with respect to the LMC). Only Pictor II and YMCA-1 show more compact trajectories.

Phoenix II shows somewhat lower but still substantial probabilities, with $p_{\rm{LMC}}$ ranging between $\sim 0.65-0.85$ across different potentials, indicating it is also a very likely LMC satellite. In comparison with the previous galaxies, its orbit is slightly less constrained, and it can  be easily noticed how a fraction of the MC realizations are  incompatible with being an LMC satellite.

Horologium I, Horologium II, and Reticulum II show more variable results, with probabilities exceeding 0.5 in 3, 1, and 4 out of 6 potentials, respectively, something that can also be seen in their orbits. For these galaxies, membership in the LMC satellite system  probability depends more on the chosen potential. We note, however, that for the rest of candidates the association probabilities are stable across potentials. Horologium I and Reticulum II can be considered to have moderate probabilities of being LMC satellites. Horologium II is the least likely of this group, though the obtained probabilities remain non-negligible.

DELVE 6 shows in all the cases the lowest probabilities, never reaching above 0.2, casting doubts on its association with the LMC. However, we note that DELVE 6 is the galaxy with the highest uncertainty in PMs among the potential LMC satellites. This likely explains its low association probability, since
uncertainties in the phase space coordinates are more likely to scatter the dwarfs away from the LMC rather than bring them closer when reconstructing their orbits (\citealt{Vasiliev2024}).

Multiple studies have explored the possibility that some dwarf neighbours of the MW are actually LMC satellites, using several different methods and criteria (e.g. \citealt{DOnghia2008, Nichols2011, Kallivayalil2018,Pardy2020, SantosSantos2021, Battaglia2022, Vasiliev2024}, and references therein). Here we limit our comparison mostly to those based on orbital integration.

\citet{ErkalBelokurov2020} rewound the orbits of dwarfs and the LMC up to 5 Gyr and derived the orbital energy of the dwarfs with respect to the LMC to verify whether they were bound to it. \citet{Pace2025} proceeded likewise in order to study the association of Pictor II with the LMC. \citet{Patel2020} reconstructed the orbits of several dwarfs under a combined potentials including the MW, LMC and SMC, and calculated the fraction of orbits that, for a given dwarf, lay within the region where the LMC's density profile is dominant (somewhat similar to the criterion we use in this paper). \citet{CorreaMagnus2022} backward-integrated the orbits of MW neighbouring dwarfs and derived their energy with respect to the LMC, computing probabilities based on the fraction of orbits that were bound in the time interval $-3<t<-1$ Gyr. However, rather than treating all realizations as equally likely, they weighted the orbits using distribution-function priors based on the adopted MW potential. \citet{Battaglia2022} and \citet{Pace2022} compared the velocities of dwarfs at their most recent close approach to the LMC with the LMC escape velocity. We also compared with the results of 
\citet{Vasiliev2024}, who employed N-body simulations of the interacting MW-LMC system (using four potentials that coincide with some of those used in this work) to study the possible origins of observed MW satellites under the scenario where the LMC had performed two pericentres around the MW. Association probabilities were estimated using three different methods: particle matching, Gaussian-mixture modelling, and orbit rewinding (see further details in \citealt{Vasiliev2024}).

Broadly speaking, all these studies agree in classifying Carina II, Carina III, Horologium I, Hydrus I,  Phoenix II, Pictor II, Reticulum II, and the SMC as likely LMC satellites, which is in  agreement with what we report here. One-to-one comparisons of likelihoods with other studies are complex, given the different datasets and particular techniques used for probability estimates. Nevertheless, the general trends are also in agreement.

\subsubsection{Orbital parameters of the LMC satellites}
\label{sec:orbparamlmc}

\begin{table}[]
\renewcommand{\arraystretch}{1.25}
\setlength{\tabcolsep}{4pt} 
    \centering
    \caption{Orbital parameters with respect to the LMC}
    \begin{tabular}{lcccc}
    \toprule
    Galaxy & $r_{\rm peri}^{\mathrm{LMC}}$ & $t_{\rm peri}^{\mathrm{LMC}}$ & $r_{\rm apo}^{\mathrm{LMC}}$ & $t_{\rm apo}^{\mathrm{LMC}}$ \\
               & [kpc] & [Gyr] & [kpc] & [Gyr]\\
    \multicolumn{1}{c}{(1)}  & (2) & (3) & (4) & (5) \\
    \midrule
    \midrule
Carina II & $9^{+9}_{-5}$ & $-0.91^{+0.04}_{-0.06}$ & $50^{+2}_{-2}$ & $-0.334^{+0.007}_{-0.013}$ \\
Carina III & $13^{+1}_{-2}$ & $-0.18^{+0.03}_{-0.04}$ & $45^{+17}_{-11}$ & $-0.8^{+0.1}_{-0.2}$ \\
DELVE 6 & $24^{+10}_{-16}$ & $-0.20^{+0.10}_{-0.34}$ & $35^{+22}_{-7}$ & $-0.8^{+0.3}_{-0.4}$ \\
Horologium I & $43^{+5}_{-5}$ & $-0.09^{+0.03}_{-0.02}$ & $48^{+9}_{-7}$ & $-0.37^{+0.04}_{-0.07}$ \\
Horologium II & $24^{+16}_{-15}$ & $-0.3^{+0.3}_{-0.3}$ & $46^{+14}_{-9}$ & $-0.8^{+0.5}_{-0.5}$ \\
Hydrus I & $9^{+2}_{-1}$ & $-0.26^{+0.01}_{-0.01}$ & $33.8^{+0.4}_{-0.4}$ & $-0.76^{+0.04}_{-0.05}$ \\
Phoenix II & $11^{+10}_{-6}$ & $-0.5^{+0.1}_{-0.3}$ & $49^{+17}_{-9}$ & $-1.3^{+0.2}_{-0.3}$ \\
Pictor II & $10^{+1}_{-1}$ & $-0.03^{+0.02}_{-0.02}$ & $17^{+3}_{-3}$ & $-0.24^{+0.03}_{-0.04}$ \\
Reticulum II & $15.0^{+0.9}_{-0.9}$ & $-0.170^{+0.007}_{-0.011}$ & $53^{+13}_{-9}$ & $-1.0^{+0.1}_{-0.3}$ \\
SMC & $8^{+4}_{-3}$ & $-0.17^{+0.02}_{-0.05}$ & $41^{+16}_{-11}$ & $-0.8^{+0.1}_{-0.2}$ \\
YMCA-1 & $7^{+4}_{-3}$ & $-0.2^{+0.1}_{-0.2}$ & $15^{+9}_{-3}$ & $-0.2^{+0.2}_{-0.3}$ \\
    \bottomrule
    \end{tabular}
    \tablefoot{Orbital parameters  with respect to the LMC for galaxies with $p_{\rm{LMC}} > 0.1$, calculated with potential L2M10. Columns show, from left to right: (1) name of the galaxy, (2) most recent pericentre and (3) time at which it took place, (4) most recent apocentre and (5) time at which it took place, all of them with respect to the LMC.}
    \label{tab:orbparamLMC}
\end{table}

The LMC and its satellites share a common motion, moving together as a gravitationally bound system. Consequently, the orbital properties of the satellites are best understood when measured with respect to their true host, the LMC, rather than the MW. In Table~\ref{tab:orbparamLMC}, we provide the orbital parameters with respect to the LMC for the galaxies for which we obtained $p_{\rm{LMC}} > 0.1$, under potential L2M10. The parameters were calculated analogously to the ones measured with respect to the MW (Section~\ref{sec:orbparam}); however, for the derivation, we only used those orbits that comply with the criterion outlined in Section~\ref{sec:potlmcsat}.

The galaxies show close approaches to the LMC, reaching usually below $\sim 15$ kpc, except for Horologium I and II and DELVE 6. Horologium I, in particular, shows a distant and very recent pericentre; however, its previous pericentre is significantly closer to the LMC (see Figure~\ref{fig:orbitLMC}), in line with the rest of potential LMC satellites. With the exception of these  galaxies, the rest seem to have passed close enough to the LMC, so that the pericentres could have taken place within the range of the extended stellar populations of the LMC (up to 18.5 kpc \citealt{Nidever2019}). The closest approaches are those of Carina II, Hydrus I, the SMC, and YMCA-1, suggesting direct collisions with the LMC disc, while Carina III, Phoenix II, and Pictor II likely collided with its outer regions. Given the low mass of all these systems except the SMC, such collisions are unlikely to have produced strong effects on the LMC. The interaction between the Magellanic Clouds, is expected to have been significantly more impactful given the mass of the SMC. We explore this interaction in Section~\ref{sec:SMCcollision}.

Regarding the maximum separation from the LMC, all the galaxies show similar values, not getting further than 50 kpc in any case, a limit set by the tidal radius of the LMC (and the criteria imposed in this work). The exceptions are Pictor II and  YMCA-1, which show the smallest maximum separation from the LMC of barely 17 and 15 kpc, respectively, exhibiting significantly tighter orbits than the rest of the LMC satellite candidates. In the case of YMCA-1, such a tight orbit, combined with having a metallicity significantly higher than that of the majority of ultra-faint dwarfs \citep{Cerny2026}, suggests that it is more plausibly a GC than a dwarf galaxy. Pictor II, on the contrary is a DM-dominated galaxy (\citealt{Pace2025}).

Among the rest of confirmed galaxies, only Hydrus I and the SMC have apocentres below 45 kpc. The potential LMC satellites have all performed both their most recent pericentres and apocentres within the last gigayear, except for Phoenix II, which shows the largest orbital period among them (when measured for the earliest extrema, see also Fig~\ref{tab:LMCprob}). 
Currently, all of them are getting away from the LMC, except for Carina II, which seems to be heading towards a new pericentre. 

The derived pericentre and apocentre times also allow us to determine the median time the satellites have to orbit the LMC to be considered as such, under the criterion we propose in this work (see Section~\ref{sec:lmcsat}). Translating this criterion into orbital timescales, candidate satellites must orbit the LMC at least a median of $\sim 0.5$ Gyr when measuring a pericentre-to-apocentre, or $\sim 1$ Gyr for a full orbit between consecutive pericentres.

Similarly to what we observed for the orbits with respect to the MW, the effect of DF on the trajectories of the LMC satellites appears very limited and unlikely to produce significant changes in the estimated orbital parameters (see Figure~\ref{fig:orbitLMC}). Furthermore, the fact that all galaxies except Phoenix II have their most recent pericentre and apocentre within the last gigayear further reduces the expected impact of DF, as there has been insufficient time for it to produce a significant variation in the orbit. The only exception is the SMC, for which DF has a more noticeable effect given its mass. Under the SMC1 model (the lightest SMC model), the orbit shows only mild differences over the last gigayear, so the impact on the derivation of orbital parameters is limited, while under the more massive SMC2 model the effect becomes significant. However, the SMC2 mass is substantially higher than recent SMC mass estimates \citep{DeLeo2024}, and should therefore be regarded as an extreme upper limit rather than a realistic scenario.

\begin{figure*}
    \centering
    \includegraphics[width=\linewidth]{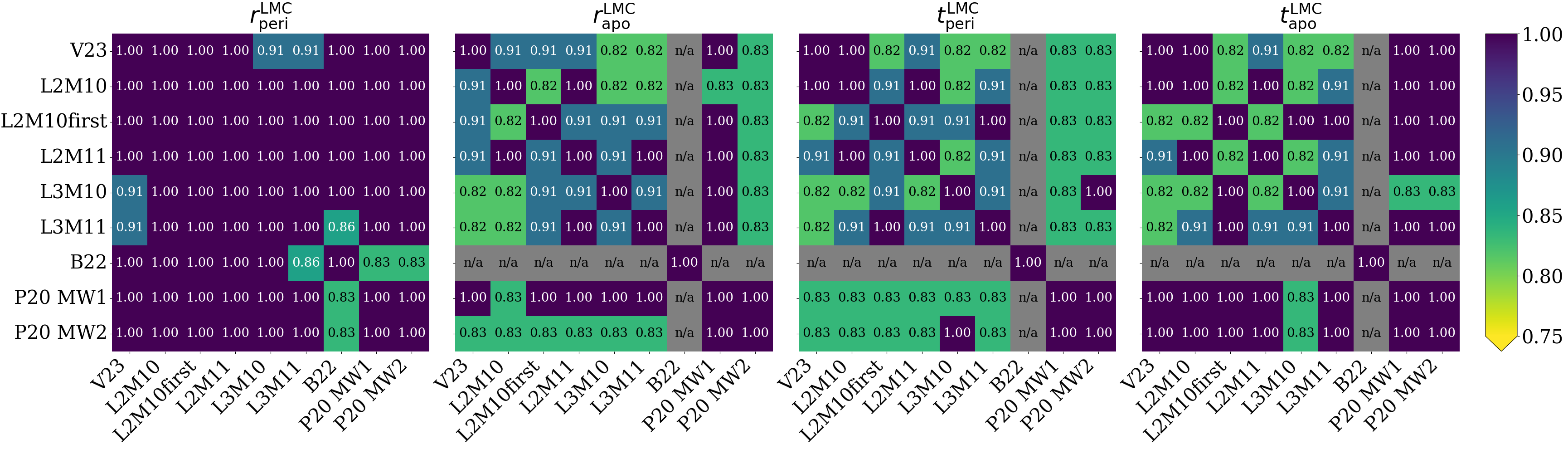}
    \caption{Compatibility of orbital parameters with respect to the LMC, for galaxies with $p_{\rm{LMC}} > 0.1$. Each panel shows a confusion matrix comparing pairwise an orbital parameter obtained with the potentials explored in this work and with those reported by \citet{Battaglia2022} (B22) and \citet{Patel2020} (P20MW1, P20MW2). The panel layout and colour scale are analogous to those in Fig.~\ref{fig:confussion_matrix_self}. Since \citet{Battaglia2022} does not report values for $r_{\rm apo}^{\mathrm{LMC}}$,  and $t_{\rm apo}^{\mathrm{LMC}}$ and uncertainties for the calculation of $t_{\rm peri}^{\mathrm{LMC}}$, the corresponding compatibility entries are labelled as n/a.}
    \label{fig:comparisonLMCparams}
\end{figure*}

Now we compare the orbital parameters obtained for the different potentials with each other, and with those presented in \citet{Battaglia2022} and \citet{Patel2020}. In Figure~\ref{fig:comparisonLMCparams} we present the confusion matrix used to summarize the agreement between the  parameters, analogous to Figure~\ref{fig:confussion_matrix_self}.

We note that for \citet{Battaglia2022} the comparisons are only possible for $r_{\rm peri}^{\mathrm{LMC}}$, because $r_{\rm apo}^{\mathrm{LMC}}$ and $t_{\rm apo}^{\mathrm{LMC}}$ are not provided, and for  $t_{\rm peri}^{\mathrm{LMC}}$ no uncertainties were reported.
In the case of \citet{Patel2020}, the comparisons are performed using the results reported based on two models relying on their fiducial LMC configuration (MW1+LMC2 and MW2+LMC2). We note that the orbit integrations in that work also account for the gravitational potential of the SMC. The orbital parameters reported in \citet{Battaglia2022} did not account for the effect of DF, whereas those of \citet{Patel2020} did include the DF of both the MW and LMC. 

When comparing between the different potentials used in this work, we find excellent agreement for the pericentric distance, only showing very mild discrepancies between V23 and L3M10 and L3M11. For the rest of the parameters, we observe a lower level of compatibility between potentials, however, the compatibility is always $\geq 80\%$, and is $100\%$ in multiple instances. In any case, the differences are mild and due to the different modellings of the potentials. 

When comparing with other works, we find good agreement across all orbital parameters. The agreement is particularly strong for $r_{\rm peri}^{\rm LMC}$, regardless of the potential considered, while compatibility with \citet{Patel2020} reaches 83\% to 100\% of cases for $t_{\rm peri}^{\rm LMC}$ and $r_{\rm apo}^{\rm LMC}$  and shows excellent agreement for $t_{\rm apo}^{\rm LMC}$.

Finally, we performed  Wilcoxon signed-rank tests to assess the presence of  systemic shifts in the  orbital parameters derived under the different potentials (even if they are compatible within uncertainties), as we did in Section~\ref{sec:orbparam} for the parameters measured with respect to the MW. However, we did not find statistically significant systematic shifts, which could be attributed to the reduced sample of galaxies for which we conducted the test.

\subsubsection{The interaction of the LMC and the SMC}
\label{sec:SMCcollision}
The SMC is the quintessential satellite of the LMC and, by far, its most massive companion. With a mass of approximately 10\% that of its host (e.g. \citealt{DeLeo2024, Watkins2024}), it is expected to exert a significant influence on the LMC, particularly during close approaches. Repeated interactions between the Magellanic Clouds are therefore likely to have played a major role in shaping the evolution of both systems, underscoring the importance of reconstructing their mutual orbital history for interpreting their present-day properties. Indeed, several distinctive structural 
features of the LMC have been attributed to such interactions, most notably in the morphology of its bar.

While the LMC bar has been proven to show comparable properties to other barred galaxies in the local Universe (in terms of bar-galaxy scaling relations, \citealt{Rathore2025a}), it is off-centred nearly 1 kpc with respect to the outer disc (\citealt{deVaucouleurs1972, ZhaoEvans2000, vanderMarel2001}), and it is tilted with respect to the disc plane by about $\sim5-15^{\circ}$ (\citealt{Choi2018}). The LMC disc also has been found to have warps (e.g. \citealt{vanderMarelCioni2001, OlsenSalyk2002, Choi2018}).  
Several studies based on observational data and/or numerical simulations have attributed these features to a recent direct collision between the SMC and the LMC disc.

\citet{Besla2012} developed two models of the LMC-SMC interaction: one involving a direct collision 100-300 Myr ago, and another in which the SMC never approached closer than 20 kpc. While the latter reproduced the large-scale structure of the Magellanic System, it poorly matched the LMC's internal properties. In contrast, the collision scenario significantly better reproduces the structure, kinematics, and bar properties of the LMC (\citealt{Besla2016, Choi2022, Rathore2025a}), as well as the PM signatures observed along the Magellanic Bridge (\citealt{Zivick2019}).

Subsequent studies have further refined the collision parameters while trying to match present day properties of the LMC: \citet{Choi2022} constrained the impact parameter to be $\sim5$ kpc and found the collision to have taken place $\sim140-160$ Myr ago, \citet{Rathore2025b} further refined the timing to $\sim150-200$ Myr ago based on the LMC bar offset. Using orbital integration, \citet{Zivick2018} independently confirmed the collision of both Clouds, reporting impact parameters of $7.5 \pm 2.5$~kpc at $147 \pm 33$~Myr ago or $9.7 \pm 4.5$~kpc at $163 \pm 36$~Myr ago, depending on the assumed MW and LMC potentials used for their integration.

Our results for the latest SMC pericentre ($r_{\rm peri}^{\mathrm{LMC}} = 8^{+4}_{-3}$ kpc $t_{\rm peri}^{\mathrm{LMC}} = -0.17^{+0.02}_{-0.05}$ Gyr) are in good agreement with the impact parameters and collision times reported in these studies, thereby confirming the existence of a recent collision between the Magellanic Clouds. This agreement offers a valuable cross-validation between independent analyses and lends confidence to the robustness of our orbital integrations. 

\begin{table*}[]
\renewcommand{\arraystretch}{1.5}
    \centering
    \caption{Pericentres of the SMC around the LMC}
    \begin{tabular}{lcccccccc}
    \toprule
    Potential & $r_{\rm peri,1}^{\mathrm{LMC}}$ & $t_{\rm peri,1}^{\mathrm{LMC}}$ &  $r_{\rm peri,2}^{\mathrm{LMC}}$ & $t_{\rm peri,2}^{\mathrm{LMC}}$ & $r_{\rm peri,3}^{\mathrm{LMC}}$ & $t_{\rm peri,3}^{\mathrm{LMC}}$ & $r_{\rm peri,4}^{\mathrm{LMC}}$ & $t_{\rm peri,4}^{\mathrm{LMC}}$ \\   
    & [kpc] & [Gyr] & [kpc] & [Gyr] & [kpc] & [Gyr] & [kpc] & [Gyr] \\
    \multicolumn{1}{c}{(1)}  & (2) & (3) & (4) & (5) & (6) & (7) & (8) & (9) \\
    \midrule
    \midrule
        V23 & $8^{+4}_{-3}$  &  $-0.17^{+0.03}_{-0.05}$  &  $11^{+9}_{-4}$  &  $-1.14^{+0.18}_{-0.44}$ & $11^{+8}_{-4}$  &  $-2.15^{+0.42}_{-0.89}$ & $11^{+5}_{-4}$  &  $-3.07^{+0.58}_{-0.97}$ \\
        L2M10 & $8^{+4}_{-3}$  &  $-0.17^{+0.02}_{-0.05}$  & $11^{+10}_{-5}$  &  $-1.35^{+0.31}_{-0.59}$  & $11^{+7}_{-4}$  &  $-2.55^{+0.65}_{-1.12}$  &  $9^{+3}_{-3}$  &  $-3.38^{+0.72}_{-1.05}$  \\
        L2M10first & $8^{+4}_{-3}$  &  $-0.16^{+0.03}_{-0.04}$ & $11^{+9}_{-4}$  &  $-1.12^{+0.18}_{-0.39}$  &  $11^{+7}_{-4}$  &  $-2.11^{+0.4}_{-0.76}$  &  $9^{+4}_{-3}$  &  $-2.87^{+0.48}_{-0.63}$\\
        L2M11 & $8^{+4}_{-3}$  &  $-0.17^{+0.03}_{-0.04}$  & $11^{+9}_{-5}$  &  $-1.26^{+0.24}_{-0.5}$  & $11^{+9}_{-5}$  &  $-2.42^{+0.56}_{-1.15}$  & $9^{+4}_{-3}$  &  $-3.27^{+0.65}_{-0.98}$ \\
        L3M10 & $8^{+4}_{-3}$  &  $-0.15^{+0.03}_{-0.04}$  & $10^{+8}_{-4}$  &  $-1.05^{+0.18}_{-0.4}$  & $10^{+7}_{-4}$  &  $-1.99^{+0.41}_{-0.82}$  &  $10^{+5}_{-4}$  &  $-2.86^{+0.58}_{-0.89}$\\  
        L3M11 & $8^{+4}_{-3}$  &  $-0.17^{+0.03}_{-0.05}$ & $12^{+11}_{-4}$  &  $-1.16^{+0.2}_{-0.47}$ & $12^{+11}_{-4}$  &  $-2.24^{+0.49}_{-1.16}$ & $11^{+4}_{-3}$  &  $-3.1^{+0.6}_{-1.0}$ \\
    \bottomrule
    \end{tabular}
    \tablefoot{Orbital parameters of the last 4 pericentres of the SMC around the LMC. Columns show, from left to right, the pericentric distance and the time of each passage, ordered from the most recent to the earliest, for all explored potentials.}
    \label{tab:SMCpassages}
\end{table*}

The reconstructed orbits of the SMC reveal that its recent collision with the LMC is unlikely to be an isolated event. In Table~\ref{tab:SMCpassages} we report the distance and time of the most recent and previous pericentres of the SMC about the LMC, for all the potentials. The pericentric distances are consistent with additional collisions with the LMC, but with more outer regions, roughly every 1 Gyr.

Interestingly, the timing of these collisions broadly coincides with episodes of intense star formation reported in the Magellanic Clouds. \citet{Massana2022} derived the SFH of the SMC finding peaks in the star formation rate approximately 3, 2, 1.1, and 0.45 Gyr ago, plus an ongoing episode of enhanced star formation. The comparison with the SFH of the LMC (\citealt{RuizLara2020}) showed that the star formation of the Magellanic Clouds is correlated over the last 3.5 Gyr, with synchronized peaks, likely driven by their mutual interaction. 

The peaks in the star formation rate at $\sim$1.1, 2, and 3 Gyr correspond well with the pericentres we find for the SMC at approximately $-1.2$, $-2.2$, and $-3.1$ Gyr (see Table~\ref{tab:SMCpassages}), while for the most recent pericentre, occurring $\sim$0.2 Gyr ago, the association is less straightforward. It could be linked to either the peak at $\sim$0.45 Gyr or the ongoing episode, given the uncertainties inherent to SFH derivations based on CMD fitting and the fact that enhanced star formation triggered by galaxy interactions can be delayed with respect to the moment of closest approach (e.g., \citealt{Moreno2015}). The ongoing episode could additionally be due to the very recent pericentre of the LMC system about the MW ($\sim$0.05 Gyr ago).

Analogous correlations between orbital histories and star formation activity have been reported for MW satellites (e.g. \citealt{RuizLara2020Sag, Rusakov2021, Bennet2024}). Nevertheless, while the temporal coincidences between the SMC pericentres and the SFH peaks are suggestive, the limitations of the orbit reconstruction prevent us from establishing a definitive causal link between the two.

Orbital parameters of the earlier SMC pericentres have  large uncertainties and should be interpreted carefully. Several approximations  in our modelling for the integration become increasingly relevant at earlier times for the SMC: the SMC is treated as a point-mass particle moving in the LMC potential, neglecting the mutual reflex motion between the two galaxies, the mass evolution of the system, and the dynamical response to collisions. Furthermore, the cumulative effect of DF on the SMC,  becomes more significant at earlier times, potentially affecting the reconstructed pericentre distances and timings for the earliest passages. These approximations  limit our ability to draw firm conclusions on the connection between the orbital and star formation histories of the Magellanic Clouds, and a more detailed modelling of the SMC-LMC interaction would be required to do so.

\subsubsection{Capture of MW satellites by the LMC}
The LMC is currently merely $\sim$50 kpc from the MW \citep{Pietrzynski2019}, and since it crossed the MW virial radius approximately 1.5 Gyr ago, it has been moving alongside its cohort of satellites through a region with a high density of dwarfs, namely the MW satellites. Given the LMC's intense gravitational potential and its close proximity to several MW dwarfs along its path, it cannot be ruled out that the LMC has captured some of them, thereby augmenting its own satellite population.

We explore this possibility by analysing those MW satellites that have recently performed close approaches to the LMC. We select dwarfs having pericentres with respect to the LMC in the last gigayear that reached within 1.5 times the tidal radius of the LMC, within 3$\sigma$. We consider this a reasonably close approach to the LMC and a conservative way to account for a potential underestimation of the tidal radius, while simultaneously enlarging the sample of galaxies available for analysis.

We find 9 galaxies that have performed such approaches to the LMC: Aquarius II, Aquarius III, Grus II, Pegasus IV, Reticulum III, Sagittarius II, Tucana III, Tucana IV and Tucana V. For the majority of them we find that their velocities relative to the LMC are well above the LMC escape velocity at all times.

\begin{figure*}
    \centering
    \includegraphics[width=\linewidth]{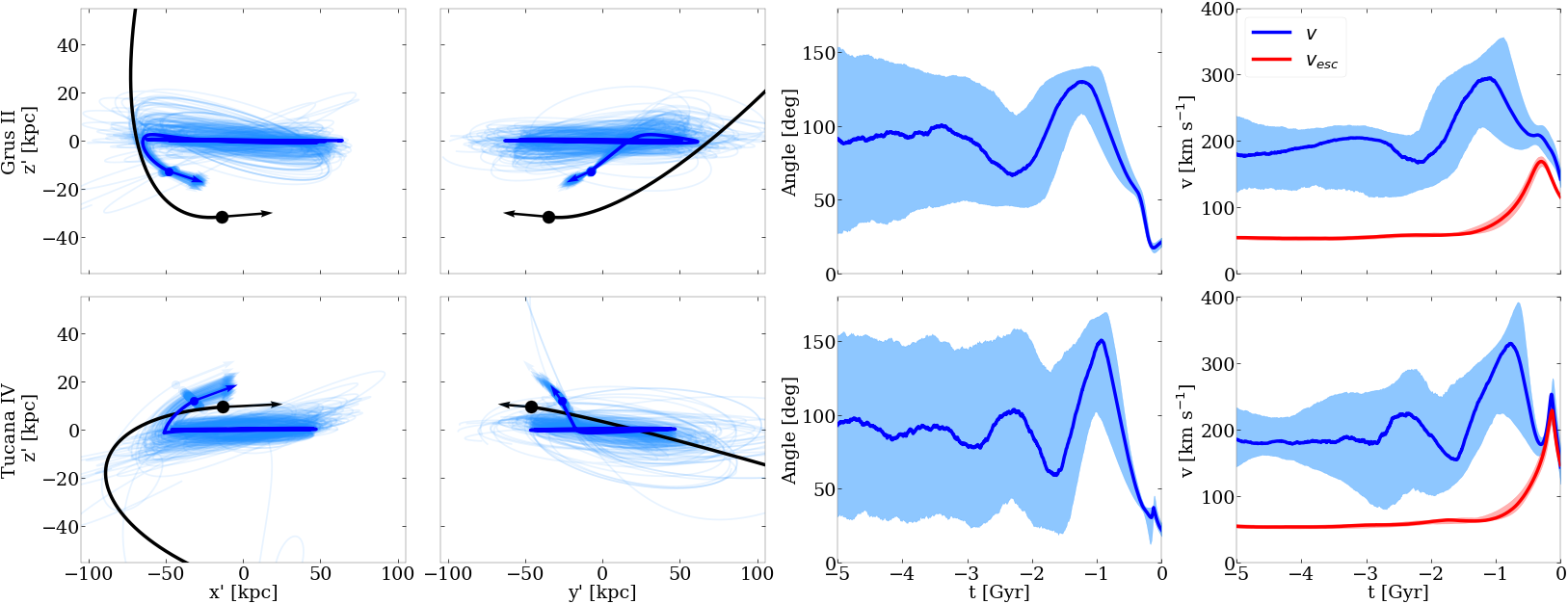}
    \caption{
Impact of the LMC passage on the trajectories of Grus II (top) and Tucana IV (bottom), under potential L2M10. For each row, the first and second panels from the left show the orbits of the corresponding dwarf and that of the LMC, represented by blue and black lines, respectively. Current positions and velocity directions are marked with solid dots and arrows. For the dwarfs, we show in light blue 100 randomly selected orbits from the MC realizations to display a sample of the ensemble of possible trajectories, and in dark blue, the orbit calculated using the nominal values of the phase-space coordinates. Note that the coordinates $x', y', z'$ do not correspond to Galactic Cartesian coordinates but are aligned with the principal axes of the nominal trajectory to better visualize the orbital deflection by showing the orbital plane in the $x', y'$ projection. The third column shows the evolution of the angle between the velocity vectors of the dwarf and the LMC. The fourth column shows the velocity of the dwarfs relative to the LMC (blue) and the escape velocity of the LMC at the position of the dwarf (red), as a function of time. Shaded regions represent the 16th and 84th percentile intervals of the corresponding measurements.}
    \label{fig:capgrstuc}
\end{figure*}

However, we find two notable exceptions: Grus II and Tucana IV.
In Figure~\ref{fig:capgrstuc} we show their trajectories, the angle between their velocity vector and that of the LMC, and their velocity relative to the LMC compared to the escape velocity in their position as a function of time, for potential L2M10 (see the analogous figures for the rest of potentials in the repository). 
Both galaxies have performed very close passages by the LMC, reaching its maximum approach of $22.9^{+2.9}_{-2.9}$ kpc for Grus II and only $5.3^{+3.0}_{-1.7}$ kpc for Tucana IV, at $-0.29^{+0.04}_{-0.05}$ Gyr and $-0.14 ^{+0.02}_{-0.03}$ Gyr, respectively. In the case of Tucana IV, and similarly to what we report for other LMC satellites (see Section~\ref{sec:orbparamlmc}), such close approach to the LMC means that it has collided with it. \citealt{Simon2020} already argued for the collision of Tucana IV with the LMC, reporting a collision parameter of $4.1^{+3.2}_{-2.2}$ kpc, and that the collision took place $119^{+26}_{-18}$ Myr ago.

Prior to the passage by the LMC, both galaxies 
displayed relatively planar orbits. Over the last 0.5 Gyr they have suffered a drastic deflection of their trajectories, reorienting towards the direction of motion of the LMC (suffering a variation in the direction of the velocity of nearly $\sim$90 degrees in the last gigayear). We observe this trend across all the considered potentials. For Grus II, the reorientation of the orbit entailed the alignment of its specific angular momentum with that of the VPOS, becoming the only newly acquired member of the structure (\citealt{MartinezGarcia2025}).

Regarding velocities relative to the LMC, Tucana IV is currently slightly below  the escape velocity across all potentials, while Grus II is only for potential V23, for the rest of potentials its velocity is above the escape. This suggests that these MW satellites could have been recently captured by the LMC, with such capture less likely and model-dependent for Grus II. However, definitive confirmation of capture is challenging because both galaxies are currently experiencing a complex gravitational interaction with  the MW and the LMC, where the dynamics are dominated by the competing influences of both  galaxies. We note that under our criterion for LMC satellite membership, these galaxies do not qualify: they have not yet performed successive pericentres (or a pericentre and an apocentre) within the LMC tidal radius. Additionally, the duration of their interaction with the LMC remains below the estimated $\sim 0.5\text{--}1\text{ Gyr}$ lower limit for bound satellites (Section~\ref{sec:orbparamlmc}).

Several studies have explored the possibility of some MW satellites having been captured by the LMC. \citet{Patel2020} found, across multiple MW + LMC and MW + LMC + SMC potentials, that Reticulum II and Phoenix II could have been recently captured. Our results do not support this scenario: the probabilities we report for these galaxies are based on a criterion of at least two passages around the LMC, qualifying them as long-term LMC satellites, and are generally above 0.6, except for Reticulum II under potentials L2M10 and L2M11.

\citealt{Battaglia2022} argued that Grus II could be a former MW satellite captured by the LMC, using a potential very similar to our V23, under which we find that Grus II's velocity is below the LMC escape velocity. For Tucana IV, they further argued that neglecting the SMC potential during orbit integration could lead to misclassifying it as a MW satellite rather than a long-term LMC satellite. \citet{Simon2020} similarly found that the probability of Tucana IV having always been an LMC satellite increased to $\sim 20\%$ when the SMC potential was included.

To investigate this further, we integrated the orbit of Tucana IV backwards in time using potentials that include the SMC's gravitational influence. For each of the six MW + LMC potentials used in this work, we considered the SMC models presented in  \citet{Patel2020} (SMC1 and SMC2), yielding a total of 12 MW + LMC + SMC combinations. Further details on the implementation and a discussion of potential caveats are provided in Appendix~\ref{sec:inttuc}.

Table~\ref{tab:TucIVprob} compares the probabilities of Tucana IV being a long-term LMC satellite under the six MW + LMC potentials and the 12 MW + LMC + SMC potentials, using the criterion described in Section~\ref{sec:potlmcsat}. We find no evidence that including the SMC increases this likelihood. The only exception is potential V23, with a heavy SMC (SMC2), which yields a slight increase relative to its MW + LMC counterpart, though the resulting probability of 0.19 is insufficient to draw firm conclusions. Overall, the probabilities of being and LMC satellite remain low across all potentials, supporting the classification of Tucana IV as a long-term MW satellite that may have only recently been captured by the LMC.

\begin{table}[]
    \centering
    \fontsize{10}{12}\selectfont
    \setlength{\tabcolsep}{3pt} 
    \caption{Probability of Tucana IV being a long-term satellite of the LMC.}
    \begin{tabular}{lccc}
    \toprule
    Base potential & MW + LMC & MW + LMC & MW + LMC\\
     & &  + SMC1 &  + SMC2 \\
     \multicolumn{1}{c}{(1)}  & (2) & (3) & (4)\\

    \midrule
    \midrule
V23         & 0.034 & 0.017 & 0.148 \\
L2M10       & 0.009 & 0.002 & 0.001 \\
L2M10first  & 0.013 & 0.006 & 0.004 \\
L2M11       & 0.005 & 0.002 & 0.0 \\
L3M10       & 0.014 & 0.011 & 0.004 \\
L3M11       & 0.008 & 0.002 & 0.001 \\
    \bottomrule
            
    \end{tabular}
    \tablefoot{Probability of Tucana IV being a long term satellite of the LMC, under different combinations of potentials, with or without SMC. Columns from left to right show: (1) the base potential name, (2) the probability of Tucana IV being a long term satellite of the LMC for each of the potentials described in Section~\ref{sec:pot} (3) for the potentials including the SMC1 model, and (4) including the SMC2 model.}
    \label{tab:TucIVprob}
\end{table}

\section{Conclusions}
\label{sec:conclusions}
We backward-integrated the orbits of 72 dwarf galaxies in the vicinity of the MW, using accurate phase-space coordinates from the literature (\citealt{Battaglia2022, Pace2024, Cerny2026}) and a suite of six time-evolving gravitational potentials that account for the mutual interaction of the MW and the LMC (\citealt{Vasiliev2023, Vasiliev2024}). We present the largest catalogue to date of orbital parameters for the MW dwarfs in terms of galaxies and potentials explored. Our main findings are:

\begin{itemize}
\item Orbital parameters with respect to the MW: We derived orbital parameters for the full dwarf sample, reporting pericentric and apocentric distances, the times at which they took place, alongside specific orbital energies at $t = 0$. The orbital parameters show excellent consistency across the six tested potentials, and are also in good agreement with previous studies by \citet{Battaglia2022} and \citet{Pace2022}. Based on the specific orbital energy, we find that the majority of galaxies in our sample are bound to the MW+LMC system, though several systems show an ambiguous binding status or are unbound (e.g. Eridanus III, Columba I, Leo T, or Phoenix). We note that in many cases, this is a result of these galaxies having poorly constrained PMs. We also estimated the infall times of the dwarfs, finding that the majority of them have spent the last 5 Gyr within the virial radius of the MW. 

\item LMC satellite population: We identified ten galaxies as potential LMC satellites based on their orbital histories in the LMC rest frame: Carina II, Carina III, Horologium I, Horologium II, Hydrus I, Phoenix II, Pictor II, Reticulum II, the SMC and YMCA-1. Hydrus I displays the highest probability ($p_{\rm LMC}= 1$, across all potentials), followed by Carina II, Carina III, Pictor II, the SMC  and YMCA-1 (probabilities typically above 0.9). Phoenix II shows substantial but lower probabilities ($\sim$0.65–0.85), while Horologium I, Horologium II, and Reticulum II exhibit intermediate probabilities and a more model-dependent behaviour. We also studied the possibility of DELVE 6 being an LMC satellite; however, the low probabilities  obtained ($< 0.2$) cast doubts on its membership. This is a result of DELVE 6 large uncertainty in its PMs.

\item Orbital parameters of LMC satellites: We report orbital parameters with respect to the LMC for its potential satellites. The likely LMC satellites display close pericentric passages, typically reaching distances below $\sim15$ kpc from the LMC centre, suggesting interactions with the LMC's extended stellar disc in multiple cases. All satellites remain within $\sim$50 kpc of the LMC at apocentre, and most have performed both their most recent pericentric and apocentric passages within the last gigayear. The orbital parameters show good agreement across different potentials and with previous studies, such as \citet{Battaglia2022} and  \citet{Patel2020}.

\item LMC-SMC interaction: Our reconstruction of the SMC orbit relative to the LMC confirms a recent direct collision, whose impact parameter and timing are in excellent agreement with observational and simulation-based constraints from \citet{Choi2022}, \citet{Rathore2025b}, and \citet{Zivick2018}, and supports the collision scenario proposed to explain the LMC's offset bar, tilted disc, and warped structure. 
When exploring the orbit of the SMC at earlier times, we find multiple pericentres whose distance reveals a series of recurrent collisions with the LMC. The timing of these collisions is broadly consistent with the star formation rate peaks reported in both Magellanic Clouds by \citet{Massana2022} and \citet{RuizLara2020}. This temporal agreement would suggest that the synchronized star formation histories of the LMC and SMC are driven by their repeated close interactions. 

\item Potential capture of MW satellites: We investigated whether the LMC has captured any MW satellites during its recent infall. Among nine MW satellites that performed close approaches to the LMC in the last gigayear, we identified Grus II and Tucana IV as notable cases showing evidence of possible capture. Both galaxies underwent drastic orbital deflections ($\sim$90° changes in velocity direction) during their recent approaches to the LMC and currently display velocities slightly below  the its escape velocity (Tucana IV across all potentials; Grus II in one of them). For Grus II, the orbital reorientation resulted in alignment with the VPOS \citep{MartinezGarcia2025}. However, definitive confirmation of capture remains challenging due to the complex gravitational interaction these galaxies currently experience between the MW and LMC. For Tucana IV, we tested whether including the SMC potential in the integration of its orbit would increase the probability of it being a long-term LMC satellite, as suggested by previous studies, but found no significant evidence supporting this scenario. We conclude that both Grus II and Tucana IV are likely former MW satellites that may be in the process of being captured by the LMC.
\end{itemize}

The good agreement of the orbital parameters derived with the different potentials, and with results from other works, together with their consistency for the SMC with predictions from the LMC--SMC direct collision scenario, supports the robustness of our results. The comprehensive catalogue of orbital parameters presented here, available in the associated repository \footnote{\url{https://github.com/ammg-astro/OrbitIntegration}}, provides a valuable resource for interpreting the observed properties of MW and LMC satellites. We note that future, more accurate PM measurements from \textit{Gaia}, \textit{HST}, or \textit{Roman} will further improve our understanding of the orbital history of dwarf galaxies in the vicinity of the MW.

\begin{acknowledgements}
The authors thank E. Vasiliev for making publicly available the code \textsc{Agama} (\citealt{AGAMA2018, AGAMA2019}), for his assistance with questions regarding its use, and discussions on this work. The authors also thank C. López-Sanjuan for his insights and discussions throughout this work.
A. M. Martínez-García, A. del Pino, and F. Nogueras-Lara acknowledge financial support from the \emph{Severo Ochoa} grant CEX2021-001131-S (MICIU/AEI/10.13039/501100011033). A. M. Martínez-García and A. del Pino acknowledge financial support from the Spanish Ministry of Science and Innovation project PID2024-155572NB-C22 (MICIU/AEI/10.13039/501100011033, FEDER, EU), and the RyC-MAX grant 20245MAX008 (CSIC). A.~del Pino also acknowledges funding from the Ram\'on y Cajal fellowship RYC2022-038448-I (MICIU/AEI/10.13039/501100011033, co-funded by the European Social Fund Plus), and CNS2025-166660.
G. Battaglia acknowledges support from the Agencia Estatal de Investigación del Ministerio de Ciencia, Innovación y Universidades (MCIU/AEI) under grant `En la frontera de la arqueología galáctica: evolución de la materia luminosa y oscura de la Vía Láctea y las galaxias enanas del Grupo Local en la era de Gaia' (FOGALERA) and the European Regional Development Fund (ERDF) with reference PID2023-150319NB-C21/10.13039/501100011033.
E. L. {\L}okas acknowledges support from the National Science Centre of Poland under grant 2025/57/B/ST9/00321.
E. Vitral acknowledges funding from the Royal Society, under the Newton International Fellowship programme (NIF\textbackslash R1\textbackslash 241973).
K. A. McKinnon acknowledges support from the University of Toronto’s Eric and Wendy Schmidt AI in Science Post-doctoral Fellowship, a program of Schmidt Sciences. 
G. F. Thomas acknowledges support from the Agencia Estatal de Investigaci\'on del Ministerio de Ciencia en Innovaci\'on (AEI-MCIN) under grant number PID2023-150319NB-C21 and the grant RYC2024-051016-I funded by MCIN/AEI/10.13039/501100011033 and by the European Social Fund Plus.
S. Cardona-Barrero  acknowledges financial support from the Spanish Ministry of Science and Innovation (MICINN) through RYC2022-035838-I and PID2021-128131NB-I00 (CoBEARD project).
B. Anguiano thanks the support by the Spanish Ministry of Science, Innovation and Universities and the State Research Agency (MICIU/AEI) with the grant RYC2022-037011-I and by the European Social Fund Plus (FSE+).
F. Nogueras-Lara gratefully acknowledges financial support from the Ramón y Cajal programme (RYC2023-044924-I), funded by MCIN/AEI/10.13039/501100011033 and FSE+; from grant PID2024-162148NA-I00, funded by MCIN/AEI/10.13039/501100011033 and the European Regional Development Fund (ERDF) “A way of making Europe”. 
A. Hidalgo-Pinilla acknowledges the financial support provided by grants PID2024-155572NB-C21 and PID2024-155572NB-C22 (MICIU/AEI/10.13039/501100011033 and ERDF/EU), and by grant PID2021-124918NB-C41 (MCIN/AEI/10.13039/501100011033 and ERDF/EU).
This work has made use of data from the European Space Agency (ESA) mission Gaia (https://www.cosmos.esa.int/gaia), processed by the Gaia Data Processing and Analysis Consortium (DPAC, https://www.cosmos.esa.int/web/gaia/dpac/consortium). Funding for the DPAC has been provided by national institutions, in particular the institutions participating in the Gaia Multilateral Agreement.
A. M. Martínez-García thanks A. Almeida for her support and discussions on this work.

\end{acknowledgements}

\bibliographystyle{aa} 
\bibliography{biblio.bib}

@ARTICLE{McConnachie2012,
       author = {{McConnachie}, Alan W.},
        title = "{The Observed Properties of Dwarf Galaxies in and around the Local Group}",
      journal = {\aj},
         year = 2012,
        month = jul,
       volume = {144},
       number = {1},
          eid = {4},
        pages = {4},
          doi = {10.1088/0004-6256/144/1/4},
archivePrefix = {arXiv},
       eprint = {1204.1562},
 primaryClass = {astro-ph.CO},
       adsurl = {https://ui.adsabs.harvard.edu/abs/2012AJ....144....4M}
}

@article{GaiaCollaboration2016,
author = {{Gaia Collaboration} and Prusti, T and {J de Bruijne}, J H and {A Brown}, A G and Vallenari, A and Babusiaux, C and {L Bailer-Jones}, C A and Bastian, U and Biermann, M and Evans, D W and Eyer, L and Jansen, F and Jordi, C and Klioner, S A and Lammers, U and Lindegren, L and Luri, X and Mignard, F and Milligan, D J and Panem, C and Poinsignon, V and Pourbaix, D and Randich, S and Sarri, G and Sartoretti, P and Siddiqui, H I and Soubiran, C and Valette, V and van Leeuwen, F and Walton, N A and Aerts, C and Arenou, F and Cropper, M and Drimmel, R and H{\o}g, E and Katz, D and Lattanzi, M G and Grebel, E K and Holland, A D and Huc, C and Passot, X and Bramante, L and Cacciari, C and Casta{\~{n}}eda, J and Chaoul, L and Cheek, N and {De Angeli}, F and Fabricius, C and Guerra, R and Hern{\'{a}}ndez, J and Jean-Antoine-Piccolo, A and Masana, E and Messineo, R and Mowlavi, N and Nienartowicz, K and Ord{\'{o}}{\~{n}}ez-Blanco, D and Panuzzo, P and Portell, J and Richards, P J and Barache, C and Barata, C and Barbier, A and Barblan, F and Baroni, M and {Barrado Navascu{\'{e}}s}, D and Barros, M and Barstow, M A and Becciani, U and Bellazzini, M and Bellei, G and {Bello Garc{\'{i}}a}, A and Belokurov, V and Bendjoya, P and Berihuete, A and Bianchi, L and Bienaym{\'{e}}, O and Billebaud, F and Blagorodnova, N and Blanco-Cuaresma, S and Boch, T and Bombrun, A and Borrachero, R and Bouquillon, S and Bourda, G and Bouy, H and Bragaglia, A and Breddels, M A and Brouillet, N and Br{\"{u}}semeister, T and Bucciarelli, B and Budnik, F and Burgess, P and Burgon, R and Burlacu, A and Busonero, D and Buzzi, R and Caffau, E and Cambras, J and Campbell, H and Cancelliere, R and Cantat-Gaudin, T and Carlucci, T and Carrasco, J M and Castellani, M and Charlot, P and Charnas, J and Charvet, P and Chassat, F and Chiavassa, A and Clotet, M and Cocozza, G and Collins, R S and Collins, P and Costigan, G and Crifo, F and {G Cross}, N J and Crosta, M and Crowley, C and Dafonte, C and Damerdji, Y and Dapergolas, A and David, P and David, M and {De Cat}, P and de Felice, F and de Laverny, P and {De Luise}, F and {De March}, R and de Martino, D and de Souza, R and Debosscher, J and del Pozo, E and Delbo, M and Delgado, A and Delgado, H E and di Marco, F and {Di Matteo}, P and Diakite, S and Distefano, E and Dolding, C and {Dos Anjos}, S and Drazinos, P and Dur{\'{a}}n, J and Dzigan, Y and Ecale, E and Edvardsson, B and Enke, H and Erdmann, M and Escolar, D and Espina, M and Evans, N W and {Eynard Bontemps}, G and Fabre, C and Fabrizio, M and Faigler, S and Falc{\~{a}}o, A J and {Farr{\`{a}}s Casas}, M and Faye, F and Federici, L and Fedorets, G and Fern{\'{a}}ndez-Hern{\'{a}}ndez, J and Fernique, P and Fienga, A and Figueras, F and Filippi, F and Findeisen, K and Fonti, A and Fouesneau, M and Fraile, E and Fraser, M and Fuchs, J and Furnell, R and Gai, M and Galleti, S and Galluccio, L and Garabato, D and Garc{\'{i}}a-Sedano, F and Gar{\'{e}}, P and Garofalo, A and Garralda, N and Gavras, P and Gerssen, J and Geyer, R and Gilmore, G and Girona, S and Giuffrida, G and Gomes, M and Gonz{\'{a}}lez-Marcos, A and Gonz{\'{a}}lez-N{\'{u}}{\~{n}}ez, J and Gonz{\'{a}}lez-Vidal, J J and Granvik, M and Guerrier, A and Guillout, P and Guiraud, J and G{\'{u}}rpide, A and Guti{\'{e}}rrez-S{\'{a}}nchez, R and Guy, L P and Haigron, R and Hatzidimitriou, D and Haywood, M and Heiter, U and Helmi, A and Hobbs, D and Hofmann, W and Holl, B and Holland, G and {S Hunt}, J A and Hypki, A and Koubsky, P and Kowalczyk, A and Krone-Martins, A and Kudryashova, M and Kull, I and Bachchan, R K and Lacoste-Seris, F and Lanza, A F and Lavigne, J-b and {Le Poncin-Lafitte}, C and Lebreton, Y and Lebzelter, T and Leccia, S and Leclerc, N and Lecoeur-Taibi, I and Lemaitre, V and Lenhardt, H and Leroux, F and Liao, S and Licata, E and {P Lindstr{\o}m}, H E and Lister, T A and Livanou, E and Lobel, A and L{\"{o}}ffler, W and L{\'{o}}pez, M and Lopez-Lozano, A and Lorenz, D and Loureiro, T and MacDonald, I and {Magalh{\~{a}}es Fernandes}, T and Managau, S and Mann, R G and Mantelet, G and Marchal, O and Pagani, C and Pagano, I and Pailler, F and Palacin, H and Palaversa, L and Parsons, P and Paulsen, T and Pecoraro, M and Pedrosa, R and Pentik{\"{a}}inen, H and Pereira, J and Pichon, B and Piersimoni, A M and Pineau, F-x and Plachy, E and Plum, G and Poujoulet, E and Pr{\v{s}}a, A and Pulone, L and Ragaini, S and Rago, S and Rambaux, N and Ramos-Lerate, M and Ranalli, P and Rauw, G and Read, A and Regibo, S and Renk, F and Reyl{\'{e}}, C and Ribeiro, R A and Rimoldini, L and Ripepi, V and Riva, A and Rixon, G and Roelens, M and Romero-G{\'{o}}mez, M and Rowell, N and Royer, F and Rudolph, A and Ruiz-Dern, L and Sadowski, G and {Sagrist{\`{a}} Sell{\'{e}}s}, T and Sahlmann, J and Salgado, J and Salguero, E and Sarasso, M and Savietto, H and Schnorhk, A and Schultheis, M and Sciacca, E and Segol, M and Segovia, J C and Segransan, D and Serpell, E and Shih, I-c and Smareglia, R and Smart, R L and Smith, C and Solano, E and Solitro, F and Sordo, R and {Soria Nieto}, S and Souchay, J and Spagna, A and Spoto, F and Stampa, U and Steele, I A and Steidelm{\"{u}}ller, H and Stephenson, C A and Stoev, H and Suess, F F and S{\"{u}}veges, M and Surdej, J and Szabados, L and Szegedi-Elek, E and Tapiador, D and Taris, F and Tauran, G and Taylor, M B and Teixeira, R and Terrett, D and Tingley, B and Trager, S C and Turon, C and Ulla, A and Utrilla, E and Valentini, G and van Elteren, A and {Van Hemelryck}, E and van Leeuwen, M and Varadi, M and Vecchiato, A and Veljanoski, J and Via, T and Vicente, D and Vogt, S and Voss, H and Votruba, V and Voutsinas, S and Walmsley, G and Weiler, M and Weingrill, K and Werner, D and Wevers, T and Whitehead, G and Wyrzykowski, {\L} and Yoldas, A and {\v{Z}}erjal, M and Zucker, S and Zurbach, C and Zwitter, T and Alecu, A and Allen, M and {Allende Prieto}, C and Amorim, A and Anglada-Escud{\'{e}}, G and Arsenijevic, V and Azaz, S and Balm, P and Beck, M and Bernstein, H-h and Bigot, L and Bijaoui, A and Blasco, C and Bonfigli, M and Bono, G and Boudreault, S and Bressan, A and Brown, S and Brunet, P-m and Bunclark, P and Buonanno, R and Butkevich, A G and Carret, C and Carrion, C and Chemin, L and Ch{\'{e}}reau, F and Corcione, L and Darmigny, E and de Boer, K S and de Teodoro, P and de Zeeuw, P T and {Delle Luche}, C and Domingues, C D and Dubath, P and Fodor, F and Fr{\'{e}}zouls, B and Fries, A and Fustes, D and Fyfe, D and Gallardo, E and Gallegos, J and Gardiol, D and Gebran, M and Gomboc, A and G{\'{o}}mez, A and Grux, E and Gueguen, A and Heyrovsky, A and Hoar, J and Iannicola, G and {Isasi Parache}, Y and Janotto, A-m and Joliet, E and Jonckheere, A and Keil, R and Kim, D-w and Klagyivik, P and Klar, J and Knude, J and Kochukhov, O and Kolka, I and Kos, J and Kutka, A and Lainey, V and LeBouquin, D and Liu, C and Loreggia, D and Makarov, V V and Marseille, M G and Martayan, C and Martinez-Rubi, O and Massart, B and Meynadier, F and Mignot, S and Munari, U and Nguyen, A-t and Nordlander, T and Ocvirk, P and {Olias Sanz}, A and Ortiz, P and Osorio, J and Oszkiewicz, D and Ouzounis, A and Palmer, M and Park, P and Pasquato, E and Peltzer, C and Peralta, J and P{\'{e}}turaud, F and Pieniluoma, T and Pigozzi, E},
doi = {10.1051/0004-6361/201629272},
journal = {\aap},
title = {{The Gaia mission}},
url = {http://www.cosmos.esa.int/gaia},
volume = {595},
year = {2016},
pages = {A1}
}

@article{Sohn2017,
author = {Sohn, Sangmo Tony and Patel, Ekta and Besla, Gurtina and van der Marel, Roeland P and Bullock, James S and Strigari, Louis E and van de Ven, Glenn and Walker, Matt G and Bellini, Andrea},
doi = {10.3847/1538-4357/aa917b},
journal = {\apj},
month = {nov},
number = {2},
pages = {93},
publisher = {American Astronomical Society},
title = {{Space Motions of the Dwarf Spheroidal Galaxies Draco and Sculptor Based on HST Proper Motions with a {\~{}}10 yr Time Baseline}},
url = {https://doi.org/10.3847{\%}2F1538-4357{\%}2Faa917b},
volume = {849},
year = {2017}
}

@article{Fritz2018,
author = {Fritz, T. K. and Battaglia, G. and Pawlowski, M. S. and Kallivayalil, N. and van der Marel, R. and Sohn, S. T. and Brook, C. and Besla, G.},
doi = {10.1051/0004-6361/201833343},
journal = {A{\&}A},
pages = {A103},
title = {{Gaia DR2 proper motions of dwarf galaxies within 420 kpc - Orbits, Milky Way mass, tidal influences, planar alignments, and group infall}},
url = {https://doi.org/10.1051/0004-6361/201833343},
volume = {619},
year = {2018}
}

@article{Sohn2013,
author = {Sohn, Sangmo Tony and Besla, Gurtina and van der Marel, Roeland P and Boylan-Kolchin, Michael and Majewski, Steven R and Bullock, James S},
doi = {10.1088/0004-637x/768/2/139},
journal = {\apj},
month = {apr},
number = {2},
pages = {139},
publisher = {American Astronomical Society},
title = {{{\{}THE{\}} {\{}SPACE{\}} {\{}MOTION{\}} {\{}OF{\}} {\{}LEO{\}} I:{\{}HUBBLE{\}} {\{}SPACE{\}} {\{}TELESCOPEPROPER{\}} {\{}MOTION{\}} {\{}AND{\}} {\{}IMPLIED{\}} {\{}ORBIT{\}}}},
url = {https://doi.org/10.1088{\%}2F0004-637x{\%}2F768{\%}2F2{\%}2F139},
volume = {768},
year = {2013}
}

@article{Fabricius2020,
	author = {Fabricius, C. and Luri, X. and Arenou, F. and Babusiaux, C. and Helmi, A. and Muraveva, T. and Reyl\'e, C. and Spoto, F. and Vallenari, A. and Antoja, T. and Balbinot, E. and Barache, C. and Bauchet, N. and Bragaglia, A. and Busonero, D. and Cantat-Gaudin, T. and Carrasco, J. M. and Diakit\'e, S. and Fabrizio, M. and Figueras, F. and Garcia-Gutierrez, A. and Garofalo, A. and Jordi, C. and Kervella, P. and Khanna, S. and Leclerc, N. and Licata, E. and Lambert, S. and Marrese, P. M. and Masip, A. and Ramos, P. and Robichon, N. and Robin, A. C. and Romero-G\'omez, M. and Rubele, S. and Weiler, M.},
	doi = {10.1051/0004-6361/202039834},
	journal = {A\&A},
	pages = {A5},
	title = {Gaia Early Data Release 3 - Catalogue validation},
	url = {https://doi.org/10.1051/0004-6361/202039834},
	volume = 649,
	year = 2021}

@article{GaiaEDR3,
	author = {{Gaia Collaboration} and {Brown, A. G. A.} and {Vallenari, A.} and {Prusti, T.} and {de Bruijne, J. H. J.} and {Babusiaux, C.} and {Biermann, M.} and {Creevey, O. L.} and {Evans, D. W.} and {Eyer, L.} and {Hutton, A.} and {Jansen, F.} and {Jordi, C.} and {Klioner, S. A.} and {Lammers, U.} and {Lindegren, L.} and {Luri, X.} and {Mignard, F.} and {Panem, C.} and {Pourbaix, D.} and {Randich, S.} and {Sartoretti, P.} and {Soubiran, C.} and {Walton, N. A.} and {Arenou, F.} and {Bailer-Jones, C. A. L.} and {Bastian, U.} and {Cropper, M.} and {Drimmel, R.} and {Katz, D.} and {Lattanzi, M. G.} and {van Leeuwen, F.} and {Bakker, J.} and {Cacciari, C.} and {Casta\~neda, J.} and {De Angeli, F.} and {Ducourant, C.} and {Fabricius, C.} and {Fouesneau, M.} and {Fr\'emat, Y.} and {Guerra, R.} and {Guerrier, A.} and {Guiraud, J.} and {Jean-Antoine Piccolo, A.} and {Masana, E.} and {Messineo, R.} and {Mowlavi, N.} and {Nicolas, C.} and {Nienartowicz, K.} and {Pailler, F.} and {Panuzzo, P.} and {Riclet, F.} and {Roux, W.} and {Seabroke, G. M.} and {Sordo, R.} and {Tanga, P.} and {Th\'evenin, F.} and {Gracia-Abril, G.} and {Portell, J.} and {Teyssier, D.} and {Altmann, M.} and {Andrae, R.} and {Bellas-Velidis, I.} and {Benson, K.} and {Berthier, J.} and {Blomme, R.} and {Brugaletta, E.} and {Burgess, P. W.} and {Busso, G.} and {Carry, B.} and {Cellino, A.} and {Cheek, N.} and {Clementini, G.} and {Damerdji, Y.} and {Davidson, M.} and {Delchambre, L.} and {Dell\'{}Oro, A.} and {Fern\'andez-Hern\'andez, J.} and {Galluccio, L.} and {Garc\'{\i}a-Lario, P.} and {Garcia-Reinaldos, M.} and {Gonz\'alez-N\'u\~nez, J.} and {Gosset, E.} and {Haigron, R.} and {Halbwachs, J.-L.} and {Hambly, N. C.} and {Harrison, D. L.} and {Hatzidimitriou, D.} and {Heiter, U.} and {Hern\'andez, J.} and {Hestroffer, D.} and {Hodgkin, S. T.} and {Holl, B.} and {Jan\ss{}en, K.} and {Jevardat de Fombelle, G.} and {Jordan, S.} and {Krone-Martins, A.} and {Lanzafame, A. C.} and {L\"offler, W.} and {Lorca, A.} and {Manteiga, M.} and {Marchal, O.} and {Marrese, P. M.} and {Moitinho, A.} and {Mora, A.} and {Muinonen, K.} and {Osborne, P.} and {Pancino, E.} and {Pauwels, T.} and {Petit, J.-M.} and {Recio-Blanco, A.} and {Richards, P. J.} and {Riello, M.} and {Rimoldini, L.} and {Robin, A. C.} and {Roegiers, T.} and {Rybizki, J.} and {Sarro, L. M.} and {Siopis, C.} and {Smith, M.} and {Sozzetti, A.} and {Ulla, A.} and {Utrilla, E.} and {van Leeuwen, M.} and {van Reeven, W.} and {Abbas, U.} and {Abreu Aramburu, A.} and {Accart, S.} and {Aerts, C.} and {Aguado, J. J.} and {Ajaj, M.} and {Altavilla, G.} and {\'Alvarez, M. A.} and {\'Alvarez Cid-Fuentes, J.} and {Alves, J.} and {Anderson, R. I.} and {Anglada Varela, E.} and {Antoja, T.} and {Audard, M.} and {Baines, D.} and {Baker, S. G.} and {Balaguer-N\'u\~nez, L.} and {Balbinot, E.} and {Balog, Z.} and {Barache, C.} and {Barbato, D.} and {Barros, M.} and {Barstow, M. A.} and {Bartolom\'e, S.} and {Bassilana, J.-L.} and {Bauchet, N.} and {Baudesson-Stella, A.} and {Becciani, U.} and {Bellazzini, M.} and {Bernet, M.} and {Bertone, S.} and {Bianchi, L.} and {Blanco-Cuaresma, S.} and {Boch, T.} and {Bombrun, A.} and {Bossini, D.} and {Bouquillon, S.} and {Bragaglia, A.} and {Bramante, L.} and {Breedt, E.} and {Bressan, A.} and {Brouillet, N.} and {Bucciarelli, B.} and {Burlacu, A.} and {Busonero, D.} and {Butkevich, A. G.} and {Buzzi, R.} and {Caffau, E.} and {Cancelliere, R.} and {C\'anovas, H.} and {Cantat-Gaudin, T.} and {Carballo, R.} and {Carlucci, T.} and {Carnerero, M. I} and {Carrasco, J. M.} and {Casamiquela, L.} and {Castellani, M.} and {Castro-Ginard, A.} and {Castro Sampol, P.} and {Chaoul, L.} and {Charlot, P.} and {Chemin, L.} and {Chiavassa, A.} and {Cioni, M.-R. L.} and {Comoretto, G.} and {Cooper, W. J.} and {Cornez, T.} and {Cowell, S.} and {Crifo, F.} and {Crosta, M.} and {Crowley, C.} and {Dafonte, C.} and {Dapergolas, A.} and {David, M.} and {David, P.} and {de Laverny, P.} and {De Luise, F.} and {De March, R.} and {De Ridder, J.} and {de Souza, R.} and {de Teodoro, P.} and {de Torres, A.} and {del Peloso, E. F.} and {del Pozo, E.} and {Delbo, M.} and {Delgado, A.} and {Delgado, H. E.} and {Delisle, J.-B.} and {Di Matteo, P.} and {Diakite, S.} and {Diener, C.} and {Distefano, E.} and {Dolding, C.} and {Eappachen, D.} and {Edvardsson, B.} and {Enke, H.} and {Esquej, P.} and {Fabre, C.} and {Fabrizio, M.} and {Faigler, S.} and {Fedorets, G.} and {Fernique, P.} and {Fienga, A.} and {Figueras, F.} and {Fouron, C.} and {Fragkoudi, F.} and {Fraile, E.} and {Franke, F.} and {Gai, M.} and {Garabato, D.} and {Garcia-Gutierrez, A.} and {Garc\'{\i}a-Torres, M.} and {Garofalo, A.} and {Gavras, P.} and {Gerlach, E.} and {Geyer, R.} and {Giacobbe, P.} and {Gilmore, G.} and {Girona, S.} and {Giuffrida, G.} and {Gomel, R.} and {Gomez, A.} and {Gonzalez-Santamaria, I.} and {Gonz\'alez-Vidal, J. J.} and {Granvik, M.} and {Guti\'errez-S\'anchez, R.} and {Guy, L. P.} and {Hauser, M.} and {Haywood, M.} and {Helmi, A.} and {Hidalgo, S. L.} and {Hilger, T.} and {Hladczuk, N.} and {Hobbs, D.} and {Holland, G.} and {Huckle, H. E.} and {Jasniewicz, G.} and {Jonker, P. G.} and {Juaristi Campillo, J.} and {Julbe, F.} and {Karbevska, L.} and {Kervella, P.} and {Khanna, S.} and {Kochoska, A.} and {Kontizas, M.} and {Kordopatis, G.} and {Korn, A. J.} and {Kostrzewa-Rutkowska, Z.} and {Kruszy\'{}nska, K.} and {Lambert, S.} and {Lanza, A. F.} and {Lasne, Y.} and {Le Campion, J.-F.} and {Le Fustec, Y.} and {Lebreton, Y.} and {Lebzelter, T.} and {Leccia, S.} and {Leclerc, N.} and {Lecoeur-Taibi, I.} and {Liao, S.} and {Licata, E.} and {Lindstr\o{}m, E. P.} and {Lister, T. A.} and {Livanou, E.} and {Lobel, A.} and {Madrero Pardo, P.} and {Managau, S.} and {Mann, R. G.} and {Marchant, J. M.} and {Marconi, M.} and {Marcos Santos, M. M. S.} and {Marinoni, S.} and {Marocco, F.} and {Marshall, D. J.} and {Martin Polo, L.} and {Mart\'{\i}n-Fleitas, J. M.} and {Masip, A.} and {Massari, D.} and {Mastrobuono-Battisti, A.} and {Mazeh, T.} and {McMillan, P. J.} and {Messina, S.} and {Michalik, D.} and {Millar, N. R.} and {Mints, A.} and {Molina, D.} and {Molinaro, R.} and {Moln\'ar, L.} and {Montegriffo, P.} and {Mor, R.} and {Morbidelli, R.} and {Morel, T.} and {Morris, D.} and {Mulone, A. F.} and {Munoz, D.} and {Muraveva, T.} and {Murphy, C. P.} and {Musella, I.} and {Noval, L.} and {Ord\'enovic, C.} and {Orr\`u, G.} and {Osinde, J.} and {Pagani, C.} and {Pagano, I.} and {Palaversa, L.} and {Palicio, P. A.} and {Panahi, A.} and {Pawlak, M.} and {Pe\~nalosa Esteller, X.} and {Penttil\"a, A.} and {Piersimoni, A. M.} and {Pineau, F.-X.} and {Plachy, E.} and {Plum, G.} and {Poggio, E.} and {Poretti, E.} and {Poujoulet, E.} and {Prsa, A.} and {Pulone, L.} and {Racero, E.} and {Ragaini, S.} and {Rainer, M.} and {Raiteri, C. M.} and {Rambaux, N.} and {Ramos, P.} and {Ramos-Lerate, M.} and {Re Fiorentin, P.} and {Regibo, S.} and {Reyl\'e, C.} and {Ripepi, V.} and {Riva, A.} and {Rixon, G.} and {Robichon, N.} and {Robin, C.} and {Roelens, M.} and {Rohrbasser, L.} and {Romero-G\'omez, M.} and {Rowell, N.} and {Royer, F.} and {Rybicki, K. A.} and {Sadowski, G.} and {Sagrist\`a Sell\'es, A.} and {Sahlmann, J.} and {Salgado, J.} and {Salguero, E.} and {Samaras, N.} and {Sanchez Gimenez, V.} and {Sanna, N.} and {Santove\~na, R.} and {Sarasso, M.} and {Schultheis, M.} and {Sciacca, E.} and {Segol, M.} and {Segovia, J. C.} and {S\'egransan, D.} and {Semeux, D.} and {Shahaf, S.} and {Siddiqui, H. I.} and {Siebert, A.} and {Siltala, L.} and {Slezak, E.} and {Smart, R. L.} and {Solano, E.} and {Solitro, F.} and {Souami, D.} and {Souchay, J.} and {Spagna, A.} and {Spoto, F.} and {Steele, I. A.} and {Steidelm\"uller, H.} and {Stephenson, C. A.} and {S\"uveges, M.} and {Szabados, L.} and {Szegedi-Elek, E.} and {Taris, F.} and {Tauran, G.} and {Taylor, M. B.} and {Teixeira, R.} and {Thuillot, W.} and {Tonello, N.} and {Torra, F.} and {Torra, J.} and {Turon, C.} and {Unger, N.} and {Vaillant, M.} and {van Dillen, E.} and {Vanel, O.} and {Vecchiato, A.} and {Viala, Y.} and {Vicente, D.} and {Voutsinas, S.} and {Weiler, M.} and {Wevers, T.} and {Wyrzykowski, L.} and {Yoldas, A.} and {Yvard, P.} and {Zhao, H.} and {Zorec, J.} and {Zucker, S.} and {Zurbach, C.} and {Zwitter, T.}},
	doi = {10.1051/0004-6361/202039657},
	journal = {A\&A},
	pages = {A1},
	title = {Gaia Early Data Release 3 - Summary of the contents and survey properties},
	url = {https://doi.org/10.1051/0004-6361/202039657},
	volume = 649,
	year = 2021}

@ARTICLE{Lindegren2020,
       author = {{Lindegren}, L. and {Klioner}, S.~A. and {Hern{\'a}ndez}, J. and {Bombrun}, A. and {Ramos-Lerate}, M. and {Steidelm{\"u}ller}, H. and {Bastian}, U. and {Biermann}, M. and {de Torres}, A. and {Gerlach}, E. and {Geyer}, R. and {Hilger}, T. and {Hobbs}, D. and {Lammers}, U. and {McMillan}, P.~J. and {Stephenson}, C.~A. and {Casta{\~n}eda}, J. and {Davidson}, M. and {Fabricius}, C. and {Gracia-Abril}, G. and {Portell}, J. and {Rowell}, N. and {Teyssier}, D. and {Torra}, F. and {Bartolom{\'e}}, S. and {Clotet}, M. and {Garralda}, N. and {Gonz{\'a}lez-Vidal}, J.~J. and {Torra}, J. and {Abbas}, U. and {Altmann}, M. and {Anglada Varela}, E. and {Balaguer-N{\'u}{\~n}ez}, L. and {Balog}, Z. and {Barache}, C. and {Becciani}, U. and {Bernet}, M. and {Bertone}, S. and {Bianchi}, L. and {Bouquillon}, S. and {Brown}, A.~G.~A. and {Bucciarelli}, B. and {Busonero}, D. and {Butkevich}, A.~G. and {Buzzi}, R. and {Cancelliere}, R. and {Carlucci}, T. and {Charlot}, P. and {Cioni}, M. -R.~L. and {Crosta}, M. and {Crowley}, C. and {del Peloso}, E.~F. and {del Pozo}, E. and {Drimmel}, R. and {Esquej}, P. and {Fienga}, A. and {Fraile}, E. and {Gai}, M. and {Garcia-Reinaldos}, M. and {Guerra}, R. and {Hambly}, N.~C. and {Hauser}, M. and {Jan{\ss}en}, K. and {Jordan}, S. and {Kostrzewa-Rutkowska}, Z. and {Lattanzi}, M.~G. and {Liao}, S. and {Licata}, E. and {Lister}, T.~A. and {L{\"o}ffler}, W. and {Marchant}, J.~M. and {Masip}, A. and {Mignard}, F. and {Mints}, A. and {Molina}, D. and {Mora}, A. and {Morbidelli}, R. and {Murphy}, C.~P. and {Pagani}, C. and {Panuzzo}, P. and {Pe{\~n}alosa Esteller}, X. and {Poggio}, E. and {Re Fiorentin}, P. and {Riva}, A. and {Sagrist{\`a} Sell{\'e}s}, A. and {Sanchez Gimenez}, V. and {Sarasso}, M. and {Sciacca}, E. and {Siddiqui}, H.~I. and {Smart}, R.~L. and {Souami}, D. and {Spagna}, A. and {Steele}, I.~A. and {Taris}, F. and {Utrilla}, E. and {van Reeven}, W. and {Vecchiato}, A.},
        title = "{Gaia Early Data Release 3. The astrometric solution}",
      journal = {\aap},
         year = 2021,
        month = may,
       volume = {649},
          eid = {A2},
        pages = {A2},
          doi = {10.1051/0004-6361/202039709},
archivePrefix = {arXiv},
       eprint = {2012.03380},
 primaryClass = {astro-ph.IM},
       adsurl = {https://ui.adsabs.harvard.edu/abs/2021A&A...649A...2L}
}

@ARTICLE{Lindegren2020b,
       author = {{Lindegren}, L. and {Bastian}, U. and {Biermann}, M. and {Bombrun}, A. and {de Torres}, A. and {Gerlach}, E. and {Geyer}, R. and {Hern{\'a}ndez}, J. and {Hilger}, T. and {Hobbs}, D. and {Klioner}, S.~A. and {Lammers}, U. and {McMillan}, P.~J. and {Ramos-Lerate}, M. and {Steidelm{\"u}ller}, H. and {Stephenson}, C.~A. and {van Leeuwen}, F.},
        title = "{Gaia Early Data Release 3. Parallax bias versus magnitude, colour, and position}",
      journal = {\aap},
         year = 2021,
        month = may,
       volume = {649},
          eid = {A4},
        pages = {A4},
          doi = {10.1051/0004-6361/202039653},
archivePrefix = {arXiv},
       eprint = {2012.01742},
 primaryClass = {astro-ph.IM},
       adsurl = {https://ui.adsabs.harvard.edu/abs/2021A&A...649A...4L}
}

@article{McConnachie2020a,
  title={Revised and new proper motions for confirmed and candidate Milky Way dwarf galaxies},
  author={McConnachie, Alan W and Venn, Kim A},
  journal={\aj},
  volume={160},
  number={3},
  pages={124},
  year={2020},
  publisher={IOP Publishing}
}

@article{McConnachie2020b,
  title={Updated Proper Motions for Local Group Dwarf Galaxies Using Gaia Early Data Release 3},
  author={McConnachie, Alan W and Venn, Kim A},
  journal={Research Notes of the AAS},
  volume={4},
  number={12},
  pages={229},
  year={2020},
  publisher={IOP Publishing}
}

@ARTICLE{Piatek2003,
       author = {{Piatek}, Slawomir and {Pryor}, Carlton and {Olszewski}, Edward W. and {Harris}, Hugh C. and {Mateo}, Mario and {Minniti}, Dante and {Tinney}, Christopher G.},
        title = "{Proper Motions of Dwarf Spheroidal Galaxies from Hubble Space Telescope Imaging. II. Measurement for Carina}",
      journal = {\aj},
         year = 2003,
        month = nov,
       volume = {126},
       number = {5},
        pages = {2346-2361},
          doi = {10.1086/378713},
       adsurl = {https://ui.adsabs.harvard.edu/abs/2003AJ....126.2346P}
}

@article{Massari2018,
  title={Three-dimensional motions in the Sculptor dwarf galaxy as a glimpse of a new era},
  author={Massari, DAVIDE and Breddels, MA and Helmi, A and Posti, L and Brown, AGA and Tolstoy, E},
  journal={Nature Astronomy},
  volume={2},
  number={2},
  pages={156--161},
  year={2018},
  publisher={Nature Publishing Group}
}

@article{Massari2020,
	author = {Massari, D. and Helmi, A. and Mucciarelli, A. and Sales, L. V. and Spina, L. and Tolstoy, E.},
	doi = {10.1051/0004-6361/201935613},
	journal = {A\&A},
	pages = {A36},
	title = {Stellar 3D kinematics in the Draco dwarf spheroidal galaxy},
	url = {https://doi.org/10.1051/0004-6361/201935613},
	volume = 633,
	year = 2020}

@ARTICLE{Battaglia2022,
       author = {{Battaglia}, G. and {Taibi}, S. and {Thomas}, G.~F. and {Fritz}, T.~K.},
        title = "{Gaia early DR3 systemic motions of Local Group dwarf galaxies and orbital properties with a massive Large Magellanic Cloud}",
      journal = {\aap},
         year = 2022,
        month = jan,
       volume = {657},
          eid = {A54},
        pages = {A54},
          doi = {10.1051/0004-6361/202141528},
archivePrefix = {arXiv},
       eprint = {2106.08819},
 primaryClass = {astro-ph.GA},
       adsurl = {https://ui.adsabs.harvard.edu/abs/2022A&A...657A..54B}
}

@ARTICLE{Pace2022,
       author = {{Pace}, Andrew B. and {Erkal}, Denis and {Li}, Ting S.},
        title = "{Proper Motions, Orbits, and Tidal Influences of Milky Way Dwarf Spheroidal Galaxies}",
      journal = {\apj},
         year = 2022,
        month = dec,
       volume = {940},
       number = {2},
          eid = {136},
        pages = {136},
          doi = {10.3847/1538-4357/ac997b},
archivePrefix = {arXiv},
       eprint = {2205.05699},
 primaryClass = {astro-ph.GA},
       adsurl = {https://ui.adsabs.harvard.edu/abs/2022ApJ...940..136P}
}

@ARTICLE{Battaglia2022b,
       author = {{Battaglia}, Giuseppina and {Nipoti}, Carlo},
        title = "{Stellar dynamics and dark matter in Local Group dwarf galaxies}",
      journal = {Nature Astronomy},
         year = 2022,
        month = may,
       volume = {6},
        pages = {659-672},
          doi = {10.1038/s41550-022-01638-7},
archivePrefix = {arXiv},
       eprint = {2205.07821},
 primaryClass = {astro-ph.GA},
       adsurl = {https://ui.adsabs.harvard.edu/abs/2022NatAs...6..659B}
}

@ARTICLE{delPino2022,
       author = {{del Pino}, Andr{\'e}s and {Libralato}, Mattia and {van der Marel}, Roeland P. and {Bennet}, Paul and {Fardal}, Mark A. and {Anderson}, Jay and {Bellini}, Andrea and {Tony Sohn}, Sangmo and {Watkins}, Laura L.},
        title = "{GaiaHub: A Method for Combining Data from the Gaia and Hubble Space Telescopes to Derive Improved Proper Motions for Faint Stars}",
      journal = {\apj},
         year = 2022,
        month = jul,
       volume = {933},
       number = {1},
          eid = {76},
        pages = {76},
          doi = {10.3847/1538-4357/ac70cf},
archivePrefix = {arXiv},
       eprint = {2205.08009},
 primaryClass = {astro-ph.GA},
       adsurl = {https://ui.adsabs.harvard.edu/abs/2022ApJ...933...76D}
}

@ARTICLE{Bullock2017,
       author = {{Bullock}, James S. and {Boylan-Kolchin}, Michael},
        title = "{Small-Scale Challenges to the {\ensuremath{\Lambda}}CDM Paradigm}",
      journal = {\araa},
         year = 2017,
        month = aug,
       volume = {55},
       number = {1},
        pages = {343-387},
          doi = {10.1146/annurev-astro-091916-055313},
archivePrefix = {arXiv},
       eprint = {1707.04256},
 primaryClass = {astro-ph.CO},
       adsurl = {https://ui.adsabs.harvard.edu/abs/2017ARA&A..55..343B}
}

@article{Vasiliev2021,
    author = {Vasiliev, Eugene and Belokurov, Vasily and Erkal, Denis},
    title = {Tango for three: Sagittarius, LMC, and the Milky Way},
    journal = {\mnras},
    volume = {501},
    number = {2},
    pages = {2279-2304},
    year = {2021},
    month = {11},
    issn = {0035-8711},
    doi = {10.1093/mnras/staa3673},
    url = {https://doi.org/10.1093/mnras/staa3673},
    eprint = {https://academic.oup.com/mnras/article-pdf/501/2/2279/35392824/staa3673.pdf},
}

@article{Vasiliev2023,
	article-number = {59},
	author = {Vasiliev, Eugene},
	doi = {10.3390/galaxies11020059},
	issn = {2075-4434},
	journal = {Galaxies},
	number = {2},
	title = {The Effect of the LMC on the Milky Way System},
	url = {https://www.mdpi.com/2075-4434/11/2/59},
	volume = {11},
	year = {2023}}

@ARTICLE{Vasiliev2024,
       author = {{Vasiliev}, Eugene},
        title = "{Dear Magellanic Clouds, welcome back!}",
      journal = {\mnras},
         year = 2024,
        month = jan,
       volume = {527},
       number = {1},
        pages = {437-456},
          doi = {10.1093/mnras/stad2612},
archivePrefix = {arXiv},
       eprint = {2306.04837},
 primaryClass = {astro-ph.GA},
       adsurl = {https://ui.adsabs.harvard.edu/abs/2024MNRAS.527..437V}
}

@ARTICLE{Pace2024,
       author = {{Pace}, Andrew B},
        title = "{The Local Volume Database: a library of the observed properties of nearby dwarf galaxies and star clusters}",
      journal = {The Open Journal of Astrophysics},
         year = 2025,
        month = sep,
       volume = {8},
          eid = {142},
        pages = {142},
          doi = {10.33232/001c.144859},
archivePrefix = {arXiv},
       eprint = {2411.07424},
 primaryClass = {astro-ph.GA},
       adsurl = {https://ui.adsabs.harvard.edu/abs/2025OJAp....8E.142P}
}

@article{Cerny2025,
	author = {Cerny, W. and Chiti, A. and Geha, M. and Mutlu-Pakdil, B. and Drlica-Wagner, A. and Tan, C. Y. and Adam{\'o}w, M. and Pace, A. B. and Simon, J. D. and Sand, D. J. and Ji, A. P. and Li, T. S. and Vivas, A. K. and Bell, E. F. and Carlin, J. L. and Carballo-Bello, J. A. and Chaturvedi, A. and Choi, Y. and Doliva-Dolinsky, A. and Gnedin, O. Y. and Limberg, G. and Mart{\'\i}nez-V{\'a}zquez, C. E. and Mau, S. and Medina, G. E. and Navabi, M. and No{\"e}l, N. E. D. and Placco, V. M. and Riley, A. H. and Roederer, I. U. and Stringfellow, G. S. and Bom, C. R. and Ferguson, P. S. and James, D. J. and Mart{\'\i}nez-Delgado, D. and Massana, P. and Nidever, D. L. and Sakowska, J. D. and Santana-Silva, L. and Sherman, N. F. and Tollerud, E. J. and (DELVE Collaboration)},
	doi = {10.3847/1538-4357/ad8eba},
	journal = {\apj},
	month = {jan},
	number = {2},
	pages = {164},
	publisher = {The American Astronomical Society},
	title = {Discovery and Spectroscopic Confirmation of Aquarius III: A Low-mass Milky Way Satellite Galaxy},
	url = {https://dx.doi.org/10.3847/1538-4357/ad8eba},
	volume = {979},
	year = {2025}}

@article{Cerny2023,
	author = {Cerny, W. and Mart{\'\i}nez-V{\'a}zquez, C. E. and Drlica-Wagner, A. and Pace, A. B. and Mutlu-Pakdil, B. and Li, T. S. and Riley, A. H. and Crnojevi{\'c}, D. and Bom, C. R. and Carballo-Bello, J. A. and Carlin, J. L. and Chiti, A. and Choi, Y. and Collins, M. L. M. and Darragh-Ford, E and Ferguson, P. S. and Geha, M. and Mart{\'\i}nez-Delgado, D. and Massana, P. and Mau, S. and Medina, G. E. and Mu{\~n}oz, R. R. and Nadler, E. O. and No{\"e}l, N. E. D. and Olsen, K. A. G. and Pieres, A. and Sakowska, J. D. and Simon, J. D. and Stringfellow, G. S. and Tollerud, E. J. and Vivas, A. K. and Walker, A. R. and Wechsler, R. H. and (DELVE Collaboration)},
	doi = {10.3847/1538-4357/acdd78},
	journal = {\apj},
	month = {jul},
	number = {1},
	pages = {1},
	publisher = {The American Astronomical Society},
	title = {Six More Ultra-faint Milky Way Companions Discovered in the DECam Local Volume Exploration Survey},
	url = {https://dx.doi.org/10.3847/1538-4357/acdd78},
	volume = {953},
	year = {2023}}

@article{Cerny2021,
	author = {Cerny, W. and Pace, A. B. and Drlica-Wagner, A. and Koposov, S. E. and Vivas, A. K. and Mau, S. and Riley, A. H. and Bom, C. R. and Carlin, J. L. and Choi, Y. and Erkal, D. and Ferguson, P. S. and James, D. J. and Li, T. S. and Mart{\'\i}nez-Delgado, D. and Mart{\'\i}nez-V{\'a}zquez, C. E. and Munoz, R. R. and Mutlu-Pakdil, B. and Olsen, K. A. G. and Pieres, A. and Sakowska, J. D. and Sand, D. J. and Simon, J. D. and Smercina, A. and Stringfellow, G. S. and Tollerud, E. J. and Adam{\'o}w, M. and Hernandez-Lang, D. and Kuropatkin, N. and Santana-Silva, L. and Tucker, D. L. and Zenteno, A. and DELVE Collaboration},
	doi = {10.3847/2041-8213/ac2d9a},
	journal = {\apjl},
	month = {oct},
	number = {2},
	pages = {L44},
	publisher = {The American Astronomical Society},
	title = {Eridanus IV: an Ultra-faint Dwarf Galaxy Candidate Discovered in the DECam Local Volume Exploration Survey},
	url = {https://dx.doi.org/10.3847/2041-8213/ac2d9a},
	volume = {920},
	year = {2021}}

@article{Tan2025a,
	author = {Tan, C. Y. and Cerny, W. and Drlica-Wagner, A. and Pace, A. B. and Geha, M. and Ji, A. P. and Li, T. S. and Adam{\'o}w, M. and Anbajagane, D. and Bom, C. R. and Carballo-Bello, J. A. and Carlin, J. L. and Chang, C. and Chaturvedi, A. and Chiti, A. and Choi, Y. and Collins, M. L. M. and Doliva-Dolinsky, A. and Ferguson, P. S. and Gruendl, R. A. and James, D. J. and Limberg, G. and Navabi, M. and Mart{\'\i}nez-Delgado, D. and Mart{\'\i}nez-V{\'a}zquez, C. E. and Medina, G. E. and Mutlu-Pakdil, B. and Nidever, D. L. and No{\"e}l, N. E. D. and Riley, A. H. and Sakowska, J. D. and Sand, D. J. and Sharp, J. and Stringfellow, G. S. and Tolley, C. and Tucker, D. L. and Vivas, A. K. and (DELVE Collaboration)},
	doi = {10.3847/1538-4357/ad9b0c},
	journal = {\apj},
	month = {jan},
	number = {2},
	pages = {176},
	publisher = {The American Astronomical Society},
	title = {A Pride of Satellites in the Constellation Leo? Discovery of the Leo VI Milky Way Satellite Ultra-faint Dwarf Galaxy with DELVE Early Data Release 3},
	url = {https://dx.doi.org/10.3847/1538-4357/ad9b0c},
	volume = {979},
	year = {2025}}

@article{Cerny2023b,
	author = {Cerny, W. and Simon, J. D. and Li, T. S. and Drlica-Wagner, A. and Pace, A. B. and Mart{\'\i}nez-V{\'a}zquez, C. E. and Riley, A. H. and Mutlu-Pakdil, B. and Mau, S. and Ferguson, P. S. and Erkal, D. and Munoz, R. R. and Bom, C. R. and Carlin, J. L. and Carollo, D. and Choi, Y. and Ji, A. P. and Manwadkar, V. and Mart{\'\i}nez-Delgado, D. and Miller, A. E. and No{\"e}l, N. E. D. and Sakowska, J. D. and Sand, D. J. and Stringfellow, G. S. and Tollerud, E. J. and Vivas, A. K. and Carballo-Bello, J. A. and Hernandez-Lang, D. and James, D. J. and Nidever, D. L. and Castellon, J. L. Nilo and Olsen, K. A. G. and Zenteno, A. and (DELVE Collaboration)},
	doi = {10.3847/1538-4357/aca1c3},
	journal = {\apj},
	month = {jan},
	number = {2},
	pages = {111},
	publisher = {The American Astronomical Society},
	title = {Pegasus IV: Discovery and Spectroscopic Confirmation of an Ultra-faint Dwarf Galaxy in the Constellation Pegasus},
	url = {https://dx.doi.org/10.3847/1538-4357/aca1c3},
	volume = {942},
	year = {2023}}

@ARTICLE{McConnachie2021,
       author = {{McConnachie}, Alan W. and {Higgs}, Clare R. and {Thomas}, Guillaume F. and {Venn}, Kim A. and {C{\^o}t{\'e}}, Patrick and {Battaglia}, Giuseppina and {Lewis}, Geraint F.},
        title = "{Solo dwarfs - III. Exploring the orbital origins of isolated Local Group galaxies with Gaia Data Release 2}",
      journal = {\mnras},
         year = 2021,
        month = feb,
       volume = {501},
       number = {2},
        pages = {2363-2377},
          doi = {10.1093/mnras/staa3740},
archivePrefix = {arXiv},
       eprint = {2012.01586},
 primaryClass = {astro-ph.GA},
       adsurl = {https://ui.adsabs.harvard.edu/abs/2021MNRAS.501.2363M}
}

@ARTICLE{Zivick2018,
       author = {{Zivick}, Paul and {Kallivayalil}, Nitya and {van der Marel}, Roeland P. and {Besla}, Gurtina and {Linden}, Sean T. and {Koz{\l}owski}, Szymon and {Fritz}, Tobias K. and {Kochanek}, C.~S. and {Anderson}, J. and {Sohn}, Sangmo Tony and {Geha}, Marla C. and {Alcock}, Charles R.},
        title = "{The Proper Motion Field of the Small Magellanic Cloud: Kinematic Evidence for Its Tidal Disruption}",
      journal = {\apj},
         year = 2018,
        month = sep,
       volume = {864},
       number = {1},
          eid = {55},
        pages = {55},
          doi = {10.3847/1538-4357/aad4b0},
archivePrefix = {arXiv},
       eprint = {1804.04110},
 primaryClass = {astro-ph.GA},
       adsurl = {https://ui.adsabs.harvard.edu/abs/2018ApJ...864...55Z}
}

@ARTICLE{GaiaDR3QSO,
       author = {{Gaia Collaboration} and {Bailer-Jones}, C.~A.~L. and {Teyssier}, D. and {Delchambre}, L. and {Ducourant}, C. and {Garabato}, D. and {Hatzidimitriou}, D. and {Klioner}, S.~A. and {Rimoldini}, L. and {Bellas-Velidis}, I. and {Carballo}, R. and {Carnerero}, M.~I. and {Diener}, C. and {Fouesneau}, M. and {Galluccio}, L. and {Gavras}, P. and {Krone-Martins}, A. and {Raiteri}, C.~M. and {Teixeira}, R. and {Brown}, A.~G.~A. and {Vallenari}, A. and {Prusti}, T. and {de Bruijne}, J.~H.~J. and {Arenou}, F. and {Babusiaux}, C. and {Biermann}, M. and {Creevey}, O.~L. and {Evans}, D.~W. and {Eyer}, L. and {Guerra}, R. and {Hutton}, A. and {Jordi}, C. and {Lammers}, U.~L. and {Lindegren}, L. and {Luri}, X. and {Mignard}, F. and {Panem}, C. and {Pourbaix}, D. and {Randich}, S. and {Sartoretti}, P. and {Soubiran}, C. and {Tanga}, P. and {Walton}, N.~A. and {Bastian}, U. and {Drimmel}, R. and {Jansen}, F. and {Katz}, D. and {Lattanzi}, M.~G. and {van Leeuwen}, F. and {Bakker}, J. and {Cacciari}, C. and {Casta{\~n}eda}, J. and {De Angeli}, F. and {Fabricius}, C. and {Fr{\'e}mat}, Y. and {Guerrier}, A. and {Heiter}, U. and {Masana}, E. and {Messineo}, R. and {Mowlavi}, N. and {Nicolas}, C. and {Nienartowicz}, K. and {Pailler}, F. and {Panuzzo}, P. and {Riclet}, F. and {Roux}, W. and {Seabroke}, G.~M. and {Sordo}, R. and {Th{\'e}venin}, F. and {Gracia-Abril}, G. and {Portell}, J. and {Altmann}, M. and {Andrae}, R. and {Audard}, M. and {Benson}, K. and {Berthier}, J. and {Blomme}, R. and {Burgess}, P.~W. and {Busonero}, D. and {Busso}, G. and {C{\'a}novas}, H. and {Carry}, B. and {Cellino}, A. and {Cheek}, N. and {Clementini}, G. and {Damerdji}, Y. and {Davidson}, M. and {de Teodoro}, P. and {Nu{\~n}ez Campos}, M. and {Dell'Oro}, A. and {Esquej}, P. and {Fern{\'a}ndez-Hern{\'a}ndez}, J. and {Fraile}, E. and {Garc{\'\i}a-Lario}, P. and {Gosset}, E. and {Haigron}, R. and {Halbwachs}, J. -L. and {Hambly}, N.~C. and {Harrison}, D.~L. and {Hern{\'a}ndez}, J. and {Hestroffer}, D. and {Hodgkin}, S.~T. and {Holl}, B. and {Jan{\ss}en}, K. and {Jevardat de Fombelle}, G. and {Jordan}, S. and {Lanzafame}, A.~C. and {L{\"o}ffler}, W. and {Marchal}, O. and {Marrese}, P.~M. and {Moitinho}, A. and {Muinonen}, K. and {Osborne}, P. and {Pancino}, E. and {Pauwels}, T. and {Recio-Blanco}, A. and {Reyl{\'e}}, C. and {Riello}, M. and {Roegiers}, T. and {Rybizki}, J. and {Sarro}, L.~M. and {Siopis}, C. and {Smith}, M. and {Sozzetti}, A. and {Utrilla}, E. and {van Leeuwen}, M. and {Abbas}, U. and {{\'A}brah{\'a}m}, P. and {Abreu Aramburu}, A. and {Aerts}, C. and {Aguado}, J.~J. and {Ajaj}, M. and {Aldea-Montero}, F. and {Altavilla}, G. and {{\'A}lvarez}, M.~A. and {Alves}, J. and {Anderson}, R.~I. and {Anglada Varela}, E. and {Antoja}, T. and {Baines}, D. and {Baker}, S.~G. and {Balaguer-N{\'u}{\~n}ez}, L. and {Balbinot}, E. and {Balog}, Z. and {Barache}, C. and {Barbato}, D. and {Barros}, M. and {Barstow}, M.~A. and {Bartolom{\'e}}, S. and {Bassilana}, J. -L. and {Bauchet}, N. and {Becciani}, U. and {Bellazzini}, M. and {Berihuete}, A. and {Bernet}, M. and {Bertone}, S. and {Bianchi}, L. and {Binnenfeld}, A. and {Blanco-Cuaresma}, S. and {Boch}, T. and {Bombrun}, A. and {Bossini}, D. and {Bouquillon}, S. and {Bragaglia}, A. and {Bramante}, L. and {Breedt}, E. and {Bressan}, A. and {Brouillet}, N. and {Brugaletta}, E. and {Bucciarelli}, B. and {Burlacu}, A. and {Butkevich}, A.~G. and {Buzzi}, R. and {Caffau}, E. and {Cancelliere}, R. and {Cantat-Gaudin}, T. and {Carlucci}, T. and {Carrasco}, J.~M. and {Casamiquela}, L. and {Castellani}, M. and {Castro-Ginard}, A. and {Chaoul}, L. and {Charlot}, P. and {Chemin}, L. and {Chiaramida}, V. and {Chiavassa}, A. and {Chornay}, N. and {Comoretto}, G. and {Contursi}, G. and {Cooper}, W.~J. and {Cornez}, T. and {Cowell}, S. and {Crifo}, F. and {Cropper}, M. and {Crosta}, M. and {Crowley}, C. and {Dafonte}, C. and {Dapergolas}, A. and {David}, P. and {de Laverny}, P.},
        title = "{Gaia Data Release 3. The extragalactic content}",
      journal = {\aap},
         year = 2023,
        month = jun,
       volume = {674},
          eid = {A41},
        pages = {A41},
          doi = {10.1051/0004-6361/202243232},
archivePrefix = {arXiv},
       eprint = {2206.05681},
 primaryClass = {astro-ph.GA},
       adsurl = {https://ui.adsabs.harvard.edu/abs/2023A&A...674A..41G}
}

@ARTICLE{VasilievBaumgardt2021,
       author = {{Vasiliev}, Eugene and {Baumgardt}, Holger},
        title = "{Gaia EDR3 view on galactic globular clusters}",
      journal = {\mnras},
         year = 2021,
        month = aug,
       volume = {505},
       number = {4},
        pages = {5978-6002},
          doi = {10.1093/mnras/stab1475},
archivePrefix = {arXiv},
       eprint = {2102.09568},
 primaryClass = {astro-ph.GA},
       adsurl = {https://ui.adsabs.harvard.edu/abs/2021MNRAS.505.5978V}
}

@ARTICLE{McMillan2017,
       author = {{McMillan}, Paul J.},
        title = "{The mass distribution and gravitational potential of the Milky Way}",
      journal = {\mnras},
         year = 2017,
        month = feb,
       volume = {465},
       number = {1},
        pages = {76-94},
          doi = {10.1093/mnras/stw2759},
archivePrefix = {arXiv},
       eprint = {1608.00971},
 primaryClass = {astro-ph.GA},
       adsurl = {https://ui.adsabs.harvard.edu/abs/2017MNRAS.465...76M}
}

@software{AGAMA2018,
       author = {{Vasiliev}, Eugene},
        title = "{AGAMA: Action-based galaxy modeling framework}",
 howpublished = {Astrophysics Source Code Library, record ascl:1805.008},
         year = 2018,
        month = may,
          eid = {ascl:1805.008},
       adsurl = {https://ui.adsabs.harvard.edu/abs/2018ascl.soft05008V}
}

@ARTICLE{AGAMA2019,
       author = {{Vasiliev}, Eugene},
        title = "{AGAMA: action-based galaxy modelling architecture}",
      journal = {\mnras},
         year = 2019,
        month = jan,
       volume = {482},
       number = {2},
        pages = {1525-1544},
          doi = {10.1093/mnras/sty2672},
archivePrefix = {arXiv},
       eprint = {1802.08239},
 primaryClass = {astro-ph.GA},
       adsurl = {https://ui.adsabs.harvard.edu/abs/2019MNRAS.482.1525V}
}

@ARTICLE{Astropy2013,
       author = {{Astropy Collaboration} and {Robitaille}, Thomas P. and {Tollerud}, Erik J. and {Greenfield}, Perry and {Droettboom}, Michael and {Bray}, Erik and {Aldcroft}, Tom and {Davis}, Matt and {Ginsburg}, Adam and {Price-Whelan}, Adrian M. and {Kerzendorf}, Wolfgang E. and {Conley}, Alexander and {Crighton}, Neil and {Barbary}, Kyle and {Muna}, Demitri and {Ferguson}, Henry and {Grollier}, Fr{\'e}d{\'e}ric and {Parikh}, Madhura M. and {Nair}, Prasanth H. and {Unther}, Hans M. and {Deil}, Christoph and {Woillez}, Julien and {Conseil}, Simon and {Kramer}, Roban and {Turner}, James E.~H. and {Singer}, Leo and {Fox}, Ryan and {Weaver}, Benjamin A. and {Zabalza}, Victor and {Edwards}, Zachary I. and {Azalee Bostroem}, K. and {Burke}, D.~J. and {Casey}, Andrew R. and {Crawford}, Steven M. and {Dencheva}, Nadia and {Ely}, Justin and {Jenness}, Tim and {Labrie}, Kathleen and {Lim}, Pey Lian and {Pierfederici}, Francesco and {Pontzen}, Andrew and {Ptak}, Andy and {Refsdal}, Brian and {Servillat}, Mathieu and {Streicher}, Ole},
        title = "{Astropy: A community Python package for astronomy}",
      journal = {\aap},
         year = 2013,
        month = oct,
       volume = {558},
          eid = {A33},
        pages = {A33},
          doi = {10.1051/0004-6361/201322068},
archivePrefix = {arXiv},
       eprint = {1307.6212},
 primaryClass = {astro-ph.IM},
       adsurl = {https://ui.adsabs.harvard.edu/abs/2013A&A...558A..33A}
}

@ARTICLE{Astropy2018,
       author = {{Astropy Collaboration} and {Price-Whelan}, A.~M. and {Sip{\H{o}}cz}, B.~M. and {G{\"u}nther}, H.~M. and {Lim}, P.~L. and {Crawford}, S.~M. and {Conseil}, S. and {Shupe}, D.~L. and {Craig}, M.~W. and {Dencheva}, N. and {Ginsburg}, A. and {VanderPlas}, J.~T. and {Bradley}, L.~D. and {P{\'e}rez-Su{\'a}rez}, D. and {de Val-Borro}, M. and {Aldcroft}, T.~L. and {Cruz}, K.~L. and {Robitaille}, T.~P. and {Tollerud}, E.~J. and {Ardelean}, C. and {Babej}, T. and {Bach}, Y.~P. and {Bachetti}, M. and {Bakanov}, A.~V. and {Bamford}, S.~P. and {Barentsen}, G. and {Barmby}, P. and {Baumbach}, A. and {Berry}, K.~L. and {Biscani}, F. and {Boquien}, M. and {Bostroem}, K.~A. and {Bouma}, L.~G. and {Brammer}, G.~B. and {Bray}, E.~M. and {Breytenbach}, H. and {Buddelmeijer}, H. and {Burke}, D.~J. and {Calderone}, G. and {Cano Rodr{\'\i}guez}, J.~L. and {Cara}, M. and {Cardoso}, J.~V.~M. and {Cheedella}, S. and {Copin}, Y. and {Corrales}, L. and {Crichton}, D. and {D'Avella}, D. and {Deil}, C. and {Depagne}, {\'E}. and {Dietrich}, J.~P. and {Donath}, A. and {Droettboom}, M. and {Earl}, N. and {Erben}, T. and {Fabbro}, S. and {Ferreira}, L.~A. and {Finethy}, T. and {Fox}, R.~T. and {Garrison}, L.~H. and {Gibbons}, S.~L.~J. and {Goldstein}, D.~A. and {Gommers}, R. and {Greco}, J.~P. and {Greenfield}, P. and {Groener}, A.~M. and {Grollier}, F. and {Hagen}, A. and {Hirst}, P. and {Homeier}, D. and {Horton}, A.~J. and {Hosseinzadeh}, G. and {Hu}, L. and {Hunkeler}, J.~S. and {Ivezi{\'c}}, {\v{Z}}. and {Jain}, A. and {Jenness}, T. and {Kanarek}, G. and {Kendrew}, S. and {Kern}, N.~S. and {Kerzendorf}, W.~E. and {Khvalko}, A. and {King}, J. and {Kirkby}, D. and {Kulkarni}, A.~M. and {Kumar}, A. and {Lee}, A. and {Lenz}, D. and {Littlefair}, S.~P. and {Ma}, Z. and {Macleod}, D.~M. and {Mastropietro}, M. and {McCully}, C. and {Montagnac}, S. and {Morris}, B.~M. and {Mueller}, M. and {Mumford}, S.~J. and {Muna}, D. and {Murphy}, N.~A. and {Nelson}, S. and {Nguyen}, G.~H. and {Ninan}, J.~P. and {N{\"o}the}, M. and {Ogaz}, S. and {Oh}, S. and {Parejko}, J.~K. and {Parley}, N. and {Pascual}, S. and {Patil}, R. and {Patil}, A.~A. and {Plunkett}, A.~L. and {Prochaska}, J.~X. and {Rastogi}, T. and {Reddy Janga}, V. and {Sabater}, J. and {Sakurikar}, P. and {Seifert}, M. and {Sherbert}, L.~E. and {Sherwood-Taylor}, H. and {Shih}, A.~Y. and {Sick}, J. and {Silbiger}, M.~T. and {Singanamalla}, S. and {Singer}, L.~P. and {Sladen}, P.~H. and {Sooley}, K.~A. and {Sornarajah}, S. and {Streicher}, O. and {Teuben}, P. and {Thomas}, S.~W. and {Tremblay}, G.~R. and {Turner}, J.~E.~H. and {Terr{\'o}n}, V. and {van Kerkwijk}, M.~H. and {de la Vega}, A. and {Watkins}, L.~L. and {Weaver}, B.~A. and {Whitmore}, J.~B. and {Woillez}, J. and {Zabalza}, V. and {Astropy Contributors}},
        title = "{The Astropy Project: Building an Open-science Project and Status of the v2.0 Core Package}",
      journal = {\aj},
         year = 2018,
        month = sep,
       volume = {156},
       number = {3},
          eid = {123},
        pages = {123},
          doi = {10.3847/1538-3881/aabc4f},
archivePrefix = {arXiv},
       eprint = {1801.02634},
 primaryClass = {astro-ph.IM},
       adsurl = {https://ui.adsabs.harvard.edu/abs/2018AJ....156..123A}
}

@ARTICLE{GRAVITY2018,
       author = {{GRAVITY Collaboration} and {Abuter}, R. and {Amorim}, A. and {Anugu}, N. and {Baub{\"o}ck}, M. and {Benisty}, M. and {Berger}, J.~P. and {Blind}, N. and {Bonnet}, H. and {Brandner}, W. and {Buron}, A. and {Collin}, C. and {Chapron}, F. and {Cl{\'e}net}, Y. and {Coud{\'e} Du Foresto}, V. and {de Zeeuw}, P.~T. and {Deen}, C. and {Delplancke-Str{\"o}bele}, F. and {Dembet}, R. and {Dexter}, J. and {Duvert}, G. and {Eckart}, A. and {Eisenhauer}, F. and {Finger}, G. and {F{\"o}rster Schreiber}, N.~M. and {F{\'e}dou}, P. and {Garcia}, P. and {Garcia Lopez}, R. and {Gao}, F. and {Gendron}, E. and {Genzel}, R. and {Gillessen}, S. and {Gordo}, P. and {Habibi}, M. and {Haubois}, X. and {Haug}, M. and {Hau{\ss}mann}, F. and {Henning}, Th. and {Hippler}, S. and {Horrobin}, M. and {Hubert}, Z. and {Hubin}, N. and {Jimenez Rosales}, A. and {Jochum}, L. and {Jocou}, K. and {Kaufer}, A. and {Kellner}, S. and {Kendrew}, S. and {Kervella}, P. and {Kok}, Y. and {Kulas}, M. and {Lacour}, S. and {Lapeyr{\`e}re}, V. and {Lazareff}, B. and {Le Bouquin}, J. -B. and {L{\'e}na}, P. and {Lippa}, M. and {Lenzen}, R. and {M{\'e}rand}, A. and {M{\"u}ler}, E. and {Neumann}, U. and {Ott}, T. and {Palanca}, L. and {Paumard}, T. and {Pasquini}, L. and {Perraut}, K. and {Perrin}, G. and {Pfuhl}, O. and {Plewa}, P.~M. and {Rabien}, S. and {Ram{\'\i}rez}, A. and {Ramos}, J. and {Rau}, C. and {Rodr{\'\i}guez-Coira}, G. and {Rohloff}, R. -R. and {Rousset}, G. and {Sanchez-Bermudez}, J. and {Scheithauer}, S. and {Sch{\"o}ller}, M. and {Schuler}, N. and {Spyromilio}, J. and {Straub}, O. and {Straubmeier}, C. and {Sturm}, E. and {Tacconi}, L.~J. and {Tristram}, K.~R.~W. and {Vincent}, F. and {von Fellenberg}, S. and {Wank}, I. and {Waisberg}, I. and {Widmann}, F. and {Wieprecht}, E. and {Wiest}, M. and {Wiezorrek}, E. and {Woillez}, J. and {Yazici}, S. and {Ziegler}, D. and {Zins}, G.},
        title = "{Detection of the gravitational redshift in the orbit of the star S2 near the Galactic centre massive black hole}",
      journal = {\aap},
         year = 2018,
        month = jul,
       volume = {615},
          eid = {L15},
        pages = {L15},
          doi = {10.1051/0004-6361/201833718},
archivePrefix = {arXiv},
       eprint = {1807.09409},
 primaryClass = {astro-ph.GA},
       adsurl = {https://ui.adsabs.harvard.edu/abs/2018A&A...615L..15G}
}

@ARTICLE{BennettBovy2019,
       author = {{Bennett}, Morgan and {Bovy}, Jo},
        title = "{Vertical waves in the solar neighbourhood in Gaia DR2}",
      journal = {\mnras},
         year = 2019,
        month = jan,
       volume = {482},
       number = {1},
        pages = {1417-1425},
          doi = {10.1093/mnras/sty2813},
archivePrefix = {arXiv},
       eprint = {1809.03507},
 primaryClass = {astro-ph.GA},
       adsurl = {https://ui.adsabs.harvard.edu/abs/2019MNRAS.482.1417B}
}

@article{Wang2020,
	author = {Wang, WenTing and Han, JiaXin and Cautun, Marius and Li, ZhaoZhou and Ishigaki, Miho N.},
	da = {2020/05/06},
	doi = {10.1007/s11433-019-1541-6},
	id = {Wang2020},
	isbn = {1869-1927},
	journal = {Science China Physics, Mechanics \& Astronomy},
	number = {10},
	pages = {109801},
	title = {The mass of our Milky Way},
	ty = {JOUR},
	url = {https://doi.org/10.1007/s11433-019-1541-6},
	volume = {63},
	year = {2020}}

@article{DrimmelPoggio2018,
	author = {Drimmel, Ronald and Poggio, Eloisa},
	doi = {10.3847/2515-5172/aaef8b},
	journal = {Research Notes of the AAS},
	month = {nov},
	number = {4},
	pages = {210},
	publisher = {The American Astronomical Society},
	title = {On the Solar Velocity},
	url = {https://dx.doi.org/10.3847/2515-5172/aaef8b},
	volume = {2},
	year = {2018}}

@ARTICLE{Sotillo-Ramos2022,
       author = {{Sotillo-Ramos}, Diego and {Pillepich}, Annalisa and {Donnari}, Martina and {Nelson}, Dylan and {Eisert}, Lukas and {Rodriguez-Gomez}, Vicente and {Joshi}, Gandhali and {Vogelsberger}, Mark and {Hernquist}, Lars},
        title = "{The merger and assembly histories of Milky Way- and M31-like galaxies with TNG50: disc survival through mergers}",
      journal = {\mnras},
         year = 2022,
        month = nov,
       volume = {516},
       number = {4},
        pages = {5404-5427},
          doi = {10.1093/mnras/stac2586},
archivePrefix = {arXiv},
       eprint = {2211.00036},
 primaryClass = {astro-ph.GA},
       adsurl = {https://ui.adsabs.harvard.edu/abs/2022MNRAS.516.5404S}
}

@ARTICLE{Santistevan2020,
       author = {{Santistevan}, Isaiah B. and {Wetzel}, Andrew and {El-Badry}, Kareem and {Bland-Hawthorn}, Joss and {Boylan-Kolchin}, Michael and {Bailin}, Jeremy and {Faucher-Gigu{\`e}re}, Claude-Andr{\'e} and {Benincasa}, Samantha},
        title = "{The formation times and building blocks of Milky Way-mass galaxies in the FIRE simulations}",
      journal = {\mnras},
         year = 2020,
        month = sep,
       volume = {497},
       number = {1},
        pages = {747-764},
          doi = {10.1093/mnras/staa1923},
archivePrefix = {arXiv},
       eprint = {2001.03178},
 primaryClass = {astro-ph.GA},
       adsurl = {https://ui.adsabs.harvard.edu/abs/2020MNRAS.497..747S}
}

@ARTICLE{Santistevan2024,
       author = {{Santistevan}, Isaiah B. and {Wetzel}, Andrew and {Tollerud}, Erik and {Sanderson}, Robyn E. and {Moreno}, Jorge and {Patel}, Ekta},
        title = "{Modelling the orbital histories of satellites of Milky Way-mass galaxies: testing static host potentials against cosmological simulations}",
      journal = {\mnras},
         year = 2024,
        month = jan,
       volume = {527},
       number = {3},
        pages = {8841-8864},
          doi = {10.1093/mnras/stad3757},
archivePrefix = {arXiv},
       eprint = {2309.05708},
 primaryClass = {astro-ph.GA},
       adsurl = {https://ui.adsabs.harvard.edu/abs/2024MNRAS.527.8841S}
}

@ARTICLE{Souza2022,
       author = {{D'Souza}, Richard and {Bell}, Eric F.},
        title = "{Uncertainties associated with the backward integration of dwarf satellites using simple parametric potentials}",
      journal = {\mnras},
         year = 2022,
        month = may,
       volume = {512},
       number = {1},
        pages = {739-760},
          doi = {10.1093/mnras/stac404},
archivePrefix = {arXiv},
       eprint = {2202.05707},
 primaryClass = {astro-ph.GA},
       adsurl = {https://ui.adsabs.harvard.edu/abs/2022MNRAS.512..739D}
}

@ARTICLE{Doliva-Dolinsky2025,
       author = {{Doliva-Dolinsky}, Amandine and {Collins}, Michelle L.~M. and {Martin}, Nicolas F.},
        title = "{The satellite galaxies of the Milky Way and Andromeda}",
      journal = {arXiv e-prints},
         year = 2025,
        month = feb,
          eid = {arXiv:2502.06948},
        pages = {arXiv:2502.06948},
          doi = {10.48550/arXiv.2502.06948},
archivePrefix = {arXiv},
       eprint = {2502.06948},
 primaryClass = {astro-ph.GA},
       adsurl = {https://ui.adsabs.harvard.edu/abs/2025arXiv250206948D},
      note = {to be published by Elsevier as a Reference Module}
}

@ARTICLE{Tsiane2025,
       author = {{Tsiane}, Kabelo and {Mau}, Sidney and {Drlica-Wagner}, Alex and {Carlin}, Jeffrey L. and {Ferguson}, Peter S. and {Bechtol}, Keith and {Nadler}, Ethan O. and {Peter}, Annika H.~G. and {Mao}, Yao-Yuan and {Thornton}, Adam J. and {LSST Dark Energy Science Collaboration}},
        title = "{Predictions for the Detectability of Milky Way Satellite Galaxies and Outer-Halo Star Clusters with the Vera C. Rubin Observatory}",
      journal = {The Open Journal of Astrophysics},
         year = 2025,
        month = jul,
       volume = {8},
          eid = {89},
        pages = {89},
          doi = {10.33232/001c.142072},
archivePrefix = {arXiv},
       eprint = {2504.16203},
 primaryClass = {astro-ph.GA},
       adsurl = {https://ui.adsabs.harvard.edu/abs/2025OJAp....8E..89T}
}

@ARTICLE{Libralato2024,
       author = {{Libralato}, M. and {Bedin}, L.~R. and {Griggio}, M. and {Massari}, D. and {Anderson}, J. and {Cuillandre}, J. -C. and {Ferguson}, A.~M.~N. and {Lan{\c{c}}on}, A. and {Larsen}, S.~S. and {Schirmer}, M. and {Annibali}, F. and {Balbinot}, E. and {Dalessandro}, E. and {Erkal}, D. and {Kuzma}, P.~B. and {Saifollahi}, T. and {Verdoes Kleijn}, G. and {K{\"u}mmel}, M. and {Nakajima}, R. and {Correnti}, M. and {Battaglia}, G. and {Altieri}, B. and {Amara}, A. and {Andreon}, S. and {Baccigalupi}, C. and {Baldi}, M. and {Balestra}, A. and {Bardelli}, S. and {Basset}, A. and {Battaglia}, P. and {Bonino}, D. and {Branchini}, E. and {Brescia}, M. and {Brinchmann}, J. and {Caillat}, A. and {Camera}, S. and {Capobianco}, V. and {Carbone}, C. and {Carretero}, J. and {Casas}, S. and {Castellano}, M. and {Castignani}, G. and {Cavuoti}, S. and {Cimatti}, A. and {Colodro-Conde}, C. and {Congedo}, G. and {Conselice}, C.~J. and {Conversi}, L. and {Copin}, Y. and {Courbin}, F. and {Courtois}, H.~M. and {Cropper}, M. and {Da Silva}, A. and {Degaudenzi}, H. and {De Lucia}, G. and {Dinis}, J. and {Dubath}, F. and {Dupac}, X. and {Dusini}, S. and {Fabricius}, M. and {Farina}, M. and {Farrens}, S. and {Faustini}, F. and {Ferriol}, S. and {Fosalba}, P. and {Frailis}, M. and {Franceschi}, E. and {Fumana}, M. and {Galeotta}, S. and {Garilli}, B. and {George}, K. and {Gillard}, W. and {Gillis}, B. and {Giocoli}, C. and {G{\'o}mez-Alvarez}, P. and {Grazian}, A. and {Grupp}, F. and {Guzzo}, L. and {Haugan}, S.~V.~H. and {Hoar}, J. and {Hoekstra}, H. and {Holmes}, W. and {Hormuth}, F. and {Hornstrup}, A. and {Hudelot}, P. and {Jahnke}, K. and {Jhabvala}, M. and {Keih{\"a}nen}, E. and {Kermiche}, S. and {Kiessling}, A. and {Kilbinger}, M. and {Kubik}, B. and {Kunz}, M. and {Kurki-Suonio}, H. and {Laureijs}, R. and {Le Mignant}, D. and {Ligori}, S. and {Lilje}, P.~B. and {Lindholm}, V. and {Lloro}, I. and {Maiorano}, E. and {Mansutti}, O. and {Marggraf}, O. and {Markovic}, K. and {Martinelli}, M. and {Martinet}, N. and {Marulli}, F. and {Massey}, R. and {Medinaceli}, E. and {Mei}, S. and {Melchior}, M. and {Mellier}, Y. and {Meneghetti}, M. and {Merlin}, E. and {Meylan}, G. and {Moresco}, M. and {Moscardini}, L. and {Neissner}, C. and {Nichol}, R.~C. and {Niemi}, S. -M. and {Nightingale}, J.~W. and {Padilla}, C. and {Paltani}, S. and {Pasian}, F. and {Pedersen}, K. and {Percival}, W.~J. and {Pettorino}, V. and {Pires}, S. and {Polenta}, G. and {Poncet}, M. and {Popa}, L.~A. and {Pozzetti}, L. and {Raison}, F. and {Rebolo}, R. and {Refregier}, A. and {Renzi}, A. and {Rhodes}, J. and {Riccio}, G. and {Romelli}, E. and {Roncarelli}, M. and {Rossetti}, E. and {Saglia}, R. and {Sakr}, Z. and {S{\'a}nchez}, A.~G. and {Sapone}, D. and {Sartoris}, B. and {Sauvage}, M. and {Schneider}, P. and {Schrabback}, T. and {Secroun}, A. and {Sefusatti}, E. and {Seidel}, G. and {Seiffert}, M. and {Serrano}, S. and {Sirignano}, C. and {Sirri}, G. and {Skottfelt}, J. and {Stanco}, L. and {Steinwagner}, J. and {Tallada-Cresp{\'\i}}, P. and {Taylor}, A.~N. and {Teplitz}, H.~I. and {Tereno}, I. and {Toledo-Moreo}, R. and {Torradeflot}, F. and {Tsyganov}, A. and {Tutusaus}, I. and {Valenziano}, L. and {Vassallo}, T. and {Veropalumbo}, A. and {Wang}, Y. and {Weller}, J. and {Zamorani}, G. and {Zucca}, E. and {Burigana}, C. and {Scottez}, V. and {Scott}, D. and {Smart}, R.~L.},
        title = "{Euclid: High-precision imaging astrometry and photometry from Early Release Observations: I. Internal kinematics of NGC6397 by combining Euclid and Gaia data}",
      journal = {\aap},
         year = 2024,
        month = dec,
       volume = {692},
          eid = {A96},
        pages = {A96},
          doi = {10.1051/0004-6361/202452295},
archivePrefix = {arXiv},
       eprint = {2411.02487},
 primaryClass = {astro-ph.SR},
       adsurl = {https://ui.adsabs.harvard.edu/abs/2024A&A...692A..96L}
}

@ARTICLE{McKinnon2024,
       author = {{McKinnon}, Kevin A. and {del Pino}, Andr{\'e}s and {Rockosi}, Constance M. and {Apfel}, Miranda and {Guhathakurta}, Puragra and {van der Marel}, Roeland P. and {Bennet}, Paul and {Fardal}, Mark A. and {Libralato}, Mattia and {Sohn}, Sangmo Tony and {Vitral}, Eduardo and {Watkins}, Laura L.},
        title = "{BP3M: Bayesian Positions, Parallaxes, and Proper Motions Derived from the Hubble Space Telescope and Gaia Data}",
      journal = {\apj},
         year = 2024,
        month = sep,
       volume = {972},
       number = {2},
          eid = {150},
        pages = {150},
          doi = {10.3847/1538-4357/ad5834},
archivePrefix = {arXiv},
       eprint = {2310.20099},
 primaryClass = {astro-ph.GA},
       adsurl = {https://ui.adsabs.harvard.edu/abs/2024ApJ...972..150M}
}

@ARTICLE{Warfield2023,
       author = {{Warfield}, Jack T. and {Kallivayalil}, Nitya and {Zivick}, Paul and {Fritz}, Tobias and {Richstein}, Hannah and {Sohn}, Sangmo Tony and {del Pino}, Andr{\'e}s and {Savino}, Alessandro and {Weisz}, Daniel R.},
        title = "{HUBPUG: proper motions for local group dwarfs observed with HST utilizing Gaia as a reference frame}",
      journal = {\mnras},
         year = 2023,
        month = feb,
       volume = {519},
       number = {1},
        pages = {1189-1200},
          doi = {10.1093/mnras/stac3647},
archivePrefix = {arXiv},
       eprint = {2209.02751},
 primaryClass = {astro-ph.GA},
       adsurl = {https://ui.adsabs.harvard.edu/abs/2023MNRAS.519.1189W}
}

@ARTICLE{Jones1994,
       author = {{Jones}, B.~F. and {Klemola}, A.~R. and {Lin}, D.~N.~C.},
        title = "{Proper Motion of The Large Magellanic Cloud And The Mass of The Galaxy. I. Observational Result}",
      journal = {\aj},
         year = 1994,
        month = apr,
       volume = {107},
        pages = {1333},
          doi = {10.1086/116947},
       adsurl = {https://ui.adsabs.harvard.edu/abs/1994AJ....107.1333J}
}

@ARTICLE{Schweitzer1995,
       author = {{Schweitzer}, A.~E. and {Cudworth}, K.~M. and {Majewski}, S.~R. and {Suntzeff}, N.~B.},
        title = "{The Absolute Proper Motion and a Membership Survey of the Sculptor Dwarf Spheroidal Galaxy}",
      journal = {\aj},
         year = 1995,
        month = dec,
       volume = {110},
        pages = {2747},
          doi = {10.1086/117727},
       adsurl = {https://ui.adsabs.harvard.edu/abs/1995AJ....110.2747S}
}

@INPROCEEDINGS{ScholzIrwin1994,
       author = {{Scholz}, R. -D. and {Irwin}, M.~J.},
        title = "{Absolute Proper Motions of the Dwarf Spheroidal Galaxies in Draco and Ursa Minor}",
    booktitle = {Astronomy from Wide-Field Imaging},
         year = 1994,
       editor = {{MacGillivray}, H.~T.},
       series = {IAU Symposium},
       volume = {161},
        month = jan,
        pages = {535},
       adsurl = {https://ui.adsabs.harvard.edu/abs/1994IAUS..161..535S}
}

@ARTICLE{Piatek2002,
       author = {{Piatek}, Slawomir and {Pryor}, Carlton and {Olszewski}, Edward W. and {Harris}, Hugh C. and {Mateo}, Mario and {Minniti}, Dante and {Monet}, David G. and {Morrison}, Heather and {Tinney}, Christopher G.},
        title = "{Proper Motions of Dwarf Spheroidal Galaxies from Hubble Space Telescope Imaging. I. Method and a Preliminary Measurement for Fornax}",
      journal = {\aj},
         year = 2002,
        month = dec,
       volume = {124},
       number = {6},
        pages = {3198-3221},
          doi = {10.1086/344767},
archivePrefix = {arXiv},
       eprint = {astro-ph/0209430},
 primaryClass = {astro-ph},
       adsurl = {https://ui.adsabs.harvard.edu/abs/2002AJ....124.3198P}
}

@ARTICLE{Piatek2005,
       author = {{Piatek}, Slawomir and {Pryor}, Carlton and {Bristow}, Paul and {Olszewski}, Edward W. and {Harris}, Hugh C. and {Mateo}, Mario and {Minniti}, Dante and {Tinney}, Christopher G.},
        title = "{Proper Motions of Dwarf Spheroidal Galaxies from Hubble Space Telescope Imaging. III. Measurement for Ursa Minor}",
      journal = {\aj},
         year = 2005,
        month = jul,
       volume = {130},
       number = {1},
        pages = {95-115},
          doi = {10.1086/430532},
archivePrefix = {arXiv},
       eprint = {astro-ph/0503620},
 primaryClass = {astro-ph},
       adsurl = {https://ui.adsabs.harvard.edu/abs/2005AJ....130...95P}
}

@ARTICLE{Kallivayalil2006,
       author = {{Kallivayalil}, Nitya and {van der Marel}, Roeland P. and {Alcock}, Charles and {Axelrod}, Tim and {Cook}, Kem H. and {Drake}, A.~J. and {Geha}, M.},
        title = "{The Proper Motion of the Large Magellanic Cloud Using HST}",
      journal = {\apj},
         year = 2006,
        month = feb,
       volume = {638},
       number = {2},
        pages = {772-785},
          doi = {10.1086/498972},
archivePrefix = {arXiv},
       eprint = {astro-ph/0508457},
 primaryClass = {astro-ph},
       adsurl = {https://ui.adsabs.harvard.edu/abs/2006ApJ...638..772K}
}

@ARTICLE{Kallivayalil2006b,
       author = {{Kallivayalil}, Nitya and {van der Marel}, Roeland P. and {Alcock}, Charles},
        title = "{Is the SMC Bound to the LMC? The Hubble Space Telescope Proper Motion of the SMC}",
      journal = {\apj},
         year = 2006,
        month = dec,
       volume = {652},
       number = {2},
        pages = {1213-1229},
          doi = {10.1086/508014},
archivePrefix = {arXiv},
       eprint = {astro-ph/0606240},
 primaryClass = {astro-ph},
       adsurl = {https://ui.adsabs.harvard.edu/abs/2006ApJ...652.1213K}
}

@ARTICLE{Julio2024,
       author = {{J{\'u}lio}, Mariana P. and {Pawlowski}, Marcel S. and {Tony Sohn}, Sangmo and {Taibi}, Salvatore and {van der Marel}, Roeland P. and {McGaugh}, Stacy S.},
        title = "{Satellite group infall into the Milky Way: Exploring the Crater-Leo case with new HST proper motions}",
      journal = {\aap},
         year = 2024,
        month = jul,
       volume = {687},
          eid = {A212},
        pages = {A212},
          doi = {10.1051/0004-6361/202449985},
archivePrefix = {arXiv},
       eprint = {2404.16110},
 primaryClass = {astro-ph.GA},
       adsurl = {https://ui.adsabs.harvard.edu/abs/2024A&A...687A.212J}
}

@ARTICLE{MartinezVazquez2015,
       author = {{Mart{\'\i}nez-V{\'a}zquez}, C.~E. and {Monelli}, M. and {Bono}, G. and {Stetson}, P.~B. and {Ferraro}, I. and {Bernard}, E.~J. and {Gallart}, C. and {Fiorentino}, G. and {Iannicola}, G. and {Udalski}, A.},
        title = "{Variable stars in Local Group Galaxies - I. Tracing the early chemical enrichment and radial gradients in the Sculptor dSph with RR Lyrae stars}",
      journal = {\mnras},
         year = 2015,
        month = dec,
       volume = {454},
       number = {2},
        pages = {1509-1516},
          doi = {10.1093/mnras/stv2014},
archivePrefix = {arXiv},
       eprint = {1508.06942},
 primaryClass = {astro-ph.GA},
       adsurl = {https://ui.adsabs.harvard.edu/abs/2015MNRAS.454.1509M}
}

@ARTICLE{Vitral2024,
       author = {{Vitral}, Eduardo and {van der Marel}, Roeland P. and {Sohn}, Sangmo Tony and {Libralato}, Mattia and {del Pino}, Andr{\'e}s and {Watkins}, Laura L. and {Bellini}, Andrea and {Walker}, Matthew G. and {Besla}, Gurtina and {Pawlowski}, Marcel S. and {Mamon}, Gary A.},
        title = "{HSTPROMO Internal Proper-motion Kinematics of Dwarf Spheroidal Galaxies. I. Velocity Anisotropy and Dark Matter Cusp Slope of Draco}",
      journal = {\apj},
         year = 2024,
        month = jul,
       volume = {970},
       number = {1},
          eid = {1},
        pages = {1},
          doi = {10.3847/1538-4357/ad571c},
archivePrefix = {arXiv},
       eprint = {2407.07769},
 primaryClass = {astro-ph.GA},
       adsurl = {https://ui.adsabs.harvard.edu/abs/2024ApJ...970....1V}
}

@ARTICLE{Vitral2025,
       author = {{Vitral}, Eduardo and {van der Marel}, Roeland P. and {Sohn}, Sangmo Tony and {Pe{\~n}arrubia}, Jorge and {Patel}, Ekta and {Watkins}, Laura L. and {Libralato}, Mattia and {McKinnon}, Kevin A. and {Bellini}, Andrea and {del Pino}, Andr{\'e}s and {Bennet}, Paul},
        title = "{HSTPROMO Internal Proper-motion Kinematics of Dwarf Spheroidal Galaxies. II. Velocity Anisotropy and Dark Matter Cusp Slope of Sculptor}",
      journal = {\apj},
         year = 2026,
        month = feb,
       volume = {998},
       number = {2},
          eid = {206},
        pages = {206},
          doi = {10.3847/1538-4357/ae1f8a},
archivePrefix = {arXiv},
       eprint = {2508.20711},
 primaryClass = {astro-ph.GA},
       adsurl = {https://ui.adsabs.harvard.edu/abs/2026ApJ...998..206V}
}

@ARTICLE{Wilson1955,
       author = {{Wilson}, A.~G.},
        title = "{Sculptor-Type Systems in the Local Group of Galaxies}",
      journal = {\pasp},
         year = 1955,
        month = feb,
       volume = {67},
       number = {394},
        pages = {27-29},
          doi = {10.1086/126754},
       adsurl = {https://ui.adsabs.harvard.edu/abs/1955PASP...67...27W}
}

@ARTICLE{Bennet2024,
       author = {{Bennet}, Paul and {Patel}, Ekta and {Sohn}, Sangmo Tony and {del Pino Molina}, Andr{\'e}s and {van der Marel}, Roeland P. and {Libralato}, Mattia and {Watkins}, Laura L. and {Aparicio}, Antonio and {Besla}, Gurtina and {Gallart}, Carme and {Fardal}, Mark A. and {Monelli}, Matteo and {Sacchi}, Elena and {Tollerud}, Erik and {Weisz}, Daniel R.},
        title = "{Proper Motions and Orbits of Distant Local Group Dwarf Galaxies from a Combination of Gaia and Hubble Data}",
      journal = {\apj},
         year = 2024,
        month = aug,
       volume = {971},
       number = {1},
          eid = {98},
        pages = {98},
          doi = {10.3847/1538-4357/ad5349},
archivePrefix = {arXiv},
       eprint = {2312.09276},
 primaryClass = {astro-ph.GA},
       adsurl = {https://ui.adsabs.harvard.edu/abs/2024ApJ...971...98B}
}

@ARTICLE{Rusakov2021,
       author = {{Rusakov}, V. and {Monelli}, M. and {Gallart}, C. and {Fritz}, T.~K. and {Ruiz-Lara}, T. and {Bernard}, E.~J. and {Cassisi}, S.},
        title = "{The bursty star formation history of the Fornax dwarf spheroidal galaxy revealed with the HST}",
      journal = {\mnras},
         year = 2021,
        month = mar,
       volume = {502},
       number = {1},
        pages = {642-661},
          doi = {10.1093/mnras/stab006},
archivePrefix = {arXiv},
       eprint = {2002.09714},
 primaryClass = {astro-ph.GA},
       adsurl = {https://ui.adsabs.harvard.edu/abs/2021MNRAS.502..642R}
}

@ARTICLE{Taibi2024,
       author = {{Taibi}, S. and {Pawlowski}, M.~S. and {Khoperskov}, S. and {Steinmetz}, M. and {Libeskind}, N.~I.},
        title = "{A portrait of the vast polar structure as a young phenomenon: Hints from its member satellites}",
      journal = {\aap},
         year = 2024,
        month = jan,
       volume = {681},
          eid = {A73},
        pages = {A73},
          doi = {10.1051/0004-6361/202347473},
archivePrefix = {arXiv},
       eprint = {2310.13521},
 primaryClass = {astro-ph.GA},
       adsurl = {https://ui.adsabs.harvard.edu/abs/2024A&A...681A..73T}
}

@ARTICLE{MartinezGarcia2025,
       author = {{Mart{\'\i}nez-Garc{\'\i}a}, Alberto Manuel and {del Pino}, Andr{\'e}s},
        title = "{The long-term stability of the Vast Polar Structure and its connection to a possible previous passage of the LMC}",
      journal = {\aap},
         year = 2025,
        month = oct,
       volume = {703},
          eid = {A1},
        pages = {A1},
          doi = {10.1051/0004-6361/202556585},
archivePrefix = {arXiv},
       eprint = {2507.18685},
 primaryClass = {astro-ph.GA},
       adsurl = {https://ui.adsabs.harvard.edu/abs/2025A&A...703A...1M}
}

@ARTICLE{Pawlowski2021,
       author = {{Pawlowski}, Marcel S.},
        title = "{Phase-Space Correlations among Systems of Satellite Galaxies}",
      journal = {Galaxies},
         year = 2021,
        month = sep,
       volume = {9},
       number = {3},
          eid = {66},
        pages = {66},
          doi = {10.3390/galaxies9030066},
archivePrefix = {arXiv},
       eprint = {2109.02654},
 primaryClass = {astro-ph.GA},
       adsurl = {https://ui.adsabs.harvard.edu/abs/2021Galax...9...66P}
}

@ARTICLE{Tan2025b,
       author = {{Tan}, C.~Y. and {Drlica-Wagner}, A. and {Pace}, A.~B. and {Cerny}, W. and {Nadler}, E.~O. and {Doliva-Dolinsky}, A. and {Anbajagane}, D. and {Li}, T.~S. and {Simon}, J.~D. and {Vivas}, A.~K. and {Walker}, A.~R. and {Adam{\'o}w}, M. and {Bechtol}, K. and {Carlin}, J.~L. and {Casey}, Q.~O. and {Chang}, C. and {Chaturvedi}, A. and {Cheng}, T.-Y. and {Chiti}, A. and {Choi}, Y. and {Crnojevi{\'c}}, D. and {Ferguson}, P.~S. and {Gruendl}, R.~A. and {Ji}, A.~P. and {Limberg}, G. and {Medina}, G.~E. and {Mutlu-Pakdil}, B. and {No{\"e}l}, N.~E.~D. and {Overdeck}, K. and {Placco}, V.~M. and {Riley}, A.~H. and {Sand}, D.~J. and {Sharp}, J. and {Sherman}, N.~F. and {Stringfellow}, G.~S. and {Wechsler}, R.~H. and {Aguena}, M. and {Allam}, S. and {Alves}, O. and {Bacon}, D. and {Brooks}, D. and {Burke}, D.~L. and {Camilleri}, R. and {Carballo-Bello}, J.~A. and {Carnero Rosell}, A. and {Carretero}, J. and {da Costa}, L.~N. and {da Silva Pereira}, M.~E. and {Davis}, T.~M. and {de Vicente}, J. and {Desai}, S. and {Everett}, S. and {Flaugher}, B. and {Frieman}, J. and {Garc{\'\i}a-Bellido}, J. and {Gruen}, D. and {Gutierrez}, G. and {Herner}, K. and {Hinton}, S.~R. and {Hollowood}, D.~L. and {James}, D.~J. and {Kuehn}, K. and {Lahav}, O. and {Lee}, S. and {Marshall}, J.~L. and {Mart{\'\i}nez-V{\'a}zquez}, C.~E. and {Massana}, P. and {Mena-Fern{\'a}ndez}, J. and {Miquel}, R. and {Muir}, J. and {Myles}, J. and {Ogando}, R.~L.~C. and {Plazas Malag{\'o}n}, A.~A. and {Porredon}, A. and {Sanchez}, E. and {Sanchez Cid}, D. and {Sevilla-Noarbe}, I. and {Smith}, M. and {Suchyta}, E. and {Swanson}, M.~E.~C. and {To}, C. and {Tollerud}, E.~J. and {Tucker}, D.~L. and {Vikram}, V. and {Weaverdyck}, N. and {Yamamoto}, M. and {Zenteno}, A. and {Delve Collaboration} and {Des Collaboration}},
        title = "{DELVE Milky Way Satellite Galaxy Census. I. Satellite Population and Survey Selection Function in DES, DELVE, and Pan-STARRS}",
      journal = {\apj},
         year = 2026,
        month = mar,
       volume = {1000},
       number = {1},
          eid = {87},
        pages = {87},
          doi = {10.3847/1538-4357/ae4479},
archivePrefix = {arXiv},
       eprint = {2509.12313},
 primaryClass = {astro-ph.GA},
       adsurl = {https://ui.adsabs.harvard.edu/abs/2026ApJ..1000...87T}
}

@ARTICLE{Nadler2024,
       author = {{Nadler}, Ethan O. and {Gluscevic}, Vera and {Driskell}, Trey and {Wechsler}, Risa H. and {Moustakas}, Leonidas A. and {Benson}, Andrew and {Mao}, Yao-Yuan},
        title = "{Forecasts for Galaxy Formation and Dark Matter Constraints from Dwarf Galaxy Surveys}",
      journal = {\apj},
         year = 2024,
        month = may,
       volume = {967},
       number = {1},
          eid = {61},
        pages = {61},
          doi = {10.3847/1538-4357/ad3bb1},
archivePrefix = {arXiv},
       eprint = {2401.10318},
 primaryClass = {astro-ph.GA},
       adsurl = {https://ui.adsabs.harvard.edu/abs/2024ApJ...967...61N}
}

@ARTICLE{ErkalBelokurov2020,
       author = {{Erkal}, Denis and {Belokurov}, Vasily A.},
        title = "{Limit on the LMC mass from a census of its satellites}",
      journal = {\mnras},
         year = 2020,
        month = jul,
       volume = {495},
       number = {3},
        pages = {2554-2563},
          doi = {10.1093/mnras/staa1238},
archivePrefix = {arXiv},
       eprint = {1907.09484},
 primaryClass = {astro-ph.GA},
       adsurl = {https://ui.adsabs.harvard.edu/abs/2020MNRAS.495.2554E}
}

@ARTICLE{CorreaMagnus2022,
       author = {{Correa Magnus}, Lilia and {Vasiliev}, Eugene},
        title = "{Measuring the Milky Way mass distribution in the presence of the LMC}",
      journal = {\mnras},
         year = 2022,
        month = apr,
       volume = {511},
       number = {2},
        pages = {2610-2630},
          doi = {10.1093/mnras/stab3726},
archivePrefix = {arXiv},
       eprint = {2110.00018},
 primaryClass = {astro-ph.GA},
       adsurl = {https://ui.adsabs.harvard.edu/abs/2022MNRAS.511.2610C}
}

@ARTICLE{Patel2020,
       author = {{Patel}, Ekta and {Kallivayalil}, Nitya and {Garavito-Camargo}, Nicolas and {Besla}, Gurtina and {Weisz}, Daniel R. and {van der Marel}, Roeland P. and {Boylan-Kolchin}, Michael and {Pawlowski}, Marcel S. and {G{\'o}mez}, Facundo A.},
        title = "{The Orbital Histories of Magellanic Satellites Using Gaia DR2 Proper Motions}",
      journal = {\apj},
         year = 2020,
        month = apr,
       volume = {893},
       number = {2},
          eid = {121},
        pages = {121},
          doi = {10.3847/1538-4357/ab7b75},
archivePrefix = {arXiv},
       eprint = {2001.01746},
 primaryClass = {astro-ph.GA},
       adsurl = {https://ui.adsabs.harvard.edu/abs/2020ApJ...893..121P}
}

@ARTICLE{Simon2020,
       author = {{Simon}, J.~D. and {Li}, T.~S. and {Erkal}, D. and {Pace}, A.~B. and {Drlica-Wagner}, A. and {James}, D.~J. and {Marshall}, J.~L. and {Bechtol}, K. and {Hansen}, T. and {Kuehn}, K. and {Lidman}, C. and {Allam}, S. and {Annis}, J. and {Avila}, S. and {Bertin}, E. and {Brooks}, D. and {Burke}, D.~L. and {Rosell}, A. Carnero and {Carrasco Kind}, M. and {Carretero}, J. and {da Costa}, L.~N. and {De Vicente}, J. and {Desai}, S. and {Doel}, P. and {Eifler}, T.~F. and {Everett}, S. and {Fosalba}, P. and {Frieman}, J. and {Garc{\'\i}a-Bellido}, J. and {Gaztanaga}, E. and {Gerdes}, D.~W. and {Gruen}, D. and {Gruendl}, R.~A. and {Gschwend}, J. and {Gutierrez}, G. and {Hollowood}, D.~L. and {Honscheid}, K. and {Krause}, E. and {Kuropatkin}, N. and {MacCrann}, N. and {Maia}, M.~A.~G. and {March}, M. and {Miquel}, R. and {Palmese}, A. and {Paz-Chinch{\'o}n}, F. and {Plazas}, A.~A. and {Reil}, K. and {Roodman}, A. and {Sanchez}, E. and {Santiago}, B. and {Scarpine}, V. and {Schubnell}, M. and {Serrano}, S. and {Smith}, M. and {Suchyta}, E. and {Tarle}, G. and {Walker}, A.~R. and {DES Collaboration}},
        title = "{Birds of a Feather? Magellan/IMACS Spectroscopy of the Ultra-faint Satellites Grus II, Tucana IV, and Tucana V}",
      journal = {\apj},
         year = 2020,
        month = apr,
       volume = {892},
       number = {2},
          eid = {137},
        pages = {137},
          doi = {10.3847/1538-4357/ab7ccb},
archivePrefix = {arXiv},
       eprint = {1911.08493},
 primaryClass = {astro-ph.GA},
       adsurl = {https://ui.adsabs.harvard.edu/abs/2020ApJ...892..137S}
}

@ARTICLE{Spergel2015,
       author = {{Spergel}, D. and {Gehrels}, N. and {Baltay}, C. and {Bennett}, D. and {Breckinridge}, J. and {Donahue}, M. and {Dressler}, A. and {Gaudi}, B.~S. and {Greene}, T. and {Guyon}, O. and {Hirata}, C. and {Kalirai}, J. and {Kasdin}, N.~J. and {Macintosh}, B. and {Moos}, W. and {Perlmutter}, S. and {Postman}, M. and {Rauscher}, B. and {Rhodes}, J. and {Wang}, Y. and {Weinberg}, D. and {Benford}, D. and {Hudson}, M. and {Jeong}, W. -S. and {Mellier}, Y. and {Traub}, W. and {Yamada}, T. and {Capak}, P. and {Colbert}, J. and {Masters}, D. and {Penny}, M. and {Savransky}, D. and {Stern}, D. and {Zimmerman}, N. and {Barry}, R. and {Bartusek}, L. and {Carpenter}, K. and {Cheng}, E. and {Content}, D. and {Dekens}, F. and {Demers}, R. and {Grady}, K. and {Jackson}, C. and {Kuan}, G. and {Kruk}, J. and {Melton}, M. and {Nemati}, B. and {Parvin}, B. and {Poberezhskiy}, I. and {Peddie}, C. and {Ruffa}, J. and {Wallace}, J.~K. and {Whipple}, A. and {Wollack}, E. and {Zhao}, F.},
        title = "{Wide-Field InfrarRed Survey Telescope-Astrophysics Focused Telescope Assets WFIRST-AFTA 2015 Report}",
      journal = {arXiv e-prints},
         year = 2015,
        month = mar,
          eid = {arXiv:1503.03757},
        pages = {arXiv:1503.03757},
          doi = {10.48550/arXiv.1503.03757},
archivePrefix = {arXiv},
       eprint = {1503.03757},
 primaryClass = {astro-ph.IM},
       adsurl = {https://ui.adsabs.harvard.edu/abs/2015arXiv150303757S}
}

@ARTICLE{Ivezic2019,
       author = {{Ivezi{\'c}}, {\v{Z}}eljko and {Kahn}, Steven M. and {Tyson}, J. Anthony and {Abel}, Bob and {Acosta}, Emily and {Allsman}, Robyn and {Alonso}, David and {AlSayyad}, Yusra and {Anderson}, Scott F. and {Andrew}, John and {Angel}, James Roger P. and {Angeli}, George Z. and {Ansari}, Reza and {Antilogus}, Pierre and {Araujo}, Constanza and {Armstrong}, Robert and {Arndt}, Kirk T. and {Astier}, Pierre and {Aubourg}, {\'E}ric and {Auza}, Nicole and {Axelrod}, Tim S. and {Bard}, Deborah J. and {Barr}, Jeff D. and {Barrau}, Aurelian and {Bartlett}, James G. and {Bauer}, Amanda E. and {Bauman}, Brian J. and {Baumont}, Sylvain and {Bechtol}, Ellen and {Bechtol}, Keith and {Becker}, Andrew C. and {Becla}, Jacek and {Beldica}, Cristina and {Bellavia}, Steve and {Bianco}, Federica B. and {Biswas}, Rahul and {Blanc}, Guillaume and {Blazek}, Jonathan and {Blandford}, Roger D. and {Bloom}, Josh S. and {Bogart}, Joanne and {Bond}, Tim W. and {Booth}, Michael T. and {Borgland}, Anders W. and {Borne}, Kirk and {Bosch}, James F. and {Boutigny}, Dominique and {Brackett}, Craig A. and {Bradshaw}, Andrew and {Brandt}, William Nielsen and {Brown}, Michael E. and {Bullock}, James S. and {Burchat}, Patricia and {Burke}, David L. and {Cagnoli}, Gianpietro and {Calabrese}, Daniel and {Callahan}, Shawn and {Callen}, Alice L. and {Carlin}, Jeffrey L. and {Carlson}, Erin L. and {Chandrasekharan}, Srinivasan and {Charles-Emerson}, Glenaver and {Chesley}, Steve and {Cheu}, Elliott C. and {Chiang}, Hsin-Fang and {Chiang}, James and {Chirino}, Carol and {Chow}, Derek and {Ciardi}, David R. and {Claver}, Charles F. and {Cohen-Tanugi}, Johann and {Cockrum}, Joseph J. and {Coles}, Rebecca and {Connolly}, Andrew J. and {Cook}, Kem H. and {Cooray}, Asantha and {Covey}, Kevin R. and {Cribbs}, Chris and {Cui}, Wei and {Cutri}, Roc and {Daly}, Philip N. and {Daniel}, Scott F. and {Daruich}, Felipe and {Daubard}, Guillaume and {Daues}, Greg and {Dawson}, William and {Delgado}, Francisco and {Dellapenna}, Alfred and {de Peyster}, Robert and {de Val-Borro}, Miguel and {Digel}, Seth W. and {Doherty}, Peter and {Dubois}, Richard and {Dubois-Felsmann}, Gregory P. and {Durech}, Josef and {Economou}, Frossie and {Eifler}, Tim and {Eracleous}, Michael and {Emmons}, Benjamin L. and {Fausti Neto}, Angelo and {Ferguson}, Henry and {Figueroa}, Enrique and {Fisher-Levine}, Merlin and {Focke}, Warren and {Foss}, Michael D. and {Frank}, James and {Freemon}, Michael D. and {Gangler}, Emmanuel and {Gawiser}, Eric and {Geary}, John C. and {Gee}, Perry and {Geha}, Marla and {Gessner}, Charles J.~B. and {Gibson}, Robert R. and {Gilmore}, D. Kirk and {Glanzman}, Thomas and {Glick}, William and {Goldina}, Tatiana and {Goldstein}, Daniel A. and {Goodenow}, Iain and {Graham}, Melissa L. and {Gressler}, William J. and {Gris}, Philippe and {Guy}, Leanne P. and {Guyonnet}, Augustin and {Haller}, Gunther and {Harris}, Ron and {Hascall}, Patrick A. and {Haupt}, Justine and {Hernandez}, Fabio and {Herrmann}, Sven and {Hileman}, Edward and {Hoblitt}, Joshua and {Hodgson}, John A. and {Hogan}, Craig and {Howard}, James D. and {Huang}, Dajun and {Huffer}, Michael E. and {Ingraham}, Patrick and {Innes}, Walter R. and {Jacoby}, Suzanne H. and {Jain}, Bhuvnesh and {Jammes}, Fabrice and {Jee}, M. James and {Jenness}, Tim and {Jernigan}, Garrett and {Jevremovi{\'c}}, Darko and {Johns}, Kenneth and {Johnson}, Anthony S. and {Johnson}, Margaret W.~G. and {Jones}, R. Lynne and {Juramy-Gilles}, Claire and {Juri{\'c}}, Mario and {Kalirai}, Jason S. and {Kallivayalil}, Nitya J. and {Kalmbach}, Bryce and {Kantor}, Jeffrey P. and {Karst}, Pierre and {Kasliwal}, Mansi M. and {Kelly}, Heather and {Kessler}, Richard and {Kinnison}, Veronica and {Kirkby}, David and {Knox}, Lloyd and {Kotov}, Ivan V. and {Krabbendam}, Victor L. and {Krughoff}, K. Simon and {Kub{\'a}nek}, Petr and {Kuczewski}, John and {Kulkarni}, Shri and {Ku}, John and {Kurita}, Nadine R. and {Lage}, Craig S. and {Lambert}, Ron and {Lange}, Travis and {Langton}, J. Brian and {Le Guillou}, Laurent and {Levine}, Deborah and {Liang}, Ming and {Lim}, Kian-Tat and {Lintott}, Chris J. and {Long}, Kevin E. and {Lopez}, Margaux and {Lotz}, Paul J. and {Lupton}, Robert H. and {Lust}, Nate B. and {MacArthur}, Lauren A. and {Mahabal}, Ashish and {Mandelbaum}, Rachel and {Markiewicz}, Thomas W. and {Marsh}, Darren S. and {Marshall}, Philip J. and {Marshall}, Stuart and {May}, Morgan and {McKercher}, Robert and {McQueen}, Michelle and {Meyers}, Joshua and {Migliore}, Myriam and {Miller}, Michelle and {Mills}, David J.},
        title = "{LSST: From Science Drivers to Reference Design and Anticipated Data Products}",
      journal = {\apj},
         year = 2019,
        month = mar,
       volume = {873},
       number = {2},
          eid = {111},
        pages = {111},
          doi = {10.3847/1538-4357/ab042c},
archivePrefix = {arXiv},
       eprint = {0805.2366},
 primaryClass = {astro-ph},
       adsurl = {https://ui.adsabs.harvard.edu/abs/2019ApJ...873..111I}
}

@ARTICLE{EuclidCollaboration2022,
       author = {{Euclid Collaboration} and {Scaramella}, R. and {Amiaux}, J. and {Mellier}, Y. and {Burigana}, C. and {Carvalho}, C.~S. and {Cuillandre}, J.-C. and {Da Silva}, A. and {Derosa}, A. and {Dinis}, J. and {Maiorano}, E. and {Maris}, M. and {Tereno}, I. and {Laureijs}, R. and {Boenke}, T. and {Buenadicha}, G. and {Dupac}, X. and {Gaspar Venancio}, L.~M. and {G{\'o}mez-{\'A}lvarez}, P. and {Hoar}, J. and {Lorenzo Alvarez}, J. and {Racca}, G.~D. and {Saavedra-Criado}, G. and {Schwartz}, J. and {Vavrek}, R. and {Schirmer}, M. and {Aussel}, H. and {Azzollini}, R. and {Cardone}, V.~F. and {Cropper}, M. and {Ealet}, A. and {Garilli}, B. and {Gillard}, W. and {Granett}, B.~R. and {Guzzo}, L. and {Hoekstra}, H. and {Jahnke}, K. and {Kitching}, T. and {Maciaszek}, T. and {Meneghetti}, M. and {Miller}, L. and {Nakajima}, R. and {Niemi}, S.~M. and {Pasian}, F. and {Percival}, W.~J. and {Pottinger}, S. and {Sauvage}, M. and {Scodeggio}, M. and {Wachter}, S. and {Zacchei}, A. and {Aghanim}, N. and {Amara}, A. and {Auphan}, T. and {Auricchio}, N. and {Awan}, S. and {Balestra}, A. and {Bender}, R. and {Bodendorf}, C. and {Bonino}, D. and {Branchini}, E. and {Brau-Nogue}, S. and {Brescia}, M. and {Candini}, G.~P. and {Capobianco}, V. and {Carbone}, C. and {Carlberg}, R.~G. and {Carretero}, J. and {Casas}, R. and {Castander}, F.~J. and {Castellano}, M. and {Cavuoti}, S. and {Cimatti}, A. and {Cledassou}, R. and {Congedo}, G. and {Conselice}, C.~J. and {Conversi}, L. and {Copin}, Y. and {Corcione}, L. and {Costille}, A. and {Courbin}, F. and {Degaudenzi}, H. and {Douspis}, M. and {Dubath}, F. and {Duncan}, C.~A.~J. and {Dusini}, S. and {Farrens}, S. and {Ferriol}, S. and {Fosalba}, P. and {Fourmanoit}, N. and {Frailis}, M. and {Franceschi}, E. and {Franzetti}, P. and {Fumana}, M. and {Gillis}, B. and {Giocoli}, C. and {Grazian}, A. and {Grupp}, F. and {Haugan}, S.~V.~H. and {Holmes}, W. and {Hormuth}, F. and {Hudelot}, P. and {Kermiche}, S. and {Kiessling}, A. and {Kilbinger}, M. and {Kohley}, R. and {Kubik}, B. and {K{\"u}mmel}, M. and {Kunz}, M. and {Kurki-Suonio}, H. and {Lahav}, O. and {Ligori}, S. and {Lilje}, P.~B. and {Lloro}, I. and {Mansutti}, O. and {Marggraf}, O. and {Markovic}, K. and {Marulli}, F. and {Massey}, R. and {Maurogordato}, S. and {Melchior}, M. and {Merlin}, E. and {Meylan}, G. and {Mohr}, J.~J. and {Moresco}, M. and {Morin}, B. and {Moscardini}, L. and {Munari}, E. and {Nichol}, R.~C. and {Padilla}, C. and {Paltani}, S. and {Peacock}, J. and {Pedersen}, K. and {Pettorino}, V. and {Pires}, S. and {Poncet}, M. and {Popa}, L. and {Pozzetti}, L. and {Raison}, F. and {Rebolo}, R. and {Rhodes}, J. and {Rix}, H.-W. and {Roncarelli}, M. and {Rossetti}, E. and {Saglia}, R. and {Schneider}, P. and {Schrabback}, T. and {Secroun}, A. and {Seidel}, G. and {Serrano}, S. and {Sirignano}, C. and {Sirri}, G. and {Skottfelt}, J. and {Stanco}, L. and {Starck}, J.~L. and {Tallada-Cresp{\'\i}}, P. and {Tavagnacco}, D. and {Taylor}, A.~N. and {Teplitz}, H.~I. and {Toledo-Moreo}, R. and {Torradeflot}, F. and {Trifoglio}, M. and {Valentijn}, E.~A. and {Valenziano}, L. and {Verdoes Kleijn}, G.~A. and {Wang}, Y. and {Welikala}, N. and {Weller}, J. and {Wetzstein}, M. and {Zamorani}, G. and {Zoubian}, J. and {Andreon}, S. and {Baldi}, M. and {Bardelli}, S. and {Boucaud}, A. and {Camera}, S. and {Di Ferdinando}, D. and {Fabbian}, G. and {Farinelli}, R. and {Galeotta}, S. and {Graci{\'a}-Carpio}, J. and {Maino}, D. and {Medinaceli}, E. and {Mei}, S. and {Neissner}, C. and {Polenta}, G. and {Renzi}, A. and {Romelli}, E. and {Rosset}, C. and {Sureau}, F. and {Tenti}, M. and {Vassallo}, T. and {Zucca}, E. and {Baccigalupi}, C. and {Balaguera-Antol{\'\i}nez}, A. and {Battaglia}, P. and {Biviano}, A. and {Borgani}, S. and {Bozzo}, E. and {Cabanac}, R. and {Cappi}, A.},
        title = "{Euclid preparation. I. The Euclid Wide Survey}",
      journal = {\aap},
         year = 2022,
        month = jun,
       volume = {662},
          eid = {A112},
        pages = {A112},
          doi = {10.1051/0004-6361/202141938},
archivePrefix = {arXiv},
       eprint = {2108.01201},
 primaryClass = {astro-ph.CO},
       adsurl = {https://ui.adsabs.harvard.edu/abs/2022A&A...662A.112E}
}

@ARTICLE{Gwyn2025,
       author = {{Gwyn}, Stephen and {McConnachie}, Alan W. and {Cuillandre}, Jean-Charles and {Chambers}, Kenneth C. and {Magnier}, Eugene A. and {de Boer}, Thomas and {Hudson}, Michael J. and {Oguri}, Masamune and {Furusawa}, Hisanori and {Hildebrandt}, Hendrik and {Carlberg}, Raymond and {Ellison}, Sara L. and {Furusawa}, Junko and {Gavazzi}, Rapha{\"e}l and {Ibata}, Rodrigo and {Mellier}, Yannick and {Osato}, Ken and {Aussel}, H. and {Baumont}, Lucie and {Bayer}, Manuel and {Boulade}, Olivier and {C{\^o}t{\'e}}, Patrick and {Chemaly}, David and {Daley}, Cail and {Duc}, Pierre-Alain and {Durret}, Florence and {Ellien}, A. and {Fabbro}, S{\'e}bastien and {Ferreira}, Leonardo and {Fitriana}, Itsna K. and {Le Floc'h}, Emeric and {Fudamoto}, Yoshinobu and {Gao}, Hua and {Goh}, L.~W.~K. and {Goto}, Tomotsugu and {Guerrini}, Sacha and {Guinot}, Axel and {H{\'e}nault-Brunet}, Vincent and {Hammer}, Francois and {Harikane}, Yuichi and {Hayashi}, Kohei and {Heesters}, Nick and {Ichikawa}, Kohei and {Kilbinger}, Martin and {Kuzma}, P.~B. and {Li}, Qinxun and {Liaudat}, Tob{\'\i}as I. and {Lin}, Chien-Cheng and {M{\"u}ller}, Oliver and {Martin}, Nicolas F. and {Matsuoka}, Yoshiki and {Medina}, Gustavo E. and {Miyatake}, Hironao and {Miyazaki}, Satoshi and {Mpetha}, Charlie T. and {Nagao}, Tohru and {Navarro}, Julio F. and {Niwano}, Masafumi and {Ogami}, Itsuki and {Okabe}, Nobuhiro and {Onoue}, Masafusa and {Paek}, Gregory S.~H. and {Parker}, Laura C. and {Patton}, David R. and {Peters}, Fabian Hervas and {Prunet}, Simon and {S{\'a}nchez-Janssen}, Rub{\'e}n and {Schultheis}, M. and {Sestito}, Federico and {Smith}, Simon E.~T. and {Starck}, J.-L. and {Starkenburg}, Else and {Stone}, Connor and {Storfer}, Christopher and {Suzuki}, Yoshihisa and {Erben}, T. and {Taibi}, Salvatore and {Thomas}, G.~F. and {Toba}, Yoshiki and {Uchiyama}, Hisakazu and {Valls-Gabaud}, David and {Venn}, Kim A. and {Van Waerbeke}, Ludovic and {Wainscoat}, Richard J. and {Wilkinson}, Scott and {Wittje}, Anna and {Yoshida}, Taketo and {Zhang}, TianFang and {Zhong}, Yuxing},
        title = "{UNIONS: The Ultraviolet Near-infrared Optical Northern Survey}",
      journal = {\aj},
         year = 2025,
        month = dec,
       volume = {170},
       number = {6},
          eid = {324},
        pages = {324},
          doi = {10.3847/1538-3881/ae03ab},
archivePrefix = {arXiv},
       eprint = {2503.13783},
 primaryClass = {astro-ph.GA},
       adsurl = {https://ui.adsabs.harvard.edu/abs/2025AJ....170..324G}
}

@ARTICLE{Pietrzynski2019,
       author = {{Pietrzy{\'n}ski}, G. and {Graczyk}, D. and {Gallenne}, A. and {Gieren}, W. and {Thompson}, I.~B. and {Pilecki}, B. and {Karczmarek}, P. and {G{\'o}rski}, M. and {Suchomska}, K. and {Taormina}, M. and {Zgirski}, B. and {Wielg{\'o}rski}, P. and {Ko{\l}aczkowski}, Z. and {Konorski}, P. and {Villanova}, S. and {Nardetto}, N. and {Kervella}, P. and {Bresolin}, F. and {Kudritzki}, R.~P. and {Storm}, J. and {Smolec}, R. and {Narloch}, W.},
        title = "{A distance to the Large Magellanic Cloud that is precise to one per cent}",
      journal = {\nat},
         year = 2019,
        month = mar,
       volume = {567},
       number = {7747},
        pages = {200-203},
          doi = {10.1038/s41586-019-0999-4},
archivePrefix = {arXiv},
       eprint = {1903.08096},
 primaryClass = {astro-ph.GA},
       adsurl = {https://ui.adsabs.harvard.edu/abs/2019Natur.567..200P}
}

@ARTICLE{Nidever2019,
       author = {{Nidever}, David L. and {Olsen}, Knut and {Choi}, Yumi and {de Boer}, Thomas J.~L. and {Blum}, Robert D. and {Bell}, Eric F. and {Zaritsky}, Dennis and {Martin}, Nicolas F. and {Saha}, Abhijit and {Conn}, Blair C. and {Besla}, Gurtina and {van der Marel}, Roeland P. and {No{\"e}l}, Noelia E.~D. and {Monachesi}, Antonela and {Stringfellow}, Guy S. and {Massana}, Pol and {Cioni}, Maria-Rosa L. and {Gallart}, Carme and {Monelli}, Matteo and {Martinez-Delgado}, David and {Mu{\~n}oz}, Ricardo R. and {Majewski}, Steven R. and {Vivas}, A. Katherina and {Walker}, Alistair R. and {Kaleida}, Catherine and {Chu}, You-Hua},
        title = "{Exploring the Very Extended Low-surface-brightness Stellar Populations of the Large Magellanic Cloud with SMASH}",
      journal = {\apj},
         year = 2019,
        month = apr,
       volume = {874},
       number = {2},
          eid = {118},
        pages = {118},
          doi = {10.3847/1538-4357/aafaf7},
archivePrefix = {arXiv},
       eprint = {1805.02671},
 primaryClass = {astro-ph.GA},
       adsurl = {https://ui.adsabs.harvard.edu/abs/2019ApJ...874..118N}
}

@ARTICLE{Massana2022,
       author = {{Massana}, P. and {Ruiz-Lara}, T. and {No{\"e}l}, N.~E.~D. and {Gallart}, C. and {Nidever}, D.~L. and {Choi}, Y. and {Sakowska}, J.~D. and {Besla}, G. and {Olsen}, K.~A.~G. and {Monelli}, M. and {Dorta}, A. and {Stringfellow}, G.~S. and {Cassisi}, S. and {Bernard}, E.~J. and {Zaritsky}, D. and {Cioni}, M.-R.~L. and {Monachesi}, A. and {van der Marel}, R.~P. and {de Boer}, T.~J.~L. and {Walker}, A.~R.},
        title = "{The synchronized dance of the magellanic clouds' star formation history}",
      journal = {\mnras},
         year = 2022,
        month = jun,
       volume = {513},
       number = {1},
        pages = {L40-L45},
          doi = {10.1093/mnrasl/slac030},
archivePrefix = {arXiv},
       eprint = {2203.09523},
 primaryClass = {astro-ph.GA},
       adsurl = {https://ui.adsabs.harvard.edu/abs/2022MNRAS.513L..40M}
}

@ARTICLE{DeLeo2024,
       author = {{De Leo}, Michele and {Read}, Justin I. and {No{\"e}l}, Noelia E.~D. and {Erkal}, Denis and {Massana}, Pol and {Carrera}, Ricardo},
        title = "{Surviving the waves: evidence for a dark matter cusp in the tidally disrupting Small Magellanic Cloud}",
      journal = {\mnras},
         year = 2024,
        month = nov,
       volume = {535},
       number = {1},
        pages = {1015-1034},
          doi = {10.1093/mnras/stae2428},
archivePrefix = {arXiv},
       eprint = {2303.08838},
 primaryClass = {astro-ph.GA},
       adsurl = {https://ui.adsabs.harvard.edu/abs/2024MNRAS.535.1015D}
}

@ARTICLE{Watkins2024,
       author = {{Watkins}, Laura L. and {van der Marel}, Roeland P. and {Bennet}, Paul},
        title = "{The Mass of the Large Magellanic Cloud from the Three-dimensional Kinematics of Its Globular Clusters}",
      journal = {\apj},
         year = 2024,
        month = mar,
       volume = {963},
       number = {2},
          eid = {84},
        pages = {84},
          doi = {10.3847/1538-4357/ad1f58},
archivePrefix = {arXiv},
       eprint = {2401.14458},
 primaryClass = {astro-ph.GA},
       adsurl = {https://ui.adsabs.harvard.edu/abs/2024ApJ...963...84W}
}

@ARTICLE{Rathore2025a,
       author = {{Rathore}, Himansh and {Choi}, Yumi and {Olsen}, Knut A.~G. and {Besla}, Gurtina},
        title = "{Precise Measurements of the LMC Bar's Geometry with Gaia DR3 and a Novel Solution to Crowding-induced Incompleteness in Star Counting}",
      journal = {\apj},
         year = 2025,
        month = jan,
       volume = {978},
       number = {1},
          eid = {55},
        pages = {55},
          doi = {10.3847/1538-4357/ad93ae},
archivePrefix = {arXiv},
       eprint = {2410.18182},
 primaryClass = {astro-ph.GA},
       adsurl = {https://ui.adsabs.harvard.edu/abs/2025ApJ...978...55R}
}

@ARTICLE{deVaucouleurs1972,
       author = {{de Vaucouleurs}, G. and {Freeman}, K.~C.},
        title = "{Structure and dynamics of barred spiral galaxies, in particular of the Magellanic type}",
      journal = {Vistas in Astronomy},
         year = 1972,
        month = jan,
       volume = {14},
       number = {1},
        pages = {163-294},
          doi = {10.1016/0083-6656(72)90026-8},
       adsurl = {https://ui.adsabs.harvard.edu/abs/1972VA.....14..163D}
}

@ARTICLE{Besla2012,
       author = {{Besla}, Gurtina and {Kallivayalil}, Nitya and {Hernquist}, Lars and {van der Marel}, Roeland P. and {Cox}, T.~J. and {Kere{\v{s}}}, Du{\v{s}}an},
        title = "{The role of dwarf galaxy interactions in shaping the Magellanic System and implications for Magellanic Irregulars}",
      journal = {\mnras},
         year = 2012,
        month = apr,
       volume = {421},
       number = {3},
        pages = {2109-2138},
          doi = {10.1111/j.1365-2966.2012.20466.x},
archivePrefix = {arXiv},
       eprint = {1201.1299},
 primaryClass = {astro-ph.GA},
       adsurl = {https://ui.adsabs.harvard.edu/abs/2012MNRAS.421.2109B}
}

@ARTICLE{Choi2022,
       author = {{Choi}, Yumi and {Olsen}, Knut A.~G. and {Besla}, Gurtina and {van der Marel}, Roeland P. and {Zivick}, Paul and {Kallivayalil}, Nitya and {Nidever}, David L.},
        title = "{The Recent LMC-SMC Collision: Timing and Impact Parameter Constraints from Comparison of Gaia LMC Disk Kinematics and N-body Simulations}",
      journal = {\apj},
         year = 2022,
        month = mar,
       volume = {927},
       number = {2},
          eid = {153},
        pages = {153},
          doi = {10.3847/1538-4357/ac4e90},
archivePrefix = {arXiv},
       eprint = {2201.04648},
 primaryClass = {astro-ph.GA},
       adsurl = {https://ui.adsabs.harvard.edu/abs/2022ApJ...927..153C}
}

@ARTICLE{Rathore2025b,
       author = {{Rathore}, Himansh and {Besla}, Gurtina and {Daniel}, Kathryne J. and {Beraldo e Silva}, Leandro},
        title = "{Response of the LMC's Bar to a Recent SMC Collision and Implications for the SMC's Dark Matter Profile}",
      journal = {\apj},
         year = 2025,
        month = jul,
       volume = {988},
       number = {1},
          eid = {79},
        pages = {79},
          doi = {10.3847/1538-4357/ade0ae},
archivePrefix = {arXiv},
       eprint = {2504.16163},
 primaryClass = {astro-ph.GA},
       adsurl = {https://ui.adsabs.harvard.edu/abs/2025ApJ...988...79R}
}

@ARTICLE{Besla2016,
       author = {{Besla}, Gurtina and {Mart{\'\i}nez-Delgado}, David and {van der Marel}, Roeland P. and {Beletsky}, Yuri and {Seibert}, Mark and {Schlafly}, Edward F. and {Grebel}, Eva K. and {Neyer}, Fabian},
        title = "{Low Surface Brightness Imaging of the Magellanic System: Imprints of Tidal Interactions between the Clouds in the Stellar Periphery}",
      journal = {\apj},
         year = 2016,
        month = jul,
       volume = {825},
       number = {1},
          eid = {20},
        pages = {20},
          doi = {10.3847/0004-637X/825/1/20},
archivePrefix = {arXiv},
       eprint = {1602.04222},
 primaryClass = {astro-ph.GA},
       adsurl = {https://ui.adsabs.harvard.edu/abs/2016ApJ...825...20B}
}

@ARTICLE{RuizLara2020,
       author = {{Ruiz-Lara}, T. and {Gallart}, C. and {Monelli}, M. and {Nidever}, D. and {Dorta}, A. and {Choi}, Y. and {Olsen}, K. and {Besla}, G. and {Bernard}, E.~J. and {Cassisi}, S. and {Massana}, P. and {No{\"e}l}, N.~E.~D. and {P{\'e}rez}, I. and {Rusakov}, V. and {Cioni}, M.-R.~L. and {Majewski}, S.~R. and {van der Marel}, R.~P. and {Mart{\'\i}nez-Delgado}, D. and {Monachesi}, A. and {Monteagudo}, L. and {Mu{\~n}oz}, R.~R. and {Stringfellow}, G.~S. and {Surot}, F. and {Vivas}, A.~K. and {Walker}, A.~R. and {Zaritsky}, D.},
        title = "{The Large Magellanic Cloud stellar content with SMASH. I. Assessing the stability of the Magellanic spiral arms}",
      journal = {\aap},
         year = 2020,
        month = jul,
       volume = {639},
          eid = {L3},
        pages = {L3},
          doi = {10.1051/0004-6361/202038392},
archivePrefix = {arXiv},
       eprint = {2006.10759},
 primaryClass = {astro-ph.GA},
       adsurl = {https://ui.adsabs.harvard.edu/abs/2020A&A...639L...3R}
}

@ARTICLE{Cerny2026,
       author = {{Cerny}, William and {Li}, Ting S. and {Pace}, Andrew B. and {Simon}, Joshua D. and {Geha}, Marla and {Ji}, Alexander P. and {Drlica-Wagner}, Alex and {Bruce}, Jordan and {Gnedin}, Oleg Y. and {Bell}, Eric F. and {Mau}, Sidney and {Escala}, Ivanna and {Bissonette}, Daisy and {Savino}, Alessandro and {Chiti}, Anirudh and {Kirby}, Evan N.},
        title = "{A Chemodynamical Census of the Milky Way's Ultra-Faint Compact Satellites. I. A First Population-Level Look at the Internal Kinematics and Metallicities of 19 Extremely-Low-Mass Halo Stellar Systems}",
      journal = {arXiv e-prints},
         year = 2026,
        month = feb,
          eid = {arXiv:2602.17652},
        pages = {arXiv:2602.17652},
          doi = {10.48550/arXiv.2602.17652},
archivePrefix = {arXiv},
       eprint = {2602.17652},
 primaryClass = {astro-ph.GA},
       adsurl = {https://ui.adsabs.harvard.edu/abs/2026arXiv260217652C},
      note = {submitted to ApJ}
}

@ARTICLE{Arroyo-Polonio2026,
       author = {{Arroyo-Polonio}, Jos{\'e} Mar{\'\i}a and {Battaglia}, Giuseppina and {Thomas}, Guillaume F.},
        title = "{Estimating the dynamical masses of dwarf galaxies in the presence of binary-star contamination}",
      journal = {\aap},
         year = 2026,
        month = apr,
       volume = {708},
          eid = {A287},
        pages = {A287},
          doi = {10.1051/0004-6361/202558720},
archivePrefix = {arXiv},
       eprint = {2603.03129},
 primaryClass = {astro-ph.GA},
       adsurl = {https://ui.adsabs.harvard.edu/abs/2026A&A...708A.287A}
}

@ARTICLE{Kumar2025,
       author = {{Kumar}, Prem and {Pawlowski}, Marcel S. and {Kanehisa}, Kosuke Jamie and {Li}, Pengfei and {J{\'u}lio}, Mariana P. and {Taibi}, Salvatore},
        title = "{The effect of measurement uncertainties on the inferred stability of planes of satellite galaxies}",
      journal = {\aap},
         year = 2025,
        month = jul,
       volume = {699},
          eid = {A363},
        pages = {A363},
          doi = {10.1051/0004-6361/202453087},
archivePrefix = {arXiv},
       eprint = {2506.01459},
 primaryClass = {astro-ph.GA},
       adsurl = {https://ui.adsabs.harvard.edu/abs/2025A&A...699A.363K}
}

@ARTICLE{delPino2021,
       author = {{del Pino}, Andr{\'{e}}s and {Fardal}, Mark A. and {van der Marel}, Roeland P. and {{\L}okas}, Ewa L. and {Mateu}, Cecilia and {Sohn}, Sangmo Tony},
        title = "{Revealing the Structure and Internal Rotation of the Sagittarius Dwarf Spheroidal Galaxy with Gaia and Machine Learning}",
      journal = {\apj},
         year = 2021,
        month = feb,
       volume = {908},
       number = {2},
          eid = {244},
        pages = {244},
          doi = {10.3847/1538-4357/abd5bf},
archivePrefix = {arXiv},
       eprint = {2011.02627},
 primaryClass = {astro-ph.GA},
       adsurl = {https://ui.adsabs.harvard.edu/abs/2021ApJ...908..244D}
}

@ARTICLE{NFW1996,
       author = {{Navarro}, Julio F. and {Frenk}, Carlos S. and {White}, Simon D.~M.},
        title = "{The Structure of Cold Dark Matter Halos}",
      journal = {\apj},
         year = 1996,
        month = may,
       volume = {462},
        pages = {563},
          doi = {10.1086/177173},
archivePrefix = {arXiv},
       eprint = {astro-ph/9508025},
 primaryClass = {astro-ph},
       adsurl = {https://ui.adsabs.harvard.edu/abs/1996ApJ...462..563N}
}

@ARTICLE{Kallivayalil2013,
       author = {{Kallivayalil}, Nitya and {van der Marel}, Roeland P. and {Besla}, Gurtina and {Anderson}, Jay and {Alcock}, Charles},
        title = "{Third-epoch Magellanic Cloud Proper Motions. I. Hubble Space Telescope/WFC3 Data and Orbit Implications}",
      journal = {\apj},
         year = 2013,
        month = feb,
       volume = {764},
       number = {2},
          eid = {161},
        pages = {161},
          doi = {10.1088/0004-637X/764/2/161},
archivePrefix = {arXiv},
       eprint = {1301.0832},
 primaryClass = {astro-ph.CO},
       adsurl = {https://ui.adsabs.harvard.edu/abs/2013ApJ...764..161K}
}

@ARTICLE{Penarrubia2016,
       author = {{Pe{\~n}arrubia}, Jorge and {G{\'o}mez}, Facundo A. and {Besla}, Gurtina and {Erkal}, Denis and {Ma}, Yin-Zhe},
        title = "{A timing constraint on the (total) mass of the Large Magellanic Cloud}",
      journal = {\mnras},
         year = 2016,
        month = feb,
       volume = {456},
       number = {1},
        pages = {L54-L58},
          doi = {10.1093/mnrasl/slv160},
archivePrefix = {arXiv},
       eprint = {1507.03594},
 primaryClass = {astro-ph.GA},
       adsurl = {https://ui.adsabs.harvard.edu/abs/2016MNRAS.456L..54P}
}

@ARTICLE{Laporte2018,
       author = {{Laporte}, Chervin F.~P. and {G{\'o}mez}, Facundo A. and {Besla}, Gurtina and {Johnston}, Kathryn V. and {Garavito-Camargo}, Nicolas},
        title = "{Response of the Milky Way's disc to the Large Magellanic Cloud in a first infall scenario}",
      journal = {\mnras},
         year = 2018,
        month = jan,
       volume = {473},
       number = {1},
        pages = {1218-1230},
          doi = {10.1093/mnras/stx2146},
archivePrefix = {arXiv},
       eprint = {1608.04743},
 primaryClass = {astro-ph.GA},
       adsurl = {https://ui.adsabs.harvard.edu/abs/2018MNRAS.473.1218L}
}

@ARTICLE{Erkal2021,
       author = {{Erkal}, Denis and {Deason}, Alis J. and {Belokurov}, Vasily and {Xue}, Xiang-Xiang and {Koposov}, Sergey E. and {Bird}, Sarah A. and {Liu}, Chao and {Simion}, Iulia T. and {Yang}, Chengqun and {Zhang}, Lan and {Zhao}, Gang},
        title = "{Detection of the LMC-induced sloshing of the Galactic halo}",
      journal = {\mnras},
         year = 2021,
        month = sep,
       volume = {506},
       number = {2},
        pages = {2677-2684},
          doi = {10.1093/mnras/stab1828},
archivePrefix = {arXiv},
       eprint = {2010.13789},
 primaryClass = {astro-ph.GA},
       adsurl = {https://ui.adsabs.harvard.edu/abs/2021MNRAS.506.2677E}
}

@ARTICLE{Shipp2021,
       author = {{Shipp}, Nora and {Erkal}, Denis and {Drlica-Wagner}, Alex and {Li}, Ting S. and {Pace}, Andrew B. and {Koposov}, Sergey E. and {Cullinane}, Lara R. and {Da Costa}, Gary S. and {Ji}, Alexander P. and {Kuehn}, Kyler and {Lewis}, Geraint F. and {Mackey}, Dougal and {Simpson}, Jeffrey D. and {Wan}, Zhen and {Zucker}, Daniel B. and {Bland-Hawthorn}, Joss and {Ferguson}, Peter S. and {Lilleengen}, Sophia and {Lilleengen}, Sophia},
        title = "{Measuring the Mass of the Large Magellanic Cloud with Stellar Streams Observed by S $^{5}$}",
      journal = {\apj},
         year = 2021,
        month = dec,
       volume = {923},
       number = {2},
          eid = {149},
        pages = {149},
          doi = {10.3847/1538-4357/ac2e93},
archivePrefix = {arXiv},
       eprint = {2107.13004},
 primaryClass = {astro-ph.GA},
       adsurl = {https://ui.adsabs.harvard.edu/abs/2021ApJ...923..149S}
}

@ARTICLE{Sawala2023,
       author = {{Sawala}, Till and {Teeriaho}, Meri and {Johansson}, Peter H.},
        title = "{The Local Group's mass: probably no more than the sum of its parts}",
      journal = {\mnras},
         year = 2023,
        month = jun,
       volume = {521},
       number = {4},
        pages = {4863-4877},
          doi = {10.1093/mnras/stad883},
archivePrefix = {arXiv},
       eprint = {2210.07250},
 primaryClass = {astro-ph.GA},
       adsurl = {https://ui.adsabs.harvard.edu/abs/2023MNRAS.521.4863S}
}

@ARTICLE{DOnghia2008,
       author = {{D'Onghia}, Elena and {Lake}, George},
        title = "{Small Dwarf Galaxies within Larger Dwarfs: Why Some Are Luminous while Most Go Dark}",
      journal = {\apjl},
         year = 2008,
        month = oct,
       volume = {686},
       number = {2},
        pages = {L61},
          doi = {10.1086/592995},
archivePrefix = {arXiv},
       eprint = {0802.0001},
 primaryClass = {astro-ph},
       adsurl = {https://ui.adsabs.harvard.edu/abs/2008ApJ...686L..61D}
}

@ARTICLE{Nichols2011,
       author = {{Nichols}, Matthew and {Colless}, James and {Colless}, Matthew and {Bland-Hawthorn}, Joss},
        title = "{Accretion of the Magellanic System onto the Galaxy}",
      journal = {\apj},
         year = 2011,
        month = dec,
       volume = {742},
       number = {2},
          eid = {110},
        pages = {110},
          doi = {10.1088/0004-637X/742/2/110},
archivePrefix = {arXiv},
       eprint = {1110.2784},
 primaryClass = {astro-ph.CO},
       adsurl = {https://ui.adsabs.harvard.edu/abs/2011ApJ...742..110N}
}

@ARTICLE{SantosSantos2021,
       author = {{Santos-Santos}, Isabel M.~E. and {Fattahi}, Azadeh and {Sales}, Laura V. and {Navarro}, Julio F.},
        title = "{Magellanic satellites in {\ensuremath{\Lambda}}CDM cosmological hydrodynamical simulations of the Local Group}",
      journal = {\mnras},
         year = 2021,
        month = jul,
       volume = {504},
       number = {3},
        pages = {4551-4567},
          doi = {10.1093/mnras/stab1020},
archivePrefix = {arXiv},
       eprint = {2011.13500},
 primaryClass = {astro-ph.GA},
       adsurl = {https://ui.adsabs.harvard.edu/abs/2021MNRAS.504.4551S}
}

@ARTICLE{Kallivayalil2018,
       author = {{Kallivayalil}, Nitya and {Sales}, Laura V. and {Zivick}, Paul and {Fritz}, Tobias K. and {Del Pino}, Andr{\'e}s and {Sohn}, Sangmo Tony and {Besla}, Gurtina and {van der Marel}, Roeland P. and {Navarro}, Julio F. and {Sacchi}, Elena},
        title = "{The Missing Satellites of the Magellanic Clouds? Gaia Proper Motions of the Recently Discovered Ultra-faint Galaxies}",
      journal = {\apj},
         year = 2018,
        month = nov,
       volume = {867},
       number = {1},
          eid = {19},
        pages = {19},
          doi = {10.3847/1538-4357/aadfee},
archivePrefix = {arXiv},
       eprint = {1805.01448},
 primaryClass = {astro-ph.GA},
       adsurl = {https://ui.adsabs.harvard.edu/abs/2018ApJ...867...19K}
}

@ARTICLE{Pardy2020,
       author = {{Pardy}, Stephen A. and {D'Onghia}, Elena and {Navarro}, Julio F. and {Grand}, Robert and {G{\'o}mez}, Facundo A. and {Marinacci}, Federico and {Pakmor}, R{\"u}diger and {Simpson}, Christine and {Springel}, Volker},
        title = "{Satellites of Satellites: The Case for Carina and Fornax}",
      journal = {\mnras},
         year = 2020,
        month = feb,
       volume = {492},
       number = {2},
        pages = {1543-1549},
          doi = {10.1093/mnras/stz3192},
archivePrefix = {arXiv},
       eprint = {1904.01028},
 primaryClass = {astro-ph.GA},
       adsurl = {https://ui.adsabs.harvard.edu/abs/2020MNRAS.492.1543P}
}

@ARTICLE{vanderMarel2001,
       author = {{van der Marel}, Roeland P.},
        title = "{Magellanic Cloud Structure from Near-Infrared Surveys. II. Star Count Maps and the Intrinsic Elongation of the Large Magellanic Cloud}",
      journal = {\aj},
         year = 2001,
        month = oct,
       volume = {122},
       number = {4},
        pages = {1827-1843},
          doi = {10.1086/323100},
archivePrefix = {arXiv},
       eprint = {astro-ph/0105340},
 primaryClass = {astro-ph},
       adsurl = {https://ui.adsabs.harvard.edu/abs/2001AJ....122.1827V}
}

@ARTICLE{Choi2018,
       author = {{Choi}, Yumi and {Nidever}, David L. and {Olsen}, Knut and {Blum}, Robert D. and {Besla}, Gurtina and {Zaritsky}, Dennis and {van der Marel}, Roeland P. and {Bell}, Eric F. and {Gallart}, Carme and {Cioni}, Maria-Rosa L. and {Johnson}, L. Clifton and {Vivas}, A. Katherina and {Saha}, Abhijit and {de Boer}, Thomas J.~L. and {No{\"e}l}, Noelia E.~D. and {Monachesi}, Antonela and {Massana}, Pol and {Conn}, Blair C. and {Martinez-Delgado}, David and {Mu{\~n}oz}, Ricardo R. and {Stringfellow}, Guy S.},
        title = "{SMASHing the LMC: A Tidally Induced Warp in the Outer LMC and a Large-scale Reddening Map}",
      journal = {\apj},
         year = 2018,
        month = oct,
       volume = {866},
       number = {2},
          eid = {90},
        pages = {90},
          doi = {10.3847/1538-4357/aae083},
archivePrefix = {arXiv},
       eprint = {1804.07765},
 primaryClass = {astro-ph.GA},
       adsurl = {https://ui.adsabs.harvard.edu/abs/2018ApJ...866...90C}
}

@ARTICLE{ZhaoEvans2000,
       author = {{Zhao}, HongSheng and {Evans}, N. Wyn},
        title = "{The So-called ``Bar'' in the Large Magellanic Cloud}",
      journal = {\apjl},
         year = 2000,
        month = dec,
       volume = {545},
       number = {1},
        pages = {L35-L38},
          doi = {10.1086/317324},
archivePrefix = {arXiv},
       eprint = {astro-ph/0009155},
 primaryClass = {astro-ph},
       adsurl = {https://ui.adsabs.harvard.edu/abs/2000ApJ...545L..35Z}
}

@ARTICLE{vanderMarelCioni2001,
       author = {{van der Marel}, Roeland P. and {Cioni}, Maria-Rosa L.},
        title = "{Magellanic Cloud Structure from Near-Infrared Surveys. I. The Viewing Angles of the Large Magellanic Cloud}",
      journal = {\aj},
         year = 2001,
        month = oct,
       volume = {122},
       number = {4},
        pages = {1807-1826},
          doi = {10.1086/323099},
archivePrefix = {arXiv},
       eprint = {astro-ph/0105339},
 primaryClass = {astro-ph},
       adsurl = {https://ui.adsabs.harvard.edu/abs/2001AJ....122.1807V}
}

@ARTICLE{OlsenSalyk2002,
       author = {{Olsen}, K.~A.~G. and {Salyk}, C.},
        title = "{A Warp in the Large Magellanic Cloud Disk?}",
      journal = {\aj},
         year = 2002,
        month = oct,
       volume = {124},
       number = {4},
        pages = {2045-2053},
          doi = {10.1086/342739},
       adsurl = {https://ui.adsabs.harvard.edu/abs/2002AJ....124.2045O}
}

@ARTICLE{Chandrasekhar1943,
       author = {{Chandrasekhar}, S.},
        title = "{Dynamical Friction. I. General Considerations: the Coefficient of Dynamical Friction.}",
      journal = {\apj},
         year = 1943,
        month = mar,
       volume = {97},
        pages = {255},
          doi = {10.1086/144517},
       adsurl = {https://ui.adsabs.harvard.edu/abs/1943ApJ....97..255C}
}

@ARTICLE{Hashimoto2003,
       author = {{Hashimoto}, Yoshikazu and {Funato}, Yoko and {Makino}, Junichiro},
        title = "{To Circularize or Not To Circularize?-Orbital Evolution of Satellite Galaxies}",
      journal = {\apj},
         year = 2003,
        month = jan,
       volume = {582},
       number = {1},
        pages = {196-201},
          doi = {10.1086/344260},
archivePrefix = {arXiv},
       eprint = {astro-ph/0208452},
 primaryClass = {astro-ph},
       adsurl = {https://ui.adsabs.harvard.edu/abs/2003ApJ...582..196H}
}

@ARTICLE{ZentnerBullock2003,
       author = {{Zentner}, Andrew R. and {Bullock}, James S.},
        title = "{Halo Substructure and the Power Spectrum}",
      journal = {\apj},
         year = 2003,
        month = nov,
       volume = {598},
       number = {1},
        pages = {49-72},
          doi = {10.1086/378797},
archivePrefix = {arXiv},
       eprint = {astro-ph/0304292},
 primaryClass = {astro-ph},
       adsurl = {https://ui.adsabs.harvard.edu/abs/2003ApJ...598...49Z}
}

@BOOK{BinneyTremaine1987,
       author = {{Binney}, James and {Tremaine}, Scott},
        title = "{Galactic dynamics}",
         year = 1987,
       adsurl = {https://ui.adsabs.harvard.edu/abs/1987gady.book.....B}
}

@ARTICLE{BekkiChiba2005,
       author = {{Bekki}, Kenji and {Chiba}, Masashi},
        title = "{Formation and evolution of the Magellanic Clouds - I. Origin of structural, kinematic and chemical properties of the Large Magellanic Cloud}",
      journal = {\mnras},
         year = 2005,
        month = jan,
       volume = {356},
       number = {2},
        pages = {680-702},
          doi = {10.1111/j.1365-2966.2004.08510.x},
archivePrefix = {arXiv},
       eprint = {astro-ph/0412318},
 primaryClass = {astro-ph},
       adsurl = {https://ui.adsabs.harvard.edu/abs/2005MNRAS.356..680B}
}

@ARTICLE{Besla2007,
       author = {{Besla}, Gurtina and {Kallivayalil}, Nitya and {Hernquist}, Lars and {Robertson}, Brant and {Cox}, T.~J. and {van der Marel}, Roeland P. and {Alcock}, Charles},
        title = "{Are the Magellanic Clouds on Their First Passage about the Milky Way?}",
      journal = {\apj},
         year = 2007,
        month = oct,
       volume = {668},
       number = {2},
        pages = {949-967},
          doi = {10.1086/521385},
archivePrefix = {arXiv},
       eprint = {astro-ph/0703196},
 primaryClass = {astro-ph},
       adsurl = {https://ui.adsabs.harvard.edu/abs/2007ApJ...668..949B}
}

@ARTICLE{Hernquist1990,
       author = {{Hernquist}, Lars},
        title = "{An Analytical Model for Spherical Galaxies and Bulges}",
      journal = {\apj},
         year = 1990,
        month = jun,
       volume = {356},
        pages = {359},
          doi = {10.1086/168845},
       adsurl = {https://ui.adsabs.harvard.edu/abs/1990ApJ...356..359H}
}

@ARTICLE{Plummer1911,
       author = {{Plummer}, H.~C.},
        title = "{On the problem of distribution in globular star clusters}",
      journal = {\mnras},
         year = 1911,
        month = mar,
       volume = {71},
        pages = {460-470},
          doi = {10.1093/mnras/71.5.460},
       adsurl = {https://ui.adsabs.harvard.edu/abs/1911MNRAS..71..460P}
}

@ARTICLE{Wolf2010,
       author = {{Wolf}, Joe and {Martinez}, Gregory D. and {Bullock}, James S. and {Kaplinghat}, Manoj and {Geha}, Marla and {Mu{\~n}oz}, Ricardo R. and {Simon}, Joshua D. and {Avedo}, Frank F.},
        title = "{Accurate masses for dispersion-supported galaxies}",
      journal = {\mnras},
         year = 2010,
        month = aug,
       volume = {406},
       number = {2},
        pages = {1220-1237},
          doi = {10.1111/j.1365-2966.2010.16753.x},
archivePrefix = {arXiv},
       eprint = {0908.2995},
 primaryClass = {astro-ph.CO},
       adsurl = {https://ui.adsabs.harvard.edu/abs/2010MNRAS.406.1220W}
}

@ARTICLE{BryanNorman1998,
       author = {{Bryan}, Greg L. and {Norman}, Michael L.},
        title = "{Statistical Properties of X-Ray Clusters: Analytic and Numerical Comparisons}",
      journal = {\apj},
         year = 1998,
        month = mar,
       volume = {495},
       number = {1},
        pages = {80-99},
          doi = {10.1086/305262},
archivePrefix = {arXiv},
       eprint = {astro-ph/9710107},
 primaryClass = {astro-ph},
       adsurl = {https://ui.adsabs.harvard.edu/abs/1998ApJ...495...80B}
}

@ARTICLE{Bayer2025,
       author = {{Bayer}, M. and {Starkenburg}, E. and {Thomas}, G.~F. and {Martin}, N.~F. and {Helmi}, A. and {Bystr{\"o}m}, A. and {de Boer}, T. and {Fern{\'a}ndez Alvar}, E. and {Gwyn}, S. and {Ibata}, R. and {Jablonka}, P. and {Kordopatis}, G. and {Matsuno}, T. and {McConnachie}, A.~W. and {Medina}, G.~E. and {Rusterucci}, S. and {S{\'a}nchez-Janssen}, R. and {Sestito}, F. and {Viswanathan}, A.},
        title = "{A Pristine-UNIONS view on the Galaxy: Kinematics of the distant spur feature of the Sagittarius stream traced by blue horizontal branch stars}",
      journal = {\aap},
         year = 2025,
        month = sep,
       volume = {701},
          eid = {A117},
        pages = {A117},
          doi = {10.1051/0004-6361/202554131},
archivePrefix = {arXiv},
       eprint = {2502.17319},
 primaryClass = {astro-ph.GA},
       adsurl = {https://ui.adsabs.harvard.edu/abs/2025A&A...701A.117B}
}

@ARTICLE{BobylevBaykova2023,
       author = {{Bobylev}, V.~V. and {Baykova}, A.~T.},
        title = "{Modern Estimates of the Mass of the Milky Way}",
      journal = {Astronomy Reports},
         year = 2023,
        month = aug,
       volume = {67},
       number = {8},
        pages = {812-823},
          doi = {10.1134/S1063772923080024},
       adsurl = {https://ui.adsabs.harvard.edu/abs/2023ARep...67..812B}
}

@ARTICLE{Libralato2024b,
       author = {{Libralato}, Mattia and {Argyriou}, Ioannis and {Dicken}, Dan and {Garc{\'\i}a Mar{\'\i}n}, Macarena and {Guillard}, Pierre and {Hines}, Dean C. and {Kavanagh}, Patrick J. and {Kendrew}, Sarah and {Law}, David R. and {Noriega-Crespo}, Alberto and {{\'A}lvarez-M{\'a}rquez}, Javier},
        title = "{High-precision Astrometry and Photometry with the JWST/MIRI Imager}",
      journal = {\pasp},
         year = 2024,
        month = mar,
       volume = {136},
       number = {3},
          eid = {034502},
        pages = {034502},
          doi = {10.1088/1538-3873/ad2551},
archivePrefix = {arXiv},
       eprint = {2311.12145},
 primaryClass = {astro-ph.IM},
       adsurl = {https://ui.adsabs.harvard.edu/abs/2024PASP..136c4502L}
}

@ARTICLE{Libralato2023,
       author = {{Libralato}, Mattia and {Bellini}, Andrea and {van der Marel}, Roeland P. and {Anderson}, Jay and {Sohn}, Sangmo Tony and {Watkins}, Laura L. and {Alderson}, Lili and {Allen}, Natalie and {Clampin}, Mark and {Glidden}, Ana and {Goyal}, Jayesh and {Hoch}, Kielan and {Huang}, Jingcheng and {Kammerer}, Jens and {Lewis}, Nikole K. and {Lin}, Zifan and {Long}, Douglas and {Louie}, Dana and {MacDonald}, Ryan J. and {Mountain}, Matt and {Pe{\~n}a-Guerrero}, Maria and {Perrin}, Marshall D. and {Pueyo}, Laurent and {Rebollido}, Isabel and {Rickman}, Emily and {Seager}, Sara and {Stevenson}, Kevin B. and {Valenti}, Jeff A. and {Valentine}, Daniel and {Wakeford}, Hannah R.},
        title = "{JWST-TST Proper Motions. I. High-precision NIRISS Calibration and Large Magellanic Cloud Kinematics}",
      journal = {\apj},
         year = 2023,
        month = jun,
       volume = {950},
       number = {2},
          eid = {101},
        pages = {101},
          doi = {10.3847/1538-4357/acd04f},
archivePrefix = {arXiv},
       eprint = {2303.00009},
 primaryClass = {astro-ph.GA},
       adsurl = {https://ui.adsabs.harvard.edu/abs/2023ApJ...950..101L}
}

@article{McKinnon2026,
doi = {10.1088/1538-3873/ae5a73},
url = {https://doi.org/10.1088/1538-3873/ae5a73},
year = {2026},
month = {apr},
publisher = {The American Astronomical Society},
volume = {138},
number = {4},
pages = {044507},
author = {McKinnon, Kevin A. and van der Marel, Roeland P.},
title = {Simulating Roman+Gaia Combined Astrometry, Parallaxes, and Proper Motions},
journal = {Publications of the Astronomical Society of the Pacific}
}

@ARTICLE{vanderMarelKallivayalil2014,
       author = {{van der Marel}, Roeland P. and {Kallivayalil}, Nitya},
        title = "{Third-epoch Magellanic Cloud Proper Motions. II. The Large Magellanic Cloud Rotation Field in Three Dimensions}",
      journal = {\apj},
         year = 2014,
        month = feb,
       volume = {781},
       number = {2},
          eid = {121},
        pages = {121},
          doi = {10.1088/0004-637X/781/2/121},
archivePrefix = {arXiv},
       eprint = {1305.4641},
 primaryClass = {astro-ph.CO},
       adsurl = {https://ui.adsabs.harvard.edu/abs/2014ApJ...781..121V}
}

@ARTICLE{Warfield2026,
       author = {{Warfield}, Jack T. and {McKinnon}, Kevin A. and {Sohn}, Sangmo Tony and {Kallivayalil}, Nitya and {Savino}, Alessandro and {van der Marel}, Roeland P. and {Pace}, Andrew B. and {Garling}, Christopher T. and {Ahvazi}, Niusha and {Bennet}, Paul and {Cohen}, Roger E. and {Correnti}, Matteo and {Fardal}, Mark A. and {McQuinn}, Kristen. B.~W. and {Newman}, Max J.~B. and {Vitral}, Eduardo},
        title = "{The Proper Motion of Draco II with HST Using Multiple Reference Frames and Methodologies}",
      journal = {\apj},
         year = 2026,
        month = feb,
       volume = {998},
       number = {1},
          eid = {3},
        pages = {3},
          doi = {10.3847/1538-4357/ae29f2},
archivePrefix = {arXiv},
       eprint = {2510.24849},
 primaryClass = {astro-ph.GA},
       adsurl = {https://ui.adsabs.harvard.edu/abs/2026ApJ...998....3W}
}

@ARTICLE{Ahvazi2025,
       author = {{Ahvazi}, Niusha and {Pace}, Andrew B. and {Garling}, Christopher T. and {Ou}, Xiaowei and {Kallivayalil}, Nitya and {Torrey}, Paul and {Benson}, Andrew and {Bhowmick}, Aklant and {Torres-Alb{\`a}}, N{\'u}ria and {Garcia}, Alex M. and {Saravia}, Alejandro and {Kho}, Jonathan and {Warfield}, Jack T. and {Atzberger}, Kaia R.},
        title = "{The abundance and properties of the lowest luminosity dwarf galaxies around the Milky Way: Insights from Semi-Analytic Models}",
      journal = {arXiv e-prints},
         year = 2025,
        month = nov,
          eid = {arXiv:2511.15808},
        pages = {arXiv:2511.15808},
          doi = {10.48550/arXiv.2511.15808},
archivePrefix = {arXiv},
       eprint = {2511.15808},
 primaryClass = {astro-ph.GA},
       adsurl = {https://ui.adsabs.harvard.edu/abs/2025arXiv251115808A}
}

@ARTICLE{Pace2025,
       author = {{Pace}, Andrew B. and {Li}, T.~S. and {Ji}, A.~P. and {Simon}, J.~D. and {Cerny}, W. and {Senkevich}, A.~M. and {Drlica-Wagner}, A. and {Bechtol}, K. and {Tan}, C.~Y. and {Chiti}, A. and {Erkal}, D. and {Mart{\'\i}nez-V{\'a}zquez}, C.~E. and {Ferguson}, P.~S. and {Kron}, R.~G. and {Atzberger}, K.~R. and {Chaturvedi}, A. and {Frieman}, J.~A. and {Kallivayalil}, N. and {Limberg}, G. and {Medina}, G.~E. and {Placco}, V.~M. and {Riley}, A.~H. and {Sand}, D.~J. and {Stringfellow}, G.~S. and {van der Marel}, R.~P. and {Carballo-Bello}, J.~A. and {Choi}, Y. and {Crnojevi{\'c}}, D. and {Massana}, P. and {Mutlu-Pakdil}, B. and {Navabi}, M. and {No{\"e}l}, N.~E.~D. and {Sakowska}, J.~D.},
        title = "{Spectroscopic Analysis of Pictor II: a very low metallicity ultra-faint dwarf galaxy bound to the Large Magellanic Cloud}",
      journal = {The Open Journal of Astrophysics},
         year = 2025,
        month = aug,
       volume = {8},
          eid = {112},
        pages = {112},
          doi = {10.33232/001c.142989},
archivePrefix = {arXiv},
       eprint = {2506.21841},
 primaryClass = {astro-ph.GA},
       adsurl = {https://ui.adsabs.harvard.edu/abs/2025OJAp....8E.112P}
}

@article{Heiger2024,
doi = {10.3847/1538-4357/ad0cf7},
url = {https://doi.org/10.3847/1538-4357/ad0cf7},
year = {2024},
month = {jan},
publisher = {The American Astronomical Society},
volume = {961},
number = {2},
pages = {234},
author = {Heiger, M. E. and Li, T. S. and Pace, A. B. and Simon, J. D. and Ji, A. P. and Chiti, A. and Bom, C. R. and Carballo-Bello, J. A. and Carlin, J. L. and Cerny, W. and Choi, Y. and Drlica-Wagner, A. and James, D. J. and Martínez-Vázquez, C. E. and Medina, G. E. and Mutlu-Pakdil, B. and Navabi, M. and Noël, N. E. D. and Sakowska, J. D. and Stringfellow, G. S. and (DELVE Collaboration)},
title = {Reading between the (Spectral) Lines: Magellan/IMACS Spectroscopy of the Ultrafaint Dwarf Galaxies Eridanus IV and Centaurus I},
journal = {\apj}
}

@ARTICLE{Zivick2019,
       author = {{Zivick}, Paul and {Kallivayalil}, Nitya and {Besla}, Gurtina and {Sohn}, Sangmo Tony and {van der Marel}, Roeland P. and {del Pino}, Andr{\'e}s and {Linden}, Sean T. and {Fritz}, Tobias K. and {Anderson}, J.},
        title = "{The Proper-motion Field along the Magellanic Bridge: A New Probe of the LMC-SMC Interaction}",
      journal = {\apj},
         year = 2019,
        month = mar,
       volume = {874},
       number = {1},
          eid = {78},
        pages = {78},
          doi = {10.3847/1538-4357/ab0554},
archivePrefix = {arXiv},
       eprint = {1811.09318},
 primaryClass = {astro-ph.GA},
       adsurl = {https://ui.adsabs.harvard.edu/abs/2019ApJ...874...78Z}
}

@ARTICLE{Newton2023,
       author = {{Newton}, Oliver and {Di Cintio}, Arianna and {Cardona-Barrero}, Salvador and {Libeskind}, Noam I. and {Hoffman}, Yehuda and {Knebe}, Alexander and {Sorce}, Jenny G. and {Steinmetz}, Matthias and {Tempel}, Elmo},
        title = "{The Undiscovered Ultradiffuse Galaxies of the Local Group}",
      journal = {\apjl},
         year = 2023,
        month = apr,
       volume = {946},
       number = {2},
          eid = {L37},
        pages = {L37},
          doi = {10.3847/2041-8213/acc2bb},
archivePrefix = {arXiv},
       eprint = {2212.05066},
 primaryClass = {astro-ph.GA},
       adsurl = {https://ui.adsabs.harvard.edu/abs/2023ApJ...946L..37N}
}

@article{Smith2023,
doi = {10.3847/1538-3881/acdd77},
url = {https://doi.org/10.3847/1538-3881/acdd77},
year = {2023},
month = {jul},
publisher = {The American Astronomical Society},
volume = {166},
number = {2},
pages = {76},
author = {Smith, Simon E. T. and Jensen, Jaclyn and Roediger, Joel and Sestito, Federico and Hayes, Christian R. and McConnachie, Alan W. and Cuillandre, Jean-Charles and Gwyn, Stephen and Magnier, Eugene and Chambers, Ken and Hammer, Francois and Hudson, Mike J. and Martin, Nicolas and Navarro, Julio and Scott, Douglas},
title = {Discovery of a New Local Group Dwarf Galaxy Candidate in UNIONS: Boötes V},
journal = {\aj}
}

@ARTICLE{Casey2025,
       author = {{Casey}, Quinn O. and {Mutlu-Pakdil}, Bur{\c{c}}in and {Sand}, David J. and {Pace}, Andrew B. and {Crnojevi{\'c}}, Denija and {Doliva-Dolinsky}, Amandine and {Cerny}, William and {Heiger}, Mairead E. and {Riley}, Alex H. and {Ji}, Alexander P. and {Limberg}, Guilherme and {Marin}, Laurella and {Mart{\'\i}nez-V{\'a}zquez}, Clara E. and {Medina}, Gustavo E. and {Li}, Ting S. and {Campana}, Sasha N. and {Chaturvedi}, Astha and {Sakowska}, Joanna D. and {Zenteno}, Alfredo and {Carballo-Bello}, Julio A. and {Navabi}, Mahdieh and {Bom}, Clecio R. and {DELVE Collaboration}},
        title = "{Deep Photometric Observations of Ultrafaint Milky Way Satellites Centaurus I and Eridanus IV}",
      journal = {\apj},
         year = 2025,
        month = may,
       volume = {984},
       number = {2},
          eid = {148},
        pages = {148},
          doi = {10.3847/1538-4357/adc67e},
archivePrefix = {arXiv},
       eprint = {2501.04772},
 primaryClass = {astro-ph.GA},
       adsurl = {https://ui.adsabs.harvard.edu/abs/2025ApJ...984..148C}
}

@ARTICLE{MartinezVazquez2021,
       author = {{Mart{\'\i}nez-V{\'a}zquez}, C.~E. and {Cerny}, W. and {Vivas}, A.~K. and {Drlica-Wagner}, A. and {Pace}, A.~B. and {Simon}, J.~D. and {Munoz}, R.~R. and {Walker}, A.~R. and {Allam}, S. and {Tucker}, D.~L. and {Adam{\'o}w}, M. and {Carlin}, J.~L. and {Choi}, Y. and {Ferguson}, P.~S. and {Ji}, A.~P. and {Kuropatkin}, N. and {Li}, T.~S. and {Mart{\'\i}nez-Delgado}, D. and {Mau}, S. and {Mutlu-Pakdil}, B. and {Nidever}, D.~L. and {Riley}, A.~H. and {Sakowska}, J.~D. and {Sand}, D.~J. and {Stringfellow}, G.~S. and {Stringfellow}, G.~S.},
        title = "{RR Lyrae Stars in the Newly Discovered Ultra-faint Dwarf Galaxy Centaurus I}",
      journal = {\aj},
         year = 2021,
        month = dec,
       volume = {162},
       number = {6},
          eid = {253},
        pages = {253},
          doi = {10.3847/1538-3881/ac2368},
archivePrefix = {arXiv},
       eprint = {2107.05688},
 primaryClass = {astro-ph.GA},
       adsurl = {https://ui.adsabs.harvard.edu/abs/2021AJ....162..253M}
}

@ARTICLE{Geha2026,
       author = {{Geha}, Marla and {Pelliccia}, Debora and {Prochaska}, J. Xavier and {Cerny}, William and {Davies}, Frederick B. and {Hennawi}, Joseph and {Holden}, Brad and {Reichwein}, Dusty and {Westfall}, Kyle B.},
        title = "{The Keck/DEIMOS Stellar Archive. I. Uniform Velocities and Metallicities for 78 Milky Way Dwarf Galaxies and Globular Clusters}",
      journal = {\apj},
         year = 2026,
        month = mar,
       volume = {999},
       number = {1},
          eid = {140},
        pages = {140},
          doi = {10.3847/1538-4357/ae290d},
archivePrefix = {arXiv},
       eprint = {2602.10200},
 primaryClass = {astro-ph.GA},
       adsurl = {https://ui.adsabs.harvard.edu/abs/2026ApJ...999..140G}
}

@ARTICLE{An2024,
       author = {{An}, Zhaozhou and {Walker}, Matthew G. and {Pace}, Andrew B.},
        title = "{Offset of M54 from the Sagittarius dwarf spheroidal galaxy}",
      journal = {\mnras},
         year = 2024,
        month = aug,
       volume = {532},
       number = {4},
        pages = {3713-3728},
          doi = {10.1093/mnras/stae1680},
archivePrefix = {arXiv},
       eprint = {2404.16184},
 primaryClass = {astro-ph.GA},
       adsurl = {https://ui.adsabs.harvard.edu/abs/2024MNRAS.532.3713A}
}

@ARTICLE{Cioni2000,
       author = {{Cioni}, M.-R.~L. and {van der Marel}, R.~P. and {Loup}, C. and {Habing}, H.~J.},
        title = "{The tip of the red giant branch and distance of the Magellanic Clouds: results from the DENIS survey}",
      journal = {\aap},
         year = 2000,
        month = jul,
       volume = {359},
        pages = {601-614},
          doi = {10.48550/arXiv.astro-ph/0003223},
archivePrefix = {arXiv},
       eprint = {astro-ph/0003223},
 primaryClass = {astro-ph},
       adsurl = {https://ui.adsabs.harvard.edu/abs/2000A&A...359..601C}
}

@ARTICLE{Harris2006,
       author = {{Harris}, Jason and {Zaritsky}, Dennis},
        title = "{Spectroscopic Survey of Red Giants in the Small Magellanic Cloud. I. Kinematics}",
      journal = {\aj},
         year = 2006,
        month = may,
       volume = {131},
       number = {5},
        pages = {2514-2524},
          doi = {10.1086/500974},
archivePrefix = {arXiv},
       eprint = {astro-ph/0601025},
 primaryClass = {astro-ph},
       adsurl = {https://ui.adsabs.harvard.edu/abs/2006AJ....131.2514H}
}

@ARTICLE{MartinezGarcia2023b,
       author = {{Mart{\'\i}nez-Garc{\'\i}a}, Alberto Manuel and {del Pino}, Andr{\'e}s and {{\L}okas}, Ewa L. and {van der Marel}, Roeland P. and {Aparicio}, Antonio},
        title = "{Internal kinematics of dwarf satellites of MW/M31-like galaxies in TNG50}",
      journal = {\mnras},
         year = 2023,
        month = dec,
       volume = {526},
       number = {3},
        pages = {3589-3600},
          doi = {10.1093/mnras/stad2941},
archivePrefix = {arXiv},
       eprint = {2307.13683},
 primaryClass = {astro-ph.GA},
       adsurl = {https://ui.adsabs.harvard.edu/abs/2023MNRAS.526.3589M}
}

@ARTICLE{MartinezGarcia2023a,
       author = {{Mart{\'\i}nez-Garc{\'\i}a}, Alberto Manuel and {del Pino}, Andr{\'e}s and {Aparicio}, Antonio},
        title = "{Tidally induced velocity gradients in the Milky Way dwarf spheroidal satellites}",
      journal = {\mnras},
         year = 2023,
        month = jan,
       volume = {518},
       number = {2},
        pages = {3083-3094},
          doi = {10.1093/mnras/stac3305},
archivePrefix = {arXiv},
       eprint = {2206.06339},
 primaryClass = {astro-ph.GA},
       adsurl = {https://ui.adsabs.harvard.edu/abs/2023MNRAS.518.3083M}
}

@ARTICLE{Bennet2025,
       author = {{Bennet}, Paul and {Patel}, Ekta and {Sohn}, Sangmo Tony and {del Pino}, Andr{\'e}s and {van der Marel}, Roeland P. and {Fardal}, Mark A. and {Spekkens}, Kristine and {Hunter}, Laura Congreve and {Besla}, Gurtina and {Watkins}, Laura L. and {Weisz}, Daniel R.},
        title = "{The Orbits of Isolated Dwarfs in the Local Group from New 3D Kinematics: Constraints on First Infall, Backsplash, and Quenching Mechanisms}",
      journal = {\apj},
         year = 2025,
        month = nov,
       volume = {993},
       number = {2},
          eid = {228},
        pages = {228},
          doi = {10.3847/1538-4357/ae0733},
archivePrefix = {arXiv},
       eprint = {2509.11299},
 primaryClass = {astro-ph.GA},
       adsurl = {https://ui.adsabs.harvard.edu/abs/2025ApJ...993..228B}
}

@ARTICLE{Sawala2023b,
       author = {{Sawala}, Till and {Cautun}, Marius and {Frenk}, Carlos and {Helly}, John and {Jasche}, Jens and {Jenkins}, Adrian and {Johansson}, Peter H. and {Lavaux}, Guilhem and {McAlpine}, Stuart and {Schaller}, Matthieu},
        title = "{The Milky Way's plane of satellites is consistent with {\ensuremath{\Lambda}}CDM}",
      journal = {Nature Astronomy},
         year = 2023,
        month = apr,
       volume = {7},
        pages = {481-491},
          doi = {10.1038/s41550-022-01856-z},
archivePrefix = {arXiv},
       eprint = {2205.02860},
 primaryClass = {astro-ph.GA},
       adsurl = {https://ui.adsabs.harvard.edu/abs/2023NatAs...7..481S}
}

@ARTICLE{Li2021,
       author = {{Li}, Hefan and {Hammer}, Francois and {Babusiaux}, Carine and {Pawlowski}, Marcel S. and {Yang}, Yanbin and {Arenou}, Frederic and {Du}, Cuihua and {Wang}, Jianling},
        title = "{Gaia EDR3 Proper Motions of Milky Way Dwarfs. I. 3D Motions and Orbits}",
      journal = {\apj},
         year = 2021,
        month = jul,
       volume = {916},
       number = {1},
          eid = {8},
        pages = {8},
          doi = {10.3847/1538-4357/ac0436},
archivePrefix = {arXiv},
       eprint = {2104.03974},
 primaryClass = {astro-ph.GA},
       adsurl = {https://ui.adsabs.harvard.edu/abs/2021ApJ...916....8L}
}

@ARTICLE{Vitral2021,
       author = {{Vitral}, Eduardo},
        title = "{BALRoGO: Bayesian Astrometric Likelihood Recovery of Galactic Objects - Global properties of over one hundred globular clusters with Gaia EDR3}",
      journal = {\mnras},
         year = 2021,
        month = jun,
       volume = {504},
       number = {1},
        pages = {1355-1369},
          doi = {10.1093/mnras/stab947},
archivePrefix = {arXiv},
       eprint = {2102.04841},
 primaryClass = {astro-ph.GA},
       adsurl = {https://ui.adsabs.harvard.edu/abs/2021MNRAS.504.1355V}
}

@ARTICLE{VasilievBelokurov2020,
       author = {{Vasiliev}, Eugene and {Belokurov}, Vasily},
        title = "{The last breath of the Sagittarius dSph}",
      journal = {\mnras},
         year = 2020,
        month = oct,
       volume = {497},
       number = {4},
        pages = {4162-4182},
          doi = {10.1093/mnras/staa2114},
archivePrefix = {arXiv},
       eprint = {2006.02929},
 primaryClass = {astro-ph.GA},
       adsurl = {https://ui.adsabs.harvard.edu/abs/2020MNRAS.497.4162V}
}

@ARTICLE{Vasiliev2018,
       author = {{Vasiliev}, Eugene},
        title = "{Internal dynamics of the Large Magellanic Cloud from Gaia DR2}",
      journal = {\mnras},
         year = 2018,
        month = nov,
       volume = {481},
       number = {1},
        pages = {L100-L104},
          doi = {10.1093/mnrasl/sly168},
archivePrefix = {arXiv},
       eprint = {1805.08157},
 primaryClass = {astro-ph.GA},
       adsurl = {https://ui.adsabs.harvard.edu/abs/2018MNRAS.481L.100V}
}

@ARTICLE{Freedman2001,
       author = {{Freedman}, Wendy L. and {Madore}, Barry F. and {Gibson}, Brad K. and {Ferrarese}, Laura and {Kelson}, Daniel D. and {Sakai}, Shoko and {Mould}, Jeremy R. and {Kennicutt}, Jr., Robert C. and {Ford}, Holland C. and {Graham}, John A. and {Huchra}, John P. and {Hughes}, Shaun M.~G. and {Illingworth}, Garth D. and {Macri}, Lucas M. and {Stetson}, Peter B.},
        title = "{Final Results from the Hubble Space Telescope Key Project to Measure the Hubble Constant}",
      journal = {\apj},
         year = 2001,
        month = may,
       volume = {553},
       number = {1},
        pages = {47-72},
          doi = {10.1086/320638},
archivePrefix = {arXiv},
       eprint = {astro-ph/0012376},
 primaryClass = {astro-ph},
       adsurl = {https://ui.adsabs.harvard.edu/abs/2001ApJ...553...47F}
}

@ARTICLE{GaiaCollaboration2021LMC,
       author = {{Gaia Collaboration} and {Luri}, X. and {Chemin}, L. and {Clementini}, G. and {Delgado}, H.~E. and {McMillan}, P.~J. and {Romero-G{\'o}mez}, M. and {Balbinot}, E. and {Castro-Ginard}, A. and {Mor}, R. and {Ripepi}, V. and {Sarro}, L.~M. and {Cioni}, M.-R.~L. and {Fabricius}, C. and {Garofalo}, A. and {Helmi}, A. and {Muraveva}, T. and {Brown}, A.~G.~A. and {Vallenari}, A. and {Prusti}, T. and {de Bruijne}, J.~H.~J. and {Babusiaux}, C. and {Biermann}, M. and {Creevey}, O.~L. and {Evans}, D.~W. and {Eyer}, L. and {Hutton}, A. and {Jansen}, F. and {Jordi}, C. and {Klioner}, S.~A. and {Lammers}, U. and {Lindegren}, L. and {Mignard}, F. and {Panem}, C. and {Pourbaix}, D. and {Randich}, S. and {Sartoretti}, P. and {Soubiran}, C. and {Walton}, N.~A. and {Arenou}, F. and {Bailer-Jones}, C.~A.~L. and {Bastian}, U. and {Cropper}, M. and {Drimmel}, R. and {Katz}, D. and {Lattanzi}, M.~G. and {van Leeuwen}, F. and {Bakker}, J. and {Casta{\~n}eda}, J. and {De Angeli}, F. and {Ducourant}, C. and {Fouesneau}, M. and {Fr{\'e}mat}, Y. and {Guerra}, R. and {Guerrier}, A. and {Guiraud}, J. and {Jean-Antoine Piccolo}, A. and {Masana}, E. and {Messineo}, R. and {Mowlavi}, N. and {Nicolas}, C. and {Nienartowicz}, K. and {Pailler}, F. and {Panuzzo}, P. and {Riclet}, F. and {Roux}, W. and {Seabroke}, G.~M. and {Sordo}, R. and {Tanga}, P. and {Th{\'e}venin}, F. and {Gracia-Abril}, G. and {Portell}, J. and {Teyssier}, D. and {Altmann}, M. and {Andrae}, R. and {Bellas-Velidis}, I. and {Benson}, K. and {Berthier}, J. and {Blomme}, R. and {Brugaletta}, E. and {Burgess}, P.~W. and {Busso}, G. and {Carry}, B. and {Cellino}, A. and {Cheek}, N. and {Damerdji}, Y. and {Davidson}, M. and {Delchambre}, L. and {Dell'Oro}, A. and {Fern{\'a}ndez-Hern{\'a}ndez}, J. and {Galluccio}, L. and {Garc{\'\i}a-Lario}, P. and {Garcia-Reinaldos}, M. and {Gonz{\'a}lez-N{\'u}{\~n}ez}, J. and {Gosset}, E. and {Haigron}, R. and {Halbwachs}, J.-L. and {Hambly}, N.~C. and {Harrison}, D.~L. and {Hatzidimitriou}, D. and {Heiter}, U. and {Hern{\'a}ndez}, J. and {Hestroffer}, D. and {Hodgkin}, S.~T. and {Holl}, B. and {Jan{\ss}en}, K. and {Jevardat de Fombelle}, G. and {Jordan}, S. and {Krone-Martins}, A. and {Lanzafame}, A.~C. and {L{\"o}ffler}, W. and {Lorca}, A. and {Manteiga}, M. and {Marchal}, O. and {Marrese}, P.~M. and {Moitinho}, A. and {Mora}, A. and {Muinonen}, K. and {Osborne}, P. and {Pancino}, E. and {Pauwels}, T. and {Recio-Blanco}, A. and {Richards}, P.~J. and {Riello}, M. and {Rimoldini}, L. and {Robin}, A.~C. and {Roegiers}, T. and {Rybizki}, J. and {Siopis}, C. and {Smith}, M. and {Sozzetti}, A. and {Ulla}, A. and {Utrilla}, E. and {van Leeuwen}, M. and {van Reeven}, W. and {Abbas}, U. and {Abreu Aramburu}, A. and {Accart}, S. and {Aerts}, C. and {Aguado}, J.~J. and {Ajaj}, M. and {Altavilla}, G. and {{\'A}lvarez}, M.~A. and {{\'A}lvarez Cid-Fuentes}, J. and {Alves}, J. and {Anderson}, R.~I. and {Anglada Varela}, E. and {Antoja}, T. and {Audard}, M. and {Baines}, D. and {Baker}, S.~G. and {Balaguer-N{\'u}{\~n}ez}, L. and {Balog}, Z. and {Barache}, C. and {Barbato}, D. and {Barros}, M. and {Barstow}, M.~A. and {Bartolom{\'e}}, S. and {Bassilana}, J.-L. and {Bauchet}, N. and {Baudesson-Stella}, A. and {Becciani}, U. and {Bellazzini}, M. and {Bernet}, M. and {Bertone}, S. and {Bianchi}, L. and {Blanco-Cuaresma}, S. and {Boch}, T. and {Bombrun}, A. and {Bossini}, D. and {Bouquillon}, S. and {Bragaglia}, A. and {Bramante}, L. and {Breedt}, E. and {Bressan}, A. and {Brouillet}, N. and {Bucciarelli}, B. and {Burlacu}, A. and {Busonero}, D. and {Butkevich}, A.~G. and {Buzzi}, R. and {Caffau}, E. and {Cancelliere}, R. and {C{\'a}novas}, H. and {Cantat-Gaudin}, T. and {Carballo}, R. and {Carlucci}, T. and {Carnerero}, M.~I. and {Carrasco}, J.~M. and {Casamiquela}, L. and {Castellani}, M. and {Castro Sampol}, P. and {Chaoul}, L. and {Charlot}, P. and {Chiavassa}, A. and {Comoretto}, G. and {Cooper}, W.~J. and {Cornez}, T. and {Cowell}, S. and {Crifo}, F.},
        title = "{Gaia Early Data Release 3. Structure and properties of the Magellanic Clouds}",
      journal = {\aap},
         year = 2021,
        month = may,
       volume = {649},
          eid = {A7},
        pages = {A7},
          doi = {10.1051/0004-6361/202039588},
archivePrefix = {arXiv},
       eprint = {2012.01771},
 primaryClass = {astro-ph.GA},
       adsurl = {https://ui.adsabs.harvard.edu/abs/2021A&A...649A...7G}
}

@ARTICLE{vanderMarel2002,
       author = {{van der Marel}, Roeland P. and {Alves}, David R. and {Hardy}, Eduardo and {Suntzeff}, Nicholas B.},
        title = "{New Understanding of Large Magellanic Cloud Structure, Dynamics, and Orbit from Carbon Star Kinematics}",
      journal = {\aj},
         year = 2002,
        month = nov,
       volume = {124},
       number = {5},
        pages = {2639-2663},
          doi = {10.1086/343775},
archivePrefix = {arXiv},
       eprint = {astro-ph/0205161},
 primaryClass = {astro-ph},
       adsurl = {https://ui.adsabs.harvard.edu/abs/2002AJ....124.2639V}
}

@ARTICLE{Sales2022,
       author = {{Sales}, Laura V. and {Wetzel}, Andrew and {Fattahi}, Azadeh},
        title = "{Baryonic solutions and challenges for cosmological models of dwarf galaxies}",
      journal = {Nature Astronomy},
         year = 2022,
        month = jun,
       volume = {6},
        pages = {897-910},
          doi = {10.1038/s41550-022-01689-w},
archivePrefix = {arXiv},
       eprint = {2206.05295},
 primaryClass = {astro-ph.GA},
       adsurl = {https://ui.adsabs.harvard.edu/abs/2022NatAs...6..897S}
}

@ARTICLE{Richstein2022,
       author = {{Richstein}, Hannah and {Patel}, Ekta and {Kallivayalil}, Nitya and {Simon}, Joshua D. and {Zivick}, Paul and {Tollerud}, Erik and {Fritz}, Tobias and {Warfield}, Jack T. and {Besla}, Gurtina and {van der Marel}, Roeland P. and {Wetzel}, Andrew and {Choi}, Yumi and {Deason}, Alis and {Geha}, Marla and {Guhathakurta}, Puragra and {Jeon}, Myoungwon and {Kirby}, Evan N. and {Libralato}, Mattia and {Sacchi}, Elena and {Sohn}, Sangmo Tony},
        title = "{Structural Parameters and Possible Association of the Ultra-faint Dwarfs Pegasus III and Pisces II from Deep Hubble Space Telescope Photometry}",
      journal = {\apj},
         year = 2022,
        month = jul,
       volume = {933},
       number = {2},
          eid = {217},
        pages = {217},
          doi = {10.3847/1538-4357/ac7226},
archivePrefix = {arXiv},
       eprint = {2204.01917},
 primaryClass = {astro-ph.GA},
       adsurl = {https://ui.adsabs.harvard.edu/abs/2022ApJ...933..217R}
}

@ARTICLE{Kim2016,
       author = {{Kim}, Dongwon and {Jerjen}, Helmut and {Geha}, Marla and {Chiti}, Anirudh and {Milone}, Antonino P. and {Da Costa}, Gary and {Mackey}, Dougal and {Frebel}, Anna and {Conn}, Blair},
        title = "{Portrait of a Dark Horse: a Photometric and Spectroscopic Study of the Ultra-faint Milky Way Satellite Pegasus III}",
      journal = {\apj},
         year = 2016,
        month = dec,
       volume = {833},
       number = {1},
          eid = {16},
        pages = {16},
          doi = {10.3847/0004-637X/833/1/16},
archivePrefix = {arXiv},
       eprint = {1608.04934},
 primaryClass = {astro-ph.GA},
       adsurl = {https://ui.adsabs.harvard.edu/abs/2016ApJ...833...16K}
}

@ARTICLE{BoylanKolchin2013,
       author = {{Boylan-Kolchin}, Michael and {Bullock}, James S. and {Sohn}, Sangmo Tony and {Besla}, Gurtina and {van der Marel}, Roeland P.},
        title = "{The Space Motion of Leo I: The Mass of the Milky Way's Dark Matter Halo}",
      journal = {\apj},
         year = 2013,
        month = may,
       volume = {768},
       number = {2},
          eid = {140},
        pages = {140},
          doi = {10.1088/0004-637X/768/2/140},
archivePrefix = {arXiv},
       eprint = {1210.6046},
 primaryClass = {astro-ph.CO},
       adsurl = {https://ui.adsabs.harvard.edu/abs/2013ApJ...768..140B}
}

@ARTICLE{RuizLara2020Sag,
       author = {{Ruiz-Lara}, Tom{\'a}s and {Gallart}, Carme and {Bernard}, Edouard J. and {Cassisi}, Santi},
        title = "{The recurrent impact of the Sagittarius dwarf on the star formation history of the Milky Way}",
      journal = {Nature Astronomy},
         year = 2020,
        month = may,
       volume = {4},
        pages = {965-973},
          doi = {10.1038/s41550-020-1097-0},
archivePrefix = {arXiv},
       eprint = {2003.12577},
 primaryClass = {astro-ph.GA},
       adsurl = {https://ui.adsabs.harvard.edu/abs/2020NatAs...4..965R}
}

@ARTICLE{Moreno2015,
       author = {{Moreno}, Jorge and {Torrey}, Paul and {Ellison}, Sara L. and {Patton}, David R. and {Bluck}, Asa F.~L. and {Bansal}, Gunjan and {Hernquist}, Lars},
        title = "{Mapping galaxy encounters in numerical simulations: the spatial extent of induced star formation}",
      journal = {\mnras},
         year = 2015,
        month = apr,
       volume = {448},
       number = {2},
        pages = {1107-1117},
          doi = {10.1093/mnras/stv094},
archivePrefix = {arXiv},
       eprint = {1501.03573},
 primaryClass = {astro-ph.GA},
       adsurl = {https://ui.adsabs.harvard.edu/abs/2015MNRAS.448.1107M}
}

@ARTICLE{Hammer2021,
       author = {{Hammer}, Francois and {Wang}, Jianling and {Pawlowski}, Marcel S. and {Yang}, Yanbin and {Bonifacio}, Piercarlo and {Li}, Hefan and {Babusiaux}, Carine and {Arenou}, Frederic},
        title = "{Gaia EDR3 Proper Motions of Milky Way Dwarfs. II Velocities, Total Energy, and Angular Momentum}",
      journal = {\apj},
         year = 2021,
        month = dec,
       volume = {922},
       number = {2},
          eid = {93},
        pages = {93},
          doi = {10.3847/1538-4357/ac27a8},
archivePrefix = {arXiv},
       eprint = {2109.11557},
 primaryClass = {astro-ph.GA},
       adsurl = {https://ui.adsabs.harvard.edu/abs/2021ApJ...922...93H}
}

\begin{appendix}

\section{Dynamical friction implementation}
\label{sec:DF}
We follow a procedure analogous to that described in \citet{Patel2020} to include the effect of DF in the integration of the nominal orbits of the studied galaxies. 

In order to account for the DF experienced by a dwarf moving through the MW DM halo, we use the \citet{Chandrasekhar1943} approximation:

\begin{equation}
    \mathbf{F}_{\mathrm{DF, MW}} = -\frac{4\pi G^2 M_{\mathrm{sat}}^2\ln\Lambda \rho(\mathbf{r},t) }{v^2} 
\left[\text{erf}(X) - \frac{2X}{\sqrt{\pi}} e^{-X^2}\right] \frac{\mathbf{v}}{v}
\end{equation}

\noindent Here $M_{\mathrm{sat}}$ is the total mass of the considered dwarf. As mentioned in Section~\ref{sec:dfest}, we integrated the nominal orbits of each dwarf using two different masses. For dSphs, we assumed  $10^9$ and $10^{10}$ M$_{\odot}$, and for the UFDs $10^8$ and $10^9$ M$_{\odot}$ \citep{Bullock2017, Sales2022}. For the SMC we adopted the masses of models SMC1 and SMC2 of \citet{Patel2020}.
$\rho(\mathbf{r},t)$ is the density of the MW DM halo in the position of the dwarf at the corresponding time.  $\ln\Lambda$ is the Coulomb logarithm, a dimensionless factor measuring the overall intensity of the drag. We adopt the Coulomb logarithm parametrization of \citet{Hashimoto2003}:

\begin{equation}
    \ln \Lambda =
\begin{cases}
\ln\!\left(\frac{r}{1.4\,a_s}\right), & r > 1.4\,a_s \\
0, & r \le 1.4\,a_s
\end{cases}
\label{eq:columblog}
\end{equation}

\noindent Here, $r$ is the distance of the dwarf to the centre of the MW, and $a_s$ is the scale radius. For the SMC, the scale radius is the Hernquist scale radius (\citealt{Hernquist1990}) of models SMC1 and SMC2 of \citet{Patel2020}. For the rest of dwarfs $a_s$ corresponds to their Plummer scale radius (\citealt{Plummer1911}). We calculate it by deriving the radius at which the mass enclosed within a Plummer sphere coincides with the dynamical mass estimated within the projected half-light radius. We take the projected half-light radii and LOSVs dispersions of the dwarfs from \citet{Battaglia2022}, \citet{Pace2024} and \citet{Cerny2026} (depending on which data set phase-space coordinates were taken from), and derive the mass enclosed within the half-light radius using the \citet{Wolf2010} estimator.  
We note that for some of the galaxies of the sample it is not possible to estimate their mass enclosed within the half-light radius since their velocity dispersions have not been measured yet. For these galaxies, we assume a constant value for the Coulomb logarithm, $\ln \Lambda = 3$ (e.g. \citealt{Vasiliev2023}). We find that using such values  yields comparable results to using formula \ref{eq:columblog}, something already remarked by \citet{Taibi2024}. Finally, $X = v / \sqrt2\, \sigma(r)$, where $v$ is the velocity of the dwarf with respect to the MW and estimate $\sigma(r)$ is the one-dimensional velocity dispersion of the MW measured at the distance $r$ between the dwarf and the MW. We estimate $\sigma(r)$ as follows. 

Instead of using the velocity dispersion approximation proposed in \citet{ZentnerBullock2003} and adopted in \citet{Patel2020}, $\sigma (r)$ is computed from the  MW potential and the MW DM halo density using the \textsc{Agama} framework. For simplicity, both the MW potential and the MW DM halo density adopted in this approximation are treated as static, and correspond to the initial total MW and DM halo parametrizations of the corresponding potential, prior to any interaction with the LMC. Neglecting the time evolution of $\sigma (r)$ due to the LMC interaction significantly simplifies the computation, and is a reasonable approximation given that Chandrasekhar formula is not particularly sensitive to the precise value of the velocity dispersion (e.g. \citealt{Vasiliev2023}).
The MW initial potential is sphericalized by constructing a multipole expansion and retaining only the spherical term ($l = 0$), and a quasi-spherical, isotropic distribution function is fitted using the DM halo density as the tracer population.  The velocity dispersion is then recovered as the square root of the second velocity moment on a logarithmic radial grid and interpolated with a spline to produce a smooth callable $\sigma (r)$, which is evaluated in each integration step. 

Similarly to the MW, the LMC also exerts a drag force on the dwarfs that get within its region of influence. In this case we apply the approximation described in \citet{BinneyTremaine1987}, and used by \citet{BekkiChiba2005} and \citet{Besla2007}:  

\begin{equation}
    \mathbf{F}_{\mathrm{DF, LMC}} = -0.428 \ln\Lambda \frac{G\, M_{\mathrm{sat}}^2}{r_{\mathrm{LMC}}^2} \frac{\mathbf{v_{\mathrm{LMC}}}}{v_{\mathrm{LMC}}}
\end{equation}

In this case, $r_{\mathrm{LMC}}$ is the distance of the dwarf to the LMC and $v_{\mathrm{LMC}}$ their relative velocity. We adopt $\ln\Lambda = 0.3$, as in \citealt{Patel2020}. 

We make use of this approximation of the DF to integrate the nominal orbits of all the dwarfs, for the two $M_{\mathrm{sat}}$ considered. In order to do so, we can no longer use the method \texttt{orbit} of \textsc{Agama}. Instead, we use method \texttt{force} to compute the acceleration experienced by a given galaxy at a given position and time due to the combined MW+LMC potential, and we add to it the acceleration terms due to the DF of the MW and the LMC. Then we proceed to the integration of the differential equations.

\section{Orbits and parameters}
\label{app:orb}
In Figures~\ref{fig:traj1},~\ref{fig:traj2}, and~\ref{fig:traj3} we show the orbits derived for the 72 galaxies studied in this work, under potential L2M10. The Figures show, for each galaxy, the Galactocentric distance as a function of time for the ensemble of orbits resulting from the MC realizations and the orbit derived based exclusively on the nominal values of the phase-space coordinates. We also include the nominal orbits integrated accounting for the effect of DF.
For the potential LMC satellites, we superimpose in their orbit representation the trajectory of the LMC and the extent of the tidal radius. In Table~\ref{tab:orbparam} we show the orbital parameters obtained with potential L2M10.

\begin{figure*}
    \centering
    \includegraphics[width=\linewidth]{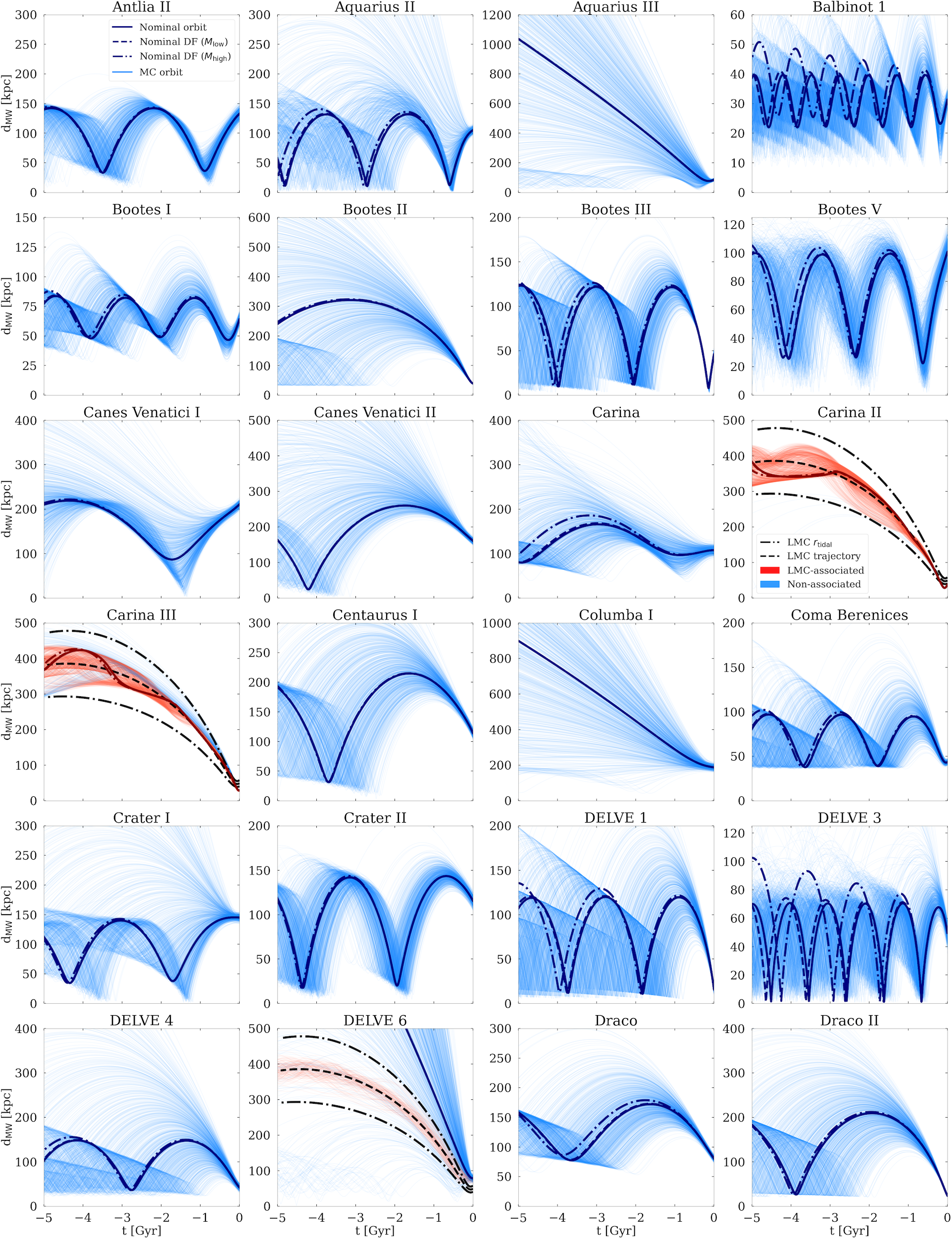}
    \caption{Reconstructed trajectories of 72 dwarfs in the vicinity of the MW using the potential L2M10. Markers coincide with those of Figure~\ref{fig:orb_sample}}
    \label{fig:traj1}
\end{figure*}

\begin{figure*}
    \centering
    \includegraphics[width=\linewidth]{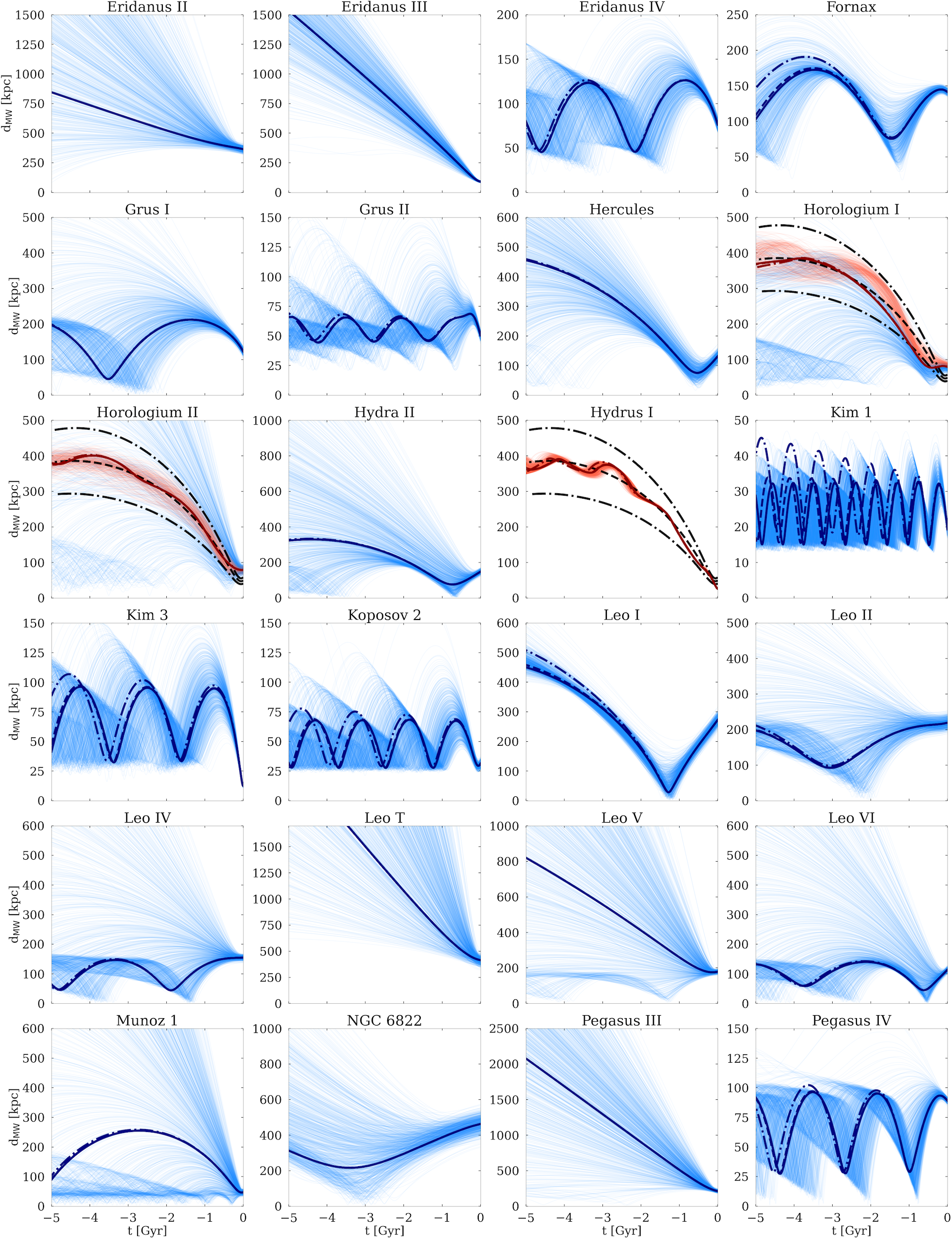}
    \caption{Continuation of Figure~\ref{fig:traj1}}
    \label{fig:traj2}
\end{figure*}

\begin{figure*}
    \centering
    \includegraphics[width=\linewidth]{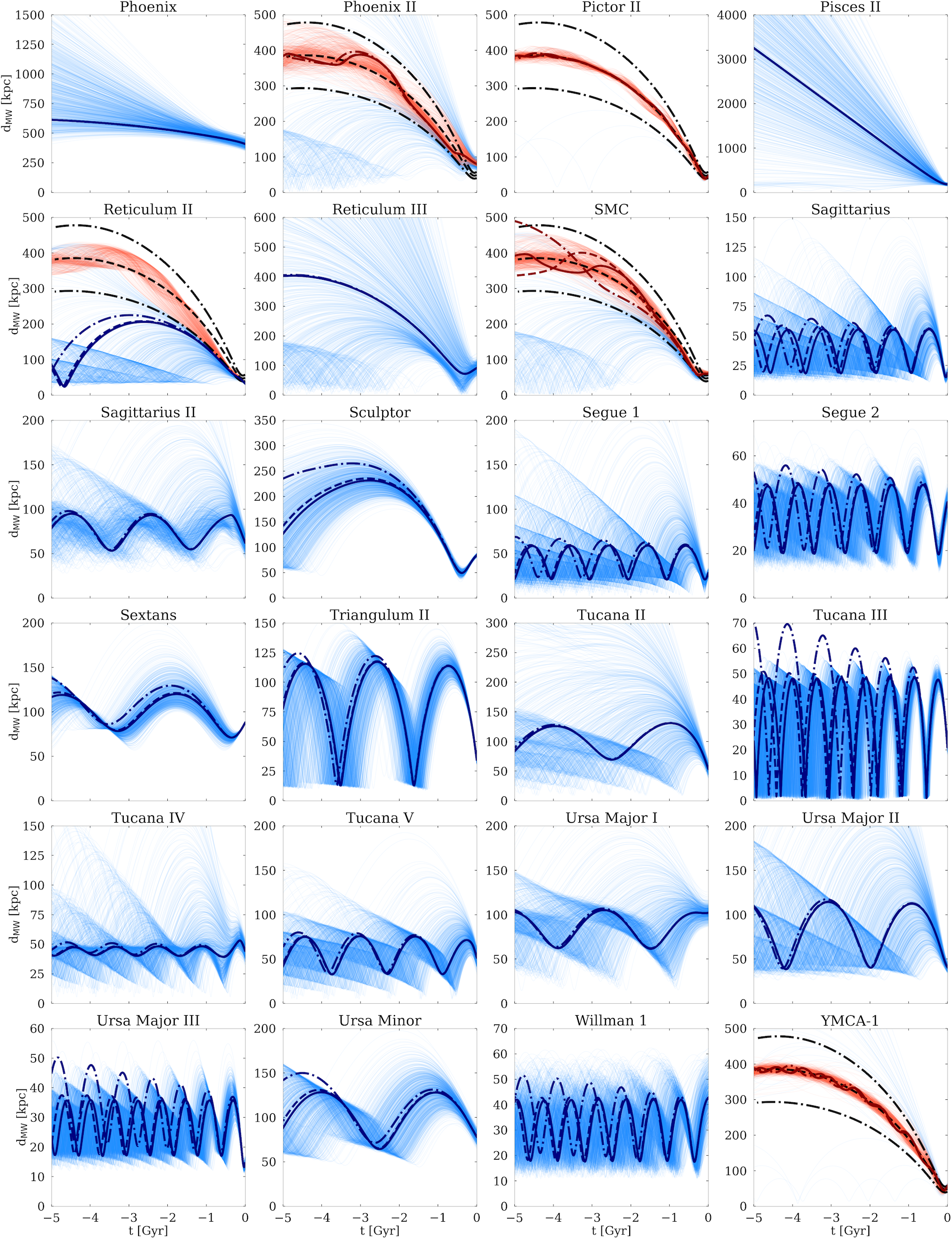}
    \caption{Continuation of Figure~\ref{fig:traj2}}
    \label{fig:traj3}
\end{figure*}

\onecolumn     
\begingroup
\renewcommand{\arraystretch}{1.43}
\setlength{\LTleft}{\fill} \setlength{\LTright}{\fill}
\begin{longtable}{l c c c c c c c c c c }
\caption{Orbital parameters with respect to the MW for potential L2M10} \label{tab:orbparam} \\
\toprule
Galaxy & $r_\mathrm{peri}$ & $t_\mathrm{peri}$ & $f_\mathrm{peri}$ & $r_\mathrm{apo}$ & $t_\mathrm{apo}$ & $f_\mathrm{apo}$ & $E / 1\times 10^{{4}}$ & $f_\mathrm{bound}$ & $t_\mathrm{infall}$ & $f_\mathrm{vir}$ \\
 & [kpc] & [Gyr] & & [kpc] & [Gyr] & & [km$^{2}$ s$^{-2}$] & & [Gyr] & \\\multicolumn{1}{c}{(1)}  & (2) & (3) & (4) & (5) & (6) & (7) & (8) & (9) & (10) & (11) \\
\midrule
\midrule
\endfirsthead
\multicolumn{11}{c}{{\textbf{\tablename\ \thetable{}} -- Continued}} \\
\toprule
Galaxy & $r_\mathrm{peri}$ & $t_\mathrm{peri}$ & $f_\mathrm{peri}$ & $r_\mathrm{apo}$ & $t_\mathrm{apo}$ & $f_\mathrm{apo}$ & $E / 1\times 10^{{4}}$ & $f_\mathrm{bound}$ & $t_\mathrm{infall}$ & $f_\mathrm{vir}$ \\
 & [kpc] & [Gyr] & & [kpc] & [Gyr] & & [km$^{2}$ s$^{-2}$] & & [Gyr] & \\\multicolumn{1}{c}{(1)}  & (2) & (3) & (4) & (5) & (6) & (7) & (8) & (9) & (10) & (11) \\
\midrule
\midrule
\endhead
\midrule
\multicolumn{11}{r}{\textit{Continued on next page...}} \\
\endfoot
\multicolumn{11}{@{} l @{} }{
\makebox[0pt][l]{
\begin{minipage}{\linewidth}
\vspace{4pt} \small \textbf{Note.} Galaxies and their orbital parameters. The table lists the orbital parameters obtained using potential L2M10, for all the studied galaxies. Columns show, from left to right: (1) galaxy name; (2) Galactocentric distance at the most recent pericentre; (3) time of the most recent pericentric passage; (4) fraction of MC realizations for which we detect the pericentre; (5) Galactocentric distance at the most recent apocentre; (6) time of the most recent apocentric passage; (7) fraction of MC realizations for which we detect the apocentre; (8) specific orbital energy at $t = 0$, (9) fraction of MC realizations for which $E<0$ at $t = 0$; (10) infall time and (11) fraction of MC realizations at the earliest integration time that are within the virial radius of the MW. For galaxies whose nominal orbit was already within the virial radius at the earliest integration time, we report $t_{\rm infall} < -5$ Gyr, indicating that their infall occurred earlier than the time span explored in this work. All quantities are measured with respect to the MW. 
The complete tables, including all galaxies and potentials, are available in the associated repository.
\end{minipage} } } \\
\endlastfoot
Antlia II & $37^{+14}_{-11}$ & $-0.89^{+0.08}_{-0.09}$ & 1.000 & $143^{+9}_{-9}$ & $-2.2^{+0.2}_{-0.3}$ & 1.000 & $-1.6^{+0.3}_{-0.3}$ & 1.000 & $<-5.0$ & 1.000 \\
Aquarius II & $39^{+58}_{-33}$ & $-0.6^{+0.2}_{-0.1}$ & 1.000 & $149^{+43}_{-32}$ & $-2.0^{+0.4}_{-0.9}$ & 0.859 & $-3.4^{+1.5}_{-0.7}$ & 0.951 & $<-5.0$ & 0.820 \\
Aquarius III & $77^{+7}_{-13}$ & $-0.12^{+0.05}_{-0.16}$ & 1.000 & $214^{+112}_{-81}$ & $-2.6^{+0.9}_{-1.4}$ & 0.173 & $1^{+4}_{-3}$ & 0.376 & $-1.08$ & 0.133 \\
Balbinot 1 & $23^{+5}_{-5}$ & $-0.18^{+0.04}_{-0.02}$ & 1.000 & $39^{+8}_{-4}$ & $-0.56^{+0.06}_{-0.07}$ & 1.000 & $-7.2^{+0.5}_{-0.5}$ & 1.000 & $<-5.0$ & 1.000 \\
Bootes I & $47^{+5}_{-5}$ & $-0.29^{+0.04}_{-0.03}$ & 1.000 & $82^{+13}_{-8}$ & $-1.15^{+0.08}_{-0.10}$ & 1.000 & $-4.4^{+0.4}_{-0.3}$ & 1.000 & $<-5.0$ & 1.000 \\
Bootes II & $33^{+1}_{-1}$ & $-4.0^{+0.9}_{-0.7}$ & 0.339 & $276^{+109}_{-72}$ & $-2.6^{+0.8}_{-1.3}$ & 0.781 & $-1.7^{+0.8}_{-0.7}$ & 0.983 & $<-5.0$ & 0.548 \\
Bootes III & $8^{+3}_{-2}$ & $-0.137^{+0.007}_{-0.010}$ & 1.000 & $122^{+17}_{-13}$ & $-1.1^{+0.1}_{-0.2}$ & 1.000 & $-4.1^{+0.3}_{-0.3}$ & 1.000 & $<-5.0$ & 1.000 \\
Bootes V & $26^{+15}_{-11}$ & $-0.62^{+0.06}_{-0.09}$ & 1.000 & $100^{+7}_{-8}$ & $-1.5^{+0.1}_{-0.2}$ & 1.000 & $-4.8^{+0.3}_{-0.2}$ & 1.000 & $<-5.0$ & 1.000 \\
Canes Venatici I & $95^{+63}_{-55}$ & $-1.6^{+0.2}_{-0.1}$ & 1.000 & $219^{+11}_{-7}$ & $-4.2^{+0.5}_{-0.5}$ & 0.734 & $-3.7^{+0.4}_{-0.3}$ & 1.000 & $<-5.0$ & 0.778 \\
Canes Venatici II & $22^{+17}_{-11}$ & $-4.1^{+0.4}_{-0.5}$ & 0.405 & $282^{+72}_{-31}$ & $-2.1^{+0.5}_{-1.2}$ & 0.776 & $-3.6^{+0.8}_{-0.3}$ & 0.997 & $<-5.0$ & 0.613 \\
Carina & $94^{+15}_{-21}$ & $-0.8^{+0.4}_{-0.1}$ & 0.882 & $107^{+6}_{-6}$ & $-0.03^{+0.02}_{-0.03}$ & 0.924 & $-1.1^{+0.6}_{-0.6}$ & 0.953 & $<-5.0$ & 0.894 \\
Carina II & $28^{+1}_{-1}$ & $-0.073^{+0.003}_{-0.000}$ & 1.000 & $389^{+24}_{-31}$ & $-3.3^{+0.5}_{-0.5}$ & 0.915 & $-2.1^{+0.3}_{-0.3}$ & 1.000 & $-1.52$ & 0.004 \\
Carina III & $29^{+1}_{-1}$ & $-0.010^{+0.000}_{-0.003}$ & 1.000 & $400^{+29}_{-40}$ & $-3.4^{+0.6}_{-0.9}$ & 0.918 & $-1.8^{+1.0}_{-0.9}$ & 0.968 & $-1.54$ & 0.000 \\
Centaurus I & $30^{+7}_{-9}$ & $-3.6^{+0.6}_{-0.8}$ & 0.795 & $221^{+51}_{-27}$ & $-1.7^{+0.4}_{-0.7}$ & 0.991 & $-1.9^{+0.3}_{-0.2}$ & 1.000 & $<-5.0$ & 0.951 \\
Columba I & $157^{+51}_{-60}$ & $-3.3^{+1.3}_{-0.6}$ & 0.062 & $198^{+49}_{-19}$ & $-1.1^{+1.0}_{-2.2}$ & 0.076 & $2^{+2}_{-2}$ & 0.027 & $-0.80$ & 0.070 \\
Coma Berenices & $43^{+1}_{-2}$ & $-0.033^{+0.003}_{-0.010}$ & 1.000 & $95^{+21}_{-17}$ & $-0.9^{+0.1}_{-0.2}$ & 1.000 & $-3.9^{+0.4}_{-0.4}$ & 1.000 & $<-5.0$ & 1.000 \\
Crater I & $48^{+41}_{-31}$ & $-1.8^{+0.4}_{-1.2}$ & 0.779 & $148^{+52}_{-5}$ & $-0.3^{+0.2}_{-1.8}$ & 0.893 & $-2.1^{+1.2}_{-0.6}$ & 0.939 & $<-5.0$ & 0.854 \\
Crater II & $21^{+7}_{-6}$ & $-2.0^{+0.2}_{-0.3}$ & 1.000 & $144^{+7}_{-7}$ & $-0.69^{+0.08}_{-0.12}$ & 1.000 & $-2.7^{+0.2}_{-0.2}$ & 1.000 & $<-5.0$ & 1.000 \\
DELVE 1 & $11^{+3}_{-2}$ & $-1.8^{+0.4}_{-0.7}$ & 0.995 & $120^{+36}_{-25}$ & $-0.9^{+0.2}_{-0.3}$ & 1.000 & $-4.7^{+0.5}_{-0.5}$ & 1.000 & $<-5.0$ & 0.999 \\
DELVE 3 & $15^{+17}_{-9}$ & $-0.8^{+0.1}_{-0.3}$ & 0.995 & $72^{+11}_{-7}$ & $-0.27^{+0.05}_{-0.14}$ & 0.999 & $-4.9^{+0.7}_{-0.5}$ & 0.999 & $<-5.0$ & 0.996 \\
DELVE 4 & $35^{+5}_{-5}$ & $-2.5^{+0.8}_{-1.2}$ & 0.815 & $144^{+90}_{-48}$ & $-1.3^{+0.5}_{-1.0}$ & 0.952 & $-4.3^{+0.8}_{-0.7}$ & 1.000 & $<-5.0$ & 0.900 \\
DELVE 6 & $78^{+5}_{-8}$ & $-0.010^{+0.007}_{-0.181}$ & 0.558 & $342^{+57}_{-147}$ & $-3^{+1}_{-1}$ & 0.242 & $5^{+8}_{-5}$ & 0.175 & $-0.78$ & 0.061 \\
Draco & $76^{+6}_{-7}$ & $-3.5^{+0.6}_{-0.7}$ & 0.924 & $171^{+32}_{-22}$ & $-1.6^{+0.3}_{-0.4}$ & 1.000 & $-4.6^{+0.2}_{-0.2}$ & 1.000 & $<-5.0$ & 0.998 \\
Draco II & $25^{+2}_{-2}$ & $-3.6^{+0.7}_{-0.8}$ & 0.789 & $207^{+64}_{-40}$ & $-1.9^{+0.4}_{-0.7}$ & 0.990 & $-4.0^{+0.4}_{-0.3}$ & 1.000 & $<-5.0$ & 0.944 \\
Eridanus II & --- & --- & 0.000 & $419^{+36}_{-27}$ & $-2.5^{+0.9}_{-1.5}$ & 0.074 & $8^{+7}_{-3}$ & 0.000 & --- & 0.013 \\
Eridanus III & $320^{+4}_{-4}$ & $-3.5^{+0.3}_{-0.3}$ & 0.002 & $360^{+73}_{-20}$ & $-2.5^{+0.2}_{-1.0}$ & 0.003 & $6^{+2}_{-2}$ & 0.000 & $-0.63$ & 0.000 \\
Eridanus IV & $45^{+7}_{-7}$ & $-2.2^{+0.4}_{-0.6}$ & 1.000 & $127^{+19}_{-12}$ & $-0.9^{+0.2}_{-0.3}$ & 1.000 & $-2.7^{+0.3}_{-0.3}$ & 1.000 & $<-5.0$ & 1.000 \\
Fornax & $77^{+27}_{-22}$ & $-1.43^{+0.09}_{-0.08}$ & 0.998 & $145^{+3}_{-3}$ & $-0.17^{+0.02}_{-0.04}$ & 0.999 & $-1.3^{+0.3}_{-0.2}$ & 1.000 & $<-5.0$ & 0.991 \\
Grus I & $39^{+21}_{-20}$ & $-3.4^{+0.6}_{-0.9}$ & 0.737 & $218^{+48}_{-21}$ & $-1.4^{+0.3}_{-0.9}$ & 0.941 & $-0.8^{+0.4}_{-0.3}$ & 0.936 & $<-5.0$ & 0.875 \\
Grus II & $43^{+25}_{-7}$ & $-1.1^{+0.6}_{-0.8}$ & 1.000 & $69^{+17}_{-5}$ & $-0.27^{+0.05}_{-0.68}$ & 1.000 & $-4.5^{+0.5}_{-0.5}$ & 1.000 & $<-5.0$ & 1.000 \\
Hercules & $76^{+20}_{-19}$ & $-0.51^{+0.09}_{-0.08}$ & 1.000 & $302^{+54}_{-60}$ & $-3.8^{+0.8}_{-0.7}$ & 0.335 & $-2.9^{+0.5}_{-0.4}$ & 1.000 & $-2.05$ & 0.154 \\
Horologium I & $75^{+268}_{-21}$ & $-0.44^{+0.05}_{-2.87}$ & 0.951 & $82^{+288}_{-8}$ & $-0.11^{+0.05}_{-2.56}$ & 0.989 & $-3^{+1}_{-1}$ & 0.981 & $-1.80$ & 0.426 \\
Horologium II & $77^{+10}_{-16}$ & $-0.08^{+0.07}_{-0.25}$ & 0.815 & $348^{+51}_{-172}$ & $-3^{+1}_{-1}$ & 0.672 & $-1^{+3}_{-2}$ & 0.562 & $-1.50$ & 0.186 \\
Hydra II & $93^{+34}_{-55}$ & $-0.6^{+0.3}_{-0.1}$ & 1.000 & $237^{+80}_{-40}$ & $-3.2^{+0.8}_{-1.2}$ & 0.454 & $0^{+2}_{-1}$ & 0.531 & $-2.52$ & 0.338 \\
Hydrus I & $353^{+7}_{-8}$ & $-3.34^{+0.08}_{-0.11}$ & 1.000 & $375^{+5}_{-6}$ & $-2.90^{+0.08}_{-0.10}$ & 1.000 & $-2.9^{+0.5}_{-0.4}$ & 1.000 & $-1.60$ & 0.000 \\
Kim 1 & $14.8^{+0.8}_{-0.7}$ & $-0.48^{+0.05}_{-0.06}$ & 1.000 & $32^{+4}_{-3}$ & $-0.22^{+0.03}_{-0.03}$ & 1.000 & $-7.9^{+0.3}_{-0.3}$ & 1.000 & $<-5.0$ & 1.000 \\
Kim 3 & $12.7^{+0.8}_{-0.6}$ & $-0.003^{+0.000}_{-0.000}$ & 1.000 & $95^{+19}_{-15}$ & $-0.8^{+0.1}_{-0.2}$ & 1.000 & $-5.5^{+0.6}_{-0.6}$ & 1.000 & $<-5.0$ & 1.000 \\
Koposov 2 & $29^{+2}_{-2}$ & $-0.06^{+0.01}_{-0.01}$ & 1.000 & $67^{+16}_{-12}$ & $-0.64^{+0.09}_{-0.12}$ & 1.000 & $-5.0^{+0.5}_{-0.5}$ & 1.000 & $<-5.0$ & 1.000 \\
Leo I & $49^{+41}_{-28}$ & $-1.31^{+0.08}_{-0.09}$ & 1.000 & --- & --- & 0.000 & $-0.1^{+0.3}_{-0.2}$ & 0.677 & $-2.57$ & 0.000 \\
Leo II & $107^{+111}_{-57}$ & $-2.4^{+2.2}_{-0.9}$ & 1.000 & $216^{+37}_{-15}$ & $-4.1^{+2.7}_{-0.6}$ & 0.489 & $-2.5^{+0.5}_{-0.3}$ & 1.000 & $<-5.0$ & 0.843 \\
Leo IV & $59^{+93}_{-39}$ & $-2^{+2}_{-1}$ & 0.672 & $156^{+68}_{-10}$ & $-2.5^{+2.5}_{-0.9}$ & 0.775 & $-2.1^{+1.6}_{-0.7}$ & 0.875 & $<-5.0$ & 0.714 \\
Leo T & --- & --- & 0.000 & --- & --- & 0.000 & $27^{+53}_{-21}$ & 0.025 & --- & 0.000 \\
Leo V & $175^{+7}_{-13}$ & $-0.10^{+0.07}_{-0.35}$ & 1.000 & $173^{+87}_{-12}$ & $-3.6^{+0.7}_{-0.8}$ & 0.199 & $1^{+5}_{-3}$ & 0.357 & $-1.00$ & 0.184 \\
Leo VI & $71^{+27}_{-46}$ & $-0.5^{+0.3}_{-0.1}$ & 1.000 & $147^{+76}_{-14}$ & $-2.1^{+0.4}_{-0.8}$ & 0.738 & $-2^{+2}_{-1}$ & 0.862 & $<-5.0$ & 0.678 \\
Muñoz 1 & $45^{+6}_{-8}$ & $-0.03^{+0.02}_{-0.17}$ & 0.992 & $81^{+138}_{-37}$ & $-0.9^{+0.3}_{-1.3}$ & 0.505 & $-3^{+5}_{-3}$ & 0.717 & $<-5.0$ & 0.467 \\
NGC 6822 & $298^{+108}_{-181}$ & $-2.8^{+1.4}_{-0.6}$ & 1.000 & --- & --- & 0.000 & $-1.9^{+1.1}_{-0.6}$ & 0.948 & $-4.39$ & 0.039 \\
Pegasus III & $110^{+123}_{-64}$ & $-3^{+1}_{-1}$ & 0.024 & $262^{+102}_{-17}$ & $-1.2^{+0.4}_{-2.4}$ & 0.044 & $8^{+11}_{-8}$ & 0.130 & $-0.25$ & 0.026 \\
Pegasus IV & $32^{+25}_{-14}$ & $-1.0^{+0.1}_{-0.3}$ & 1.000 & $94^{+3}_{-2}$ & $-0.20^{+0.04}_{-0.11}$ & 1.000 & $-5.0^{+0.4}_{-0.3}$ & 1.000 & $<-5.0$ & 1.000 \\
Phoenix & --- & --- & 0.000 & $541^{+33}_{-26}$ & $-4.0^{+0.7}_{-0.6}$ & 0.157 & $4.5^{+1.1}_{-0.6}$ & 0.000 & --- & 0.000 \\
Phoenix II & $98^{+265}_{-26}$ & $-0.4^{+0.1}_{-3.4}$ & 0.742 & $361^{+40}_{-283}$ & $-2.9^{+2.7}_{-1.0}$ & 0.902 & $-1^{+2}_{-2}$ & 0.633 & $-1.78$ & 0.215 \\
Pictor II & $40^{+2}_{-2}$ & $-0.075^{+0.005}_{-0.008}$ & 1.000 & $385^{+3}_{-26}$ & $-4.0^{+1.0}_{-0.4}$ & 1.000 & $-4.0^{+0.7}_{-0.6}$ & 1.000 & $-1.63$ & 0.003 \\
Pisces II & $105^{+53}_{-63}$ & $-2.2^{+0.5}_{-1.3}$ & 0.016 & $200^{+83}_{-18}$ & $-0.8^{+0.3}_{-2.0}$ & 0.025 & $21^{+28}_{-16}$ & 0.048 & $-0.26$ & 0.023 \\
Reticulum II & $37^{+330}_{-3}$ & $-3^{+1}_{-1}$ & 0.819 & $222^{+190}_{-128}$ & $-2.7^{+1.7}_{-1.0}$ & 0.986 & $-5.7^{+0.6}_{-0.6}$ & 1.000 & $<-5.0$ & 0.515 \\
Reticulum III & $70^{+25}_{-40}$ & $-0.3^{+0.2}_{-0.1}$ & 1.000 & $222^{+125}_{-58}$ & $-2.6^{+0.8}_{-1.6}$ & 0.522 & $-1^{+4}_{-2}$ & 0.642 & $-1.86$ & 0.400 \\
SMC & $57^{+7}_{-6}$ & $-0.23^{+0.01}_{-0.01}$ & 0.984 & $61^{+4}_{-3}$ & $-0.08^{+0.03}_{-0.08}$ & 1.000 & $-3.1^{+1.0}_{-0.8}$ & 1.000 & $-1.58$ & 0.214 \\
Sagittarius & $15^{+2}_{-2}$ & $0^{+0.0}_{-0.5}$ & 1.000 & $56^{+18}_{-12}$ & $-0.48^{+0.10}_{-0.15}$ & 1.000 & $-6.0^{+0.9}_{-0.9}$ & 1.000 & $<-5.0$ & 1.000 \\
Sagittarius II & $60^{+10}_{-12}$ & $-1.7^{+0.3}_{-0.6}$ & 0.981 & $96^{+28}_{-12}$ & $-0.5^{+0.2}_{-0.5}$ & 0.994 & $-3.2^{+1.1}_{-1.0}$ & 0.993 & $<-5.0$ & 0.990 \\
Sculptor & $49^{+4}_{-5}$ & $-0.39^{+0.02}_{-0.01}$ & 1.000 & $231^{+26}_{-20}$ & $-2.8^{+0.2}_{-0.3}$ & 1.000 & $-2.7^{+0.2}_{-0.2}$ & 1.000 & $<-5.0$ & 0.976 \\
Segue 1 & $21^{+4}_{-5}$ & $-0.08^{+0.02}_{-0.02}$ & 1.000 & $59^{+34}_{-17}$ & $-0.6^{+0.1}_{-0.3}$ & 1.000 & $-6^{+1}_{-1}$ & 1.000 & $<-5.0$ & 0.998 \\
Segue 2 & $18^{+4}_{-3}$ & $-0.23^{+0.01}_{-0.01}$ & 1.000 & $47^{+4}_{-4}$ & $-0.63^{+0.06}_{-0.06}$ & 1.000 & $-6.7^{+0.4}_{-0.3}$ & 1.000 & $<-5.0$ & 1.000 \\
Sextans & $71^{+3}_{-4}$ & $-0.35^{+0.06}_{-0.06}$ & 1.000 & $120^{+16}_{-10}$ & $-1.78^{+0.02}_{-0.04}$ & 1.000 & $-2.3^{+0.3}_{-0.2}$ & 1.000 & $<-5.0$ & 1.000 \\
Triangulum II & $13^{+2}_{-2}$ & $-1.6^{+0.2}_{-0.2}$ & 1.000 & $114^{+8}_{-8}$ & $-0.73^{+0.07}_{-0.07}$ & 1.000 & $-3.7^{+0.2}_{-0.2}$ & 1.000 & $<-5.0$ & 1.000 \\
Tucana II & $60^{+29}_{-24}$ & $-2.1^{+0.8}_{-1.5}$ & 0.786 & $126^{+73}_{-36}$ & $-0.9^{+0.4}_{-1.1}$ & 0.915 & $-2^{+1}_{-1}$ & 0.947 & $<-5.0$ & 0.863 \\
Tucana III & $1.1^{+0.3}_{-0.2}$ & $-0.54^{+0.03}_{-0.03}$ & 1.000 & $47^{+3}_{-2}$ & $-0.24^{+0.01}_{-0.01}$ & 1.000 & $-7.6^{+0.2}_{-0.2}$ & 1.000 & $<-5.0$ & 1.000 \\
Tucana IV & $39^{+15}_{-15}$ & $-0.55^{+0.18}_{-0.07}$ & 0.994 & $53^{+6}_{-4}$ & $-0.15^{+0.02}_{-0.04}$ & 0.999 & $-5.3^{+1.1}_{-1.0}$ & 1.000 & $<-5.0$ & 0.991 \\
Tucana V & $34^{+15}_{-11}$ & $-0.9^{+0.2}_{-0.4}$ & 1.000 & $72^{+15}_{-10}$ & $-0.27^{+0.08}_{-0.19}$ & 1.000 & $-4.0^{+0.9}_{-0.9}$ & 1.000 & $<-5.0$ & 1.000 \\
Ursa Major I & $89^{+18}_{-46}$ & $-0.9^{+0.8}_{-0.5}$ & 0.997 & $104^{+15}_{-7}$ & $-0.7^{+0.5}_{-1.5}$ & 1.000 & $-4.3^{+0.5}_{-0.4}$ & 1.000 & $<-5.0$ & 0.999 \\
Ursa Major II & $40^{+3}_{-4}$ & $-1.9^{+0.6}_{-1.1}$ & 0.945 & $109^{+65}_{-33}$ & $-0.9^{+0.3}_{-0.6}$ & 0.989 & $-3.5^{+1.1}_{-0.9}$ & 0.997 & $<-5.0$ & 0.976 \\
Ursa Major III & $13.2^{+0.8}_{-0.8}$ & $-0.027^{+0.003}_{-0.000}$ & 1.000 & $36^{+4}_{-4}$ & $-0.31^{+0.03}_{-0.03}$ & 1.000 & $-7.6^{+0.4}_{-0.4}$ & 1.000 & $<-5.0$ & 1.000 \\
Ursa Minor & $64^{+7}_{-7}$ & $-2.5^{+0.4}_{-0.5}$ & 1.000 & $129^{+17}_{-14}$ & $-1.1^{+0.2}_{-0.3}$ & 1.000 & $-4.8^{+0.2}_{-0.2}$ & 1.000 & $<-5.0$ & 1.000 \\
Willman 1 & $18^{+6}_{-3}$ & $-0.36^{+0.06}_{-0.11}$ & 1.000 & $42^{+7}_{-6}$ & $-0.7^{+0.1}_{-0.2}$ & 1.000 & $-6.7^{+0.7}_{-0.6}$ & 1.000 & $<-5.0$ & 1.000 \\
YMCA-1 & $45^{+4}_{-4}$ & $-0.09^{+0.01}_{-0.02}$ & 1.000 & $312^{+36}_{-49}$ & $-2.1^{+0.5}_{-0.6}$ & 0.973 & $-2^{+2}_{-2}$ & 0.858 & $-1.57$ & 0.003 \\

\bottomrule
\end{longtable}
\endgroup
\twocolumn

\section{Important information and recommendations for data users}
\label{app:rec}
In this Appendix, we summarize key considerations for using our orbital parameter catalogue, including recommendations for comparing these results with other observational studies and cosmological simulations.

\begin{itemize}
    \item Parameter robustness: Pericentre distances and times are the best constrained orbital parameters, with typical median uncertainties of $\sigma_{r_{\mathrm{peri}}} \sim 7\,\mathrm{kpc}$ and $\sigma_{t_{\mathrm{peri}}} \sim 0.16\,\mathrm{Gyr}$. In contrast, apocentric parameters have larger uncertainties ($\sigma_{r_{\mathrm{apo}}} \sim 18\text{--}26\,\mathrm{kpc}$ and $\sigma_{t_{\mathrm{apo}}} \sim 0.3\text{--}0.5\,\mathrm{Gyr}$).
    \item Sensitivity to the potential : As shown in Section~\ref{sec:orbparam}, the derivation of orbital parameters is significantly more affected by observational uncertainties than by the choice of potential. However, certain parameters are more sensitive to the potential choice than others. Quantifying this sensitivity using the ratio $\sigma_{\mathrm{pot}} / \sigma_{\mathrm{obs}}$ (see Section~\ref{sec:orbparam}), we find that $t_{\rm peri}$ is the least sensitive orbital parameter to the potential choice ($\sigma_{\mathrm{pot}} / \sigma_{\mathrm{obs}} = 0.16$), followed by  $r_{\rm peri}$, $t_{\rm apo}$ and $r_{\rm apo}$ ($\sigma_{\mathrm{pot}} / \sigma_{\mathrm{obs}}$ = 0.22, 0.25 and 0.25, respectively).
    \item Orbital parameters for individual galaxies: When adopting an orbital parameter for a particular galaxy, we recommend the readers to visually inspect its orbit, in order to assess how well constrained it is. We also recommend verifying this numerically using the values of $f_{\mathrm{peri}}$ and $f_{\mathrm{apo}}$, and require these fractions to be at least 0.5, which implies that the corresponding pericentrer or apocentre is recovered in at least half of the MC realizations. Unless a specific MW--LMC potential configuration is preferred, we recommend averaging parameters across all potentials while accounting for both observational uncertainties and the spread across potential models.
    \item Pericentre time uncertainties: We note that for a reduced number of galaxies, the $t_\mathrm{peri}$ uncertainties can be 0.0 (e.g. for L2M10, Carina II, Carina III, Sagittarius, or Ursa Major III). This occurs because their most recent pericentres are both very recent and well determined, and  fall within the same integration interval in a large number of MC realizations. This is a direct consequence of the chosen time resolution. We therefore advise the reader to adopt as  uncertainty the total integration time for the corresponding potential divided by the number of time intervals (1500), in such cases. 
    \item Aquarius II and Reticulum: the PMs for these two galaxies were obtained from \citet{Battaglia2022}, who advised exercising caution when interpreting their kinematics based on the distribution of member stars in PM space and colour--magnitude diagrams. We extend this caveat to the orbital parameters derived for these two galaxies in our catalogue.
    \item LMC satellites orbits with respect to the MW: When analysing the Galactocentric trajectories of LMC satellites, we recommend adopting the orbital parameters of the LMC itself. Because LMC satellites move as a compact group following their host's trajectory, relative extrema in their individual Galactocentric distances often reflect local orbital motion around the LMC rather than true pericentric or apocentric passages around the MW.
    \item Infall times: Reliable infall time estimates require selecting galaxies with $f_{\mathrm{vir}} \approx 0$, confirming that the vast majority of MC realizations were located outside the MW virial radius at the earliest integration time. Furthermore, users comparing these values with external observational works or cosmological simulations must ensure that consistent virial radius definitions are applied.
    \item Comparisons with cosmological simulations: Evaluating orbit integration results against cosmological simulations requires special care to ensure consistent and meaningful comparisons. Pericentric parameters ($r_{\mathrm{peri}}$ and $t_{\mathrm{peri}}$) are the most robustly derived and least sensitive to the underlying potential models, and should therefore be favoured when conducting comparisons. Additionally, host halo masses for simulated MW (and LMC) analogues should be similar to those considered in this work (see Section~\ref{sec:pot}), as infall times and  binding status are sensitive to total halo mass. In particular, infall times in this work are constrained by our $5\,\mathrm{Gyr}$ backward integration window; galaxies with derived infall times at  $-5\,\mathrm{Gyr}$ should be treated as lower limits on lookback time when compared against cosmological simulations. Furthermore, virial radius definitions must be aligned between this catalogue ($\Delta_{\mathrm{c}} \approx 100$; \citealt{BryanNorman1998}) and the target simulation suite to ensure consistent infall criteria. Finally, comparisons involving LMC satellite populations depend on both the adopted LMC mass model and the satellite identification methodology. We advise performing comparisons considering MW--LMC mass configurations similar to our model suite. When selecting LMC satellite candidates across our catalogue, we recommend averaging the estimated association probabilities ($p_{\mathrm{LMC}}$) across potentials and adopting a threshold of $p_{\mathrm{LMC}} \ge 0.5$, unless a specific MW--LMC configuration is explicitly preferred. As discussed in Appendix~\ref{app:LMCprobtest}, we report association probabilities based on multiple identification criteria (e.g., energy of the satellites with respect to the LMC  distance of the satellites relative to the LMC prior to crossing the MW virial radius), enabling a more direct comparison with various simulation identification methodologies in the literature (e.g. \citealt{SantosSantos2021}).
\end{itemize}

\section{Sensitivity of LMC satellite probabilities to the association criterion}
\label{app:LMCprobtest}

\begin{table*}[]
    \centering
    \caption{LMC satellites probability based on their energy with respect to the LMC}
    \begin{tabular}{lcccccc}
    \toprule
       Satellite &   V23 & L2M10 & L2M10first & L2M11 & L3M10 & L3M11 \\
       \multicolumn{1}{c}{(1)}  & (2) & (3) & (4) & (5) & (6) & (7) \\
    \midrule
    \midrule
Carina II     & 1.0 & 0.997 & 1.0 & 0.992 & 1.0 & 0.996\\
Carina III    & 0.992 & 0.985 & 0.999 & 0.940 & 0.996 & 0.994\\
DELVE 6       & 0.228 & 0.191 & 0.258 & 0.164 & 0.275 & 0.247\\
Horologium I  & 0.789 & 0.609 & 0.778 & 0.578 & 0.805 & 0.730\\
Horologium II & 0.624 & 0.545 & 0.650 & 0.446 & 0.677 & 0.622\\
Hydrus I      & 1.0 & 1.0 & 1.0 & 1.0 & 1.0 & 1.0\\
Phoenix II    & 0.832 & 0.704 & 0.862 & 0.743 & 0.858 & 0.788\\
Pictor II     & 0.998 & 0.997 & 1.0 & 0.995 & 0.998 & 1.0\\
Reticulum II  & 0.769 & 0.492 & 0.718 & 0.437 & 0.821 & 0.699\\
SMC          & 0.956  & 0.806 & 0.912 & 0.800 & 0.955 & 0.844\\
YMCA-1       & 0.938 & 0.972 & 0.970 & 0.928 & 0.965 & 0.969\\

    \bottomrule
    \end{tabular}
    \tablefoot{Columns match those of Table~\ref{tab:LMCprob}.}
    \label{tab:LMCprobE}
\end{table*}

\begin{table*}[]
    \centering
    \caption{LMC satellites probability based on adopting as threshold 1.5 times the tidal radius approximation}
    \begin{tabular}{lcccccc}
    \toprule
       Satellite &   V23 & L2M10 & L2M10first & L2M11 & L3M10 & L3M11 \\
       \multicolumn{1}{c}{(1)}  & (2) & (3) & (4) & (5) & (6) & (7) \\
    \midrule
    \midrule
Carina II &0.998 & 0.995 & 0.999 & 0.980 & 1.0 & 0.995 \\
Carina III &0.927 & 0.876 & 0.926 & 0.778 & 0.966 & 0.979 \\
DELVE 6 &0.188 & 0.147 & 0.171 & 0.124 & 0.208 & 0.205 \\
Horologium I &0.708 & 0.479 & 0.534 & 0.446 & 0.678 & 0.602 \\
Horologium II &0.514 & 0.414 & 0.445 & 0.365 & 0.563 & 0.532 \\
Hydrus I &1.0 & 1.0 & 1.0 & 1.0 & 1.0 & 1.0 \\
Phoenix II &0.797 & 0.69 & 0.808 & 0.738 & 0.844 & 0.779 \\
Pictor II &0.997 & 0.996 & 0.999 & 0.995 & 0.998 & 0.999 \\
Reticulum II &0.756 & 0.467 & 0.673 & 0.405 & 0.812 & 0.647 \\
SMC &0.934 & 0.780 & 0.867 & 0.794 & 0.939 & 0.834 \\
YMCA-1 &0.925 & 0.963 & 0.955 & 0.918 & 0.956 & 0.965 \\ 
    \bottomrule
    \end{tabular}
    \tablefoot{Columns match those of Table~\ref{tab:LMCprob}.}
    \label{tab:LMCprob1.5rt}
\end{table*}

\begin{table*}[]
    \centering
    \caption{LMC satellites probability based on their distance with respect to the LMC at the moment it trespassed the MW virial radius}
    \begin{tabular}{lcccccc}
    \toprule
       Satellite &   V23 & L2M10 & L2M10first & L2M11 & L3M10 & L3M11 \\
       \multicolumn{1}{c}{(1)}  & (2) & (3) & (4) & (5) & (6) & (7) \\
    \midrule
    \midrule
CarinaII     & 0.999 & 0.980 & 0.983 & 0.421 & 0.996 & 0.961 \\
CarinaIII    & 0.783 & 0.784 & 0.936 & 0.786 & 0.920 & 0.959 \\
DELVE6       & 0.158 & 0.135 & 0.191 & 0.120 & 0.186 & 0.200 \\
HorologiumI  & 0.539 & 0.344 & 0.556 & 0.266 & 0.563 & 0.516 \\
HorologiumII & 0.475 & 0.372 & 0.520 & 0.353 & 0.536 & 0.525 \\
HydrusI      &   1.0 &   1.0 &   1.0 &   1.0 &   1.0 &   1.0 \\
PhoenixII    & 0.696 & 0.568 & 0.778 & 0.659 & 0.775 & 0.735 \\
PictorII     & 0.997 & 0.996 & 0.999 & 0.995 & 0.998 & 0.999 \\
ReticulumII  & 0.692 & 0.375 & 0.678 & 0.365 & 0.783 & 0.630 \\
SMC          & 0.909 & 0.717 & 0.876 & 0.769 & 0.930 & 0.820 \\
YMCA-1       & 0.902 & 0.956 & 0.954 & 0.910 & 0.945 & 0.953 \\
    \bottomrule
    \end{tabular}
    \tablefoot{Columns match those of Table~\ref{tab:LMCprob}.}
    \label{tab:LMCprobinfall}
\end{table*}

In Section~\ref{sec:potlmcsat} we derived the association probabilities of dwarf galaxies with the LMC. As noted there, the adopted criterion slightly underestimates these probabilities due to the approximation of the tidal radius. Here we test alternative approaches to assess the robustness of our results.

First, we use the specific energy of the galaxies with respect to the LMC as an association criterion. For a given potential, for each galaxy we estimate $p_{\rm{LMC}}$ as the fraction of MC realizations for which $E_{\mathrm{LMC}} < 0$ at the earliest integration time, when the MW is sufficiently far from the LMC that its influence on the satellites is very limited. This method is similar to the one used in \citet{ErkalBelokurov2020}. The resulting probabilities are listed in Table~\ref{tab:LMCprobE}.

Second, we apply a less restrictive version of our original criterion by requiring two pericentres, or one pericentre and one apocentre, within 1.5 times the tidal radius, in order to account for the potential underestimation of the latter. The factor of 1.5 was chosen as a conservative correction. The resulting probabilities are listed in Table~\ref{tab:LMCprob1.5rt}.

Finally, we apply an alternative criterion based on the distance of each dwarf to the LMC at the moment the LMC crossed the MW virial radius. This is physically motivated by the idea that galaxies already within the LMC tidal radius at that time are genuine pre-existing companions of the LMC, having been part of its group prior to infall. For each galaxy, we compute the probability of association as the fraction of MC realizations  which lie within the LMC tidal radius at the time of the LMC infall. The resulting probabilities are listed in Table~\ref{tab:LMCprobinfall}.

The energy-based criterion tends to yield somewhat higher probabilities, with the differences being mild for galaxies with high $p_{\rm{LMC}}$ and larger for those with lower values. The relaxed tidal radius criterion produces slightly higher probabilities than our fiducial one, but the differences are negligible. The criterion based on the distance of galaxies to the LMC prior to its infall yields similar probabilities, with the only notable exception of Carina II for potential L2M11 (where its probability is nearly half of that derived with the criterion we adopted in this work). We note that when we modify this criterion to estimate probabilities based on the fraction of orbits found within 1.5 times the tidal radius at the LMC infall time, the derived probabilities are slightly larger than the fiducial one.

Neither alternative criterion leads to changes in the list of galaxies identified as candidate LMC satellites, and although the numerical probabilities may vary slightly depending on the adopted criterion, the overall trends remain consistent, supporting the robustness of our classification.

\section{Integration of Tucana IV including the SMC potential}
\label{sec:inttuc}

\begin{table*}[]
    \centering
    \caption{Probability of Tucana IV being a long-term satellite of the LMC, based on its energy with respect to the LMC}
    \begin{tabular}{lccc}
    \toprule
    Base potential & MW + LMC & MW + LMC + SMC1 & MW + LMC + SMC2 \\
     \multicolumn{1}{c}{(1)}  & (2) & (3) & (4)\\

    \midrule
    \midrule
V23         & 0.054 & 0.037 & 0.307 \\
L2M10       & 0.009 & 0.005 & 0.009 \\
L2M10first  & 0.024 & 0.022 & 0.076 \\
L2M11       & 0.123 & 0.112 & 0.087 \\
L3M10       & 0.026 & 0.017 & 0.021 \\
L3M11       & 0.184 & 0.175 & 0.168 \\
    \bottomrule
            
    \end{tabular}
    \tablefoot{Probability of Tucana IV being a long term satellite of the LMC, under different combinations of potentials, with or without SMC, based on the energy criterion described in Appendix~\ref{app:LMCprobtest}. Columns coincide with those of Table~\ref{tab:TucIVprob}.}
    \label{tab:TucIVprobE}
\end{table*}

The orbit  of Tucana IV shows a  significant reorientation due to the  strong gravitational force that it experiences during its approximation to the LMC, regardless of the considered potential. The velocity of Tucana IV relative to the LMC is slightly below  the escape velocity under all the potentials, suggesting that it could be currently bound to it. However, several works  have pointed that the SMC could play a significant role in the orbit of Tucana IV (\citealt{Simon2020, Battaglia2022}), increasing the probability that it is a long lasting satellite of the LMC rather than a possibly captured MW satellite.

In order to try to shed light on the origin of Tucana IV, we reconstructed its orbit including the effect of the SMC. 
We considered the two static models of the SMC from \citet{Patel2020}. The SMC potentials are modelled as Hernquist haloes, given that the SMC baryonic content is expected to be significantly lower than that of the LMC, as a result of their repeated interaction. The first SMC model (SMC1) is described by a Hernquist halo with scale radius of 2.5 kpc, virial radius of 45 kpc and  halo mass of $0.5 \times 10^{10} M_{\odot}$, whereas the second (SMC2) has a scale radius of 8.6 kpc, virial radius of 81 kpc, and halo mass of  $3 \times 10^{10} M_{\odot}$ (see \citealt{Patel2020} for full details). 

The potentials including the MW, LMC, and SMC are constructed as follows. For each of the potentials described in Section~\ref{sec:pot}, we incorporate an SMC model that moves along its orbit throughout the integration time. The SMC orbit is derived for each MW + LMC potential by integrating the nominal orbit accounting for the effect of DF based on the SMC mass of the corresponding model. This yields a final suite of 12 potentials, encompassing all possible combinations of MW, LMC, and SMC models. Finally, Tucana IV orbits are integrated under each of these potentials, following the procedure outlined in Section~\ref{sec:orbintmeth}.

We note that this analysis relies on several approximations. The SMC trajectory is based on the nominal orbit only, neglecting the uncertainties in its phase-space coordinates and treating it as a point-mass particle. We further neglect the reflex motion of the LMC due to the SMC interaction and the mass loss of the SMC over time. Additionally, the LMC model is limited to its DM component; given the recent and close approach of Tucana IV to the inner LMC regions, including the baryonic component could affect the results. 

We expect these limitations to have a minimal impact on our results. Tucana IV has only recently entered the vicinity of the LMC (for reference, at a lookback time of 
$\sim$0.5 Gyr it was still at $\sim$75 kpc from it), meaning that the duration of its interaction with the SMC has been very brief. At earlier times, when the approximations in the SMC orbit become increasingly relevant, Tucana IV was still far from the LMC system and therefore largely unaffected by the details of the SMC trajectory. Moreover, the dominant driver of the orbital evolution of Tucana IV is the LMC itself, with the SMC playing a secondary role.

A further consideration concerns the estimation of the probability of association with the LMC, which as discussed in Section~\ref{app:LMCprobtest}, the adopted criterion tends to yield underestimated probabilities. We therefore re-derived the probabilities for Tucana IV under the MW+LMC+SMC potentials using the energy-based criterion of Appendix~\ref{app:LMCprobtest} (Table~\ref{tab:TucIVprobE}). The resulting probabilities are generally larger, but remain low enough to confidently exclude Tucana IV as a long-term LMC satellite. The highest probability obtained is 0.307, for the V23+SMC2 combination, which corresponds to the lowest LMC mass and the most massive SMC model. We note, however, that SMC2 has a mass significantly above current SMC estimates, making it an unrealistically extreme case. Even so, the results suggest that the SMC could play a relevant role only under very specific and unlikely mass configurations.

\end{appendix}
\end{document}